\documentclass[aps,pra,reprint,groupedaddress,longbibliography]{revtex4-1}

\usepackage{graphicx,color}
\usepackage{amsmath, amssymb}
\usepackage{longtable}
\usepackage{natbib}
\usepackage{bm}
\usepackage{hyperref}

\allowdisplaybreaks[1]

\begin{document}
\title{
Spin-orbital-lattice entangled states in cubic $d^1$ double perovskites
}
\author{Naoya Iwahara}
\email[]{naoya.iwahara@gmail.com}
\author{Veacheslav Vieru}
\author{Liviu F. Chibotaru}
\email[]{liviu.chibotaru@gmail.com}
\affiliation{Theory of Nanomaterials Group,
University of Leuven,
Celestijnenlaan 200F, B-3001 Leuven, Belgium}
\date{\today}

\begin{abstract}
Interplay of spin-orbit coupling and vibronic coupling on heavy $d^1$ site of cubic double perovskites is investigated by {\it ab initio} calculations. 
The stabilization energy of spin-orbital-lattice entangled states is found comparable to or larger than the exchange interactions, suggesting the presence of Jahn-Teller dynamics in the systems.  
In Ba$_2$YMoO$_6$, the pseudo Jahn-Teller coupling enhances the mixing of the ground and excited spin-orbit multiplet states, which results in strong temperature dependence of effective magnetic moments.
The entanglement of the spin and lattice degrees of freedom induces a strong magneto-elastic response. 
This multiferroic effect is at the origin of the recently reported breaking of local point symmetry accompanying the development of magnetic ordering in Ba$_2$NaOsO$_6$.
\end{abstract}

\maketitle

\section{Introduction}
The geometrically frustrated systems with strong spin-orbit coupling on metal sites are of great interest in the context of unconventional electronic phases
\cite{Witczak-Krempa2014, Rau2016}. 
The double perovskites containing heavy $d^1$ metal ions are candidates for spin liquid systems, the reason for which they have been intensively investigated
\cite{Stitzer2002, Cussen2006, Erickson2007, Xiang2007, deVries2010, Aharen2010, Carlo2011, Steele2011, deVries2013, Coomer2013, Qu2013, Gangopadhyay2015, Marjerrison2016, Xu2016, Ahn2017, Lu2017, Liu2017}. 
Although the interplay of spin and orbital degrees of freedom has been widely studied theoretically \cite{Chen2010, Dodds2011, Ishizuka2014, Natori2016, Natori2017, Romhanyi2017, Svoboda2017}, the understanding on the role of lattice degrees of freedom in these systems is lacking. 
In cubic $d^1$ double perovskite Ba$_2$YMoO$_6$, in spite of four-fold degeneracy of the local ground multiplet ($\Gamma_8$ or effective $J = \frac{3}{2}$), the Jahn-Teller (JT) distortion \cite{Jahn1938} has not been observed in neutron diffraction measurements down to 2.7 K, which was called ``violation of the JT theorem'' \cite{Aharen2010}. 
Similarly, the x-ray diffraction shows that cubic symmetry of Ba$_2A$OsO$_6$ ($A = $ Li, Na) \cite{Stitzer2002} is retained even at 5 K \cite{Erickson2007}, while recent NMR spectra of Ba$_2$NaOsO$_6$ suggest the development of ``broken local point symmetry'' around and below Curie temperature ($\approx 10$ K) \cite{Liu2017, Lu2017}.
The absence of the clear-cut JT distortion is most likely explained by either quenching of the JT effect 
or the presence of the dynamical JT effect. 
The signs for the latter are seen, for example, in alkali-doped fullerides \cite{Chibotaru2005, Iwahara2013, Iwahara2015} 
and various metal compounds \cite{Krimmel2005, Nakatsuji2012, Kamazawa2017, Nirmala2017}.
Since the JT effect can give nontrivial influence on electronic properties, the knowledge of its relevance at local metal sites is indispensable for understanding the nature of these materials. 

In this work, on the example of three cubic $d^1$ double perovskites (Ba$_2A$OsO$_6$, $A=$ Li, Na, and Ba$_2$YMoO$_6$),
the local electronic properties generated by the interplay of spin-orbit interaction and vibronic coupling is studied. 
With the use of coupling parameters derived {\it ab initio}, the spin-orbital-lattice coupled states were accurately calculated. 
The dynamical JT stabilization comparable to or larger than Curie-Weiss constants indicates the persistence of vibronic dynamics in the crystals. 
The analysis of the local magnetic moment and response to the magnetic field reveals the reasons for the large increase of effective moment with temperature in Ba$_2$YMoO$_6$ and for the local symmetry breaking in Ba$_2$NaOsO$_6$.

\section{Electronic and vibronic model for $d^1$ systems}
The electronic structure of a $d^1$ metal ion at octahedral site is described by ligand field $\hat{H}_\text{LF}$, spin-orbit interaction $\hat{H}_\text{SO}$ and vibronic coupling $\hat{H}_\text{JT}$: 
\begin{eqnarray}
 \hat{H} &=& \hat{H}_\text{LF} + \hat{H}_\text{SO} + \hat{H}_0 + \hat{H}_\text{JT} + \hat{H}_\text{Zee},
\label{Eq:h}
\end{eqnarray}
where, $\hat{H}_0$ is the Hamiltonian for harmonic oscillation, and $\hat{H}_\text{Zee}$ is the Zeeman interaction in applied magnetic field $\bm{B}$.
The typical energy scales of $\hat{H}_\text{LF}$, $\hat{H}_\text{SO}$, $\hat{H}_\text{JT}$, and $\hat{H}_\text{Zee}$ under $|\bm{B}| \approx$ 10 T are several eV, 0.1 eV, 0.01 eV, and $10^{-4}$-$10^{-3}$ eV, respectively, and should be treated in this order.

The ligand field $\hat{H}_\text{LF}$ splits the atomic $d$ level into $e_g$ and $t_{2g}$, the latter being stabilized in octahedral environment \cite{Sugano1970}.
Due to the large ligand-field splitting, the low-energy states are well described in the space of $t_{2g}^1$ electron configurations.
Since the orbital angular momentum on sites $\hat{\bm{l}}$ is not quenched, the spin-orbit coupling is operative already in the first order \cite{Kotani1960, Sugano1970}:
\begin{eqnarray}
\hat{H}_\text{SO} = \lambda_\text{SO} \tilde{\bm{l}} \cdot \hat{\bm{s}}.
\label{Eq:hSO}
\end{eqnarray}
Here, $\lambda_\text{SO} > 0$ is spin-orbit coupling parameter, $\tilde{\bm{l}}$ is $\tilde{l}=1$ effective orbital angular momentum operator of the $t_{2g}$ orbitals, and $\hat{\bm{s}}$ is the electron spin.
$\tilde{\bm{l}}$ behaves as $-\hat{\bm{l}}_p$, where $\hat{\bm{l}}_p$ is the orbital angular momentum for $p$ orbitals \cite{Kotani1960, Sugano1970}.
The spin-orbit coupling $\hat{H}_\text{SO}$ splits the six-fold $t_{2g}^1$ configurations into $\Gamma_7$ ($J = \frac{1}{2}$) and $\Gamma_8$ ($J = \frac{3}{2}$) multiplets \cite{Koster1963}.
The latter is the ground state separated from the former by $\frac{3}{2}\lambda_\text{SO}$.

Because of the unquenched orbital momentum, the magnetic moment on the metal sites becomes
\begin{eqnarray}
 \hat{\bm{m}} &=& -\mu_\text{B} \left( \langle l\rangle \tilde{\bm{l}} + g_e \hat{\bm{s}} \right),
\label{Eq:mu_t2}
\end{eqnarray}
where, $\mu_\text{B}$ is Bohr magneton, $\langle l\rangle$ is the expectation value of $\hat{\bm{l}}$, and $g_e$ is the electron's $g$-factor.
Since $\tilde{\bm{l}}$ is opposite to $\hat{\bm{l}}_p$, the orbital and spin contributions partially cancel each other \cite{Kotani1949, Kotani1960, Sugano1970}. 
This cancellation is almost complete in the atomic limit, $\langle l \rangle = 1$, while it is not in crystals because of covalency effects ($\langle l \rangle < 1$).

The $t_{2g}$ orbitals also couple to the $e_g$ and $t_{2g}$ lattice vibrations \cite{Jahn1937, Bersuker1989, Kaplan1995}:
\begin{eqnarray}
 \hat{H}_\text{JT} &=& 
 \sum_{k} \sum_{n\Lambda \lambda} \sum_{\Lambda_1, \Lambda_2 \cdots \Lambda_k} 
 \frac{1}{k!} v^{\Lambda_1 \Lambda_2 \cdots \Lambda_k}_{n\Lambda}
\nonumber\\
 &\times&
  \{\hat{Q}_{\Lambda_1} \otimes \hat{Q}_{\Lambda_2} \otimes \cdots \otimes \hat{Q}_{\Lambda_k} \}_{n\Lambda \lambda}
 \hat{\tau}_{\Lambda \lambda}.
\label{Eq:hJT}
\end{eqnarray}
Here, $\Lambda$ ($\Lambda_i$) is $e_g$ or $t_{2g}$, $\lambda$ is its component, $n$ distinguishes the repeated representation, 
$\hat{Q}_{\Lambda \lambda}$ is the (mass-weighted) normal coordinate, $\{\hat{Q}_{\Lambda_1} \otimes \hat{Q}_{\Lambda_2} \otimes \cdots \otimes \hat{Q}_{\Lambda_k} \}_{\Lambda \lambda}$ is the symmetrized product of coordinates, $v^{\Lambda_1 \Lambda_2 \cdots \Lambda_k}_{n\Lambda}$ is the $k$-th order orbital vibronic coupling parameter, 
and $\hat{\tau}_{\Lambda \lambda}$ are the matrices of Clebsch-Gordan coefficients.

For the details of the model Hamiltonian, see Appendix \ref{A:H}.

\begin{table}[tb]
\begin{ruledtabular} 
\caption{
Spin-orbit coupling parameter $\lambda_\text{SO}$ (meV), orbital angular momenta $\langle l \rangle$, vibrational frequency $\omega_\Lambda$ (meV) and vibronic coupling parameters $v^{\Lambda_1 \cdots \Lambda_k}_{n\Lambda}$ (a.u.).
(1),(2) for $\langle l\rangle$ stand for post HF and DFT values, respectively.
}
\label{Table:V}
\begin{tabular}{cccc}
 & Ba$_2$LiOsO$_6$ & Ba$_2$NaOsO$_6$ & Ba$_2$YMoO$_6$ \\
\hline
$\lambda_\text{SO}$     & 379.9 &  384.5  &   88.1 \\
$\langle l \rangle$ (1) & 0.772 &  0.779  &  0.859 \\
$\langle l \rangle$ (2) & 0.551 &  0.562  &  0.643 \\
$\omega_E$              & 99.10 & 100.94  & 100.50 \\
$\omega_{T_2}$          & 50.37 &  49.43  &  46.52 \\
$v_E$              & $-3.2998 \times 10^{-4}$ & $-2.6436 \times 10^{-4}$ & $ 1.0199 \times 10^{-4}$ \\
$v_{T_2}$          & $ 0.4963 \times 10^{-4}$ & $ 0.4945 \times 10^{-4}$ & $ 0.7334 \times 10^{-4}$ \\
$v^{EE}_E$         & $-1.6604 \times 10^{-5}$ & $-1.2693 \times 10^{-5}$ & $ 0.0821 \times 10^{-5}$ \\
$v^{T_2T_2}_E$     & $-0.0850 \times 10^{-5}$ & $-0.0842 \times 10^{-5}$ & $-0.0646 \times 10^{-5}$ \\
$v^{T_2T_2}_{T_2}$ & $-0.0011 \times 10^{-5}$ & $-0.0006 \times 10^{-5}$ & $ 0.0002 \times 10^{-5}$ \\
$v^{ET_2}_{T_2}$   & $-0.0388 \times 10^{-5}$ & $-0.0289 \times 10^{-5}$ & $ 0.0248 \times 10^{-5}$ \\
$v^{EEE}_{A_1}$    & $ 0.1151 \times 10^{-6}$ & $ 0.1032 \times 10^{-6}$ & $ 0.0009 \times 10^{-6}$ \\
$v^{EEE}_{E}$      & $ 2.7380 \times 10^{-6}$ & $ 2.0917 \times 10^{-6}$ & $ 0.0290 \times 10^{-6}$ \\
$v^{EEEE}_{A_1}$   & $-0.3158 \times 10^{-7}$ & $-0.2011 \times 10^{-7}$ & $ 0.0112 \times 10^{-7}$ \\
$v^{EEEE}_{1E}$    & $ 2.5417 \times 10^{-7}$ & $ 1.6786 \times 10^{-7}$ & $-0.0362 \times 10^{-7}$ \\
$v^{EEEE}_{2E}$    & $ 0.6240 \times 10^{-7}$ & $ 0.4204 \times 10^{-7}$ & $-0.0004 \times 10^{-7}$ \\
\end{tabular}
\end{ruledtabular} 
\end{table}

\section{Ab initio derivation of coupling parameters}
\label{Sec:Abinitio}
The spin-orbit $\lambda_\text{SO}$ and vibronic $v^{\Lambda_1 \cdots \Lambda_k}_{n\Lambda}$ coupling parameters were derived from the cluster calculations with post Hartree-Fock (HF) methods, while $\langle l \rangle$'s were extracted by both (1) post HF and (2) density functional theory (DFT) calculations (see Appendices \ref{A:Abinitio} and \ref{A:DFT}).
The obtained parameters are listed in Table \ref{Table:V}.

The vibronic coupling in Ba$_2A$OsO$_6$ is stronger than in Ba$_2$YMoO$_6$.
Particularly, the nonlinear vibronic coupling to the $e_g$ modes is about 10-100 times stronger in the former compounds.
Moreover, the vibronic coupling parameters of Ba$_2A$OsO$_6$ differ from each other. 
The strength of the vibronic coupling is determined by overlap between the $t_{2g}$ orbital with the distribution of the vibronic operator. 
The former depends on the type of $d$ orbital ($4d$ or $5d$) and also environment such as metal-oxygen bond length
\footnote{
The Os-O distance in Ba$_2$LiOsO$_6$ is longer by 0.019 {\AA} than that in Ba$_2$NaOsO$_6$.
}, resulting in different vibronic coupling parameters.

In all compounds, the expectation value of the orbital angular momentum, $\langle l \rangle$, is reduced from unity due to the delocalization of the $t_{2g}$ electron over ligands.
As expected, the DFT values are smaller by 25-30 \% than post HF values since the latter underestimates metal-ligand covalency. 
The present DFT value of $\langle l \rangle$ for Ba$_2$NaOsO$_6$ is close to the previous calculation, 0.536 \cite{Ahn2017}.
The spin-orbit coupling parameters $\lambda_\text{SO}$ from the {\it ab initio} calculations are also in good agreement with previous calculations \cite{Xu2016}.
However, as in the case of $\langle l \rangle$, $\lambda_\text{SO}$ may be overestimated by post HF calculations. 
Since both $\langle l \rangle$ and $\lambda_\text{SO}$ are mainly contributed by the $d$ metal orbitals, the covalent reduction of the latter should be similar to the former.

\begin{figure}[tb]
\begin{center}
\begin{tabular}{lll}
(a) \\
\multicolumn{3}{c}{
\includegraphics[height=4.5cm]{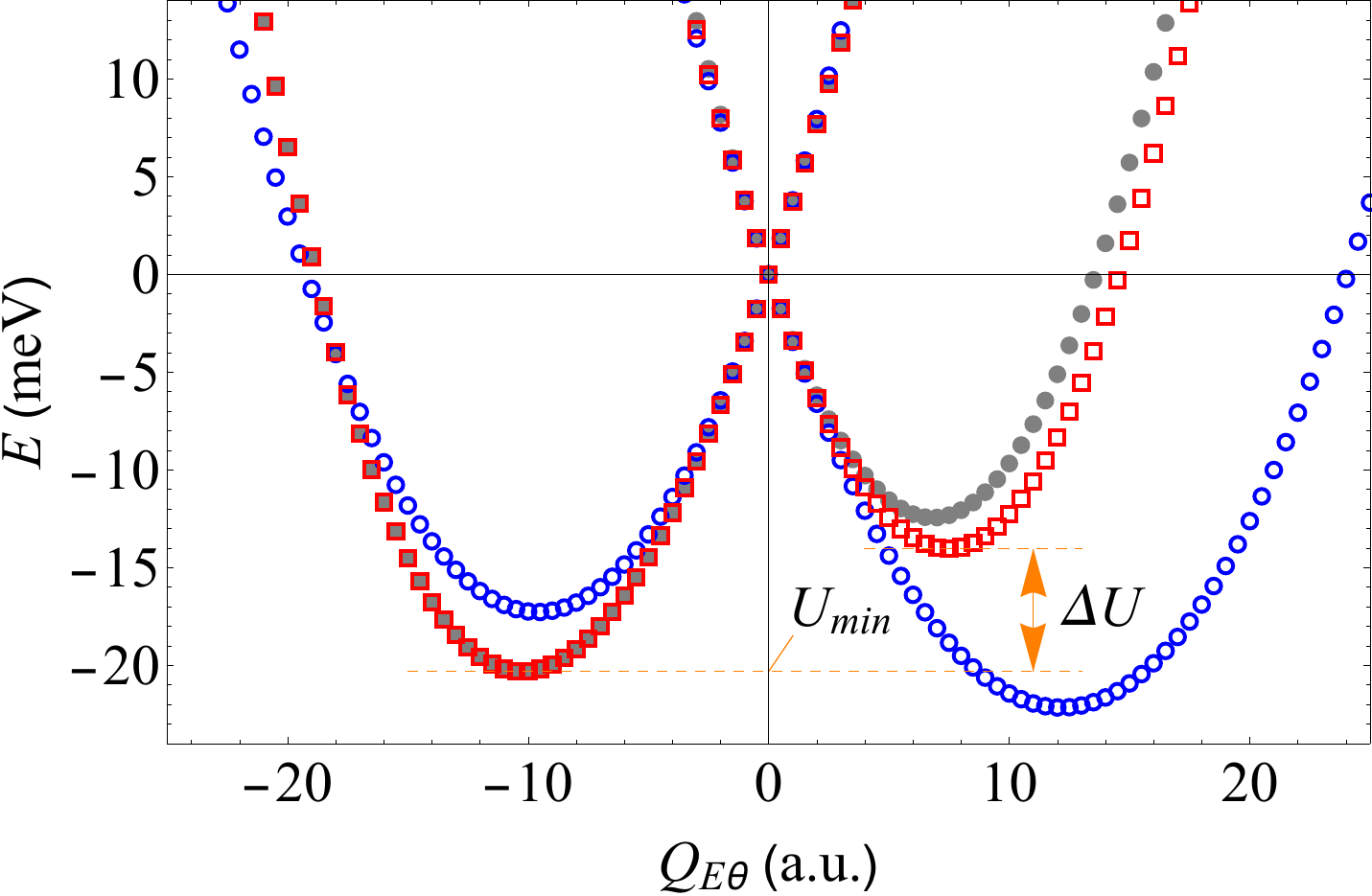}
}
\\
(b) & ~~~~ &(c) \\
\multicolumn{1}{c}{
\includegraphics[height=2.5cm]{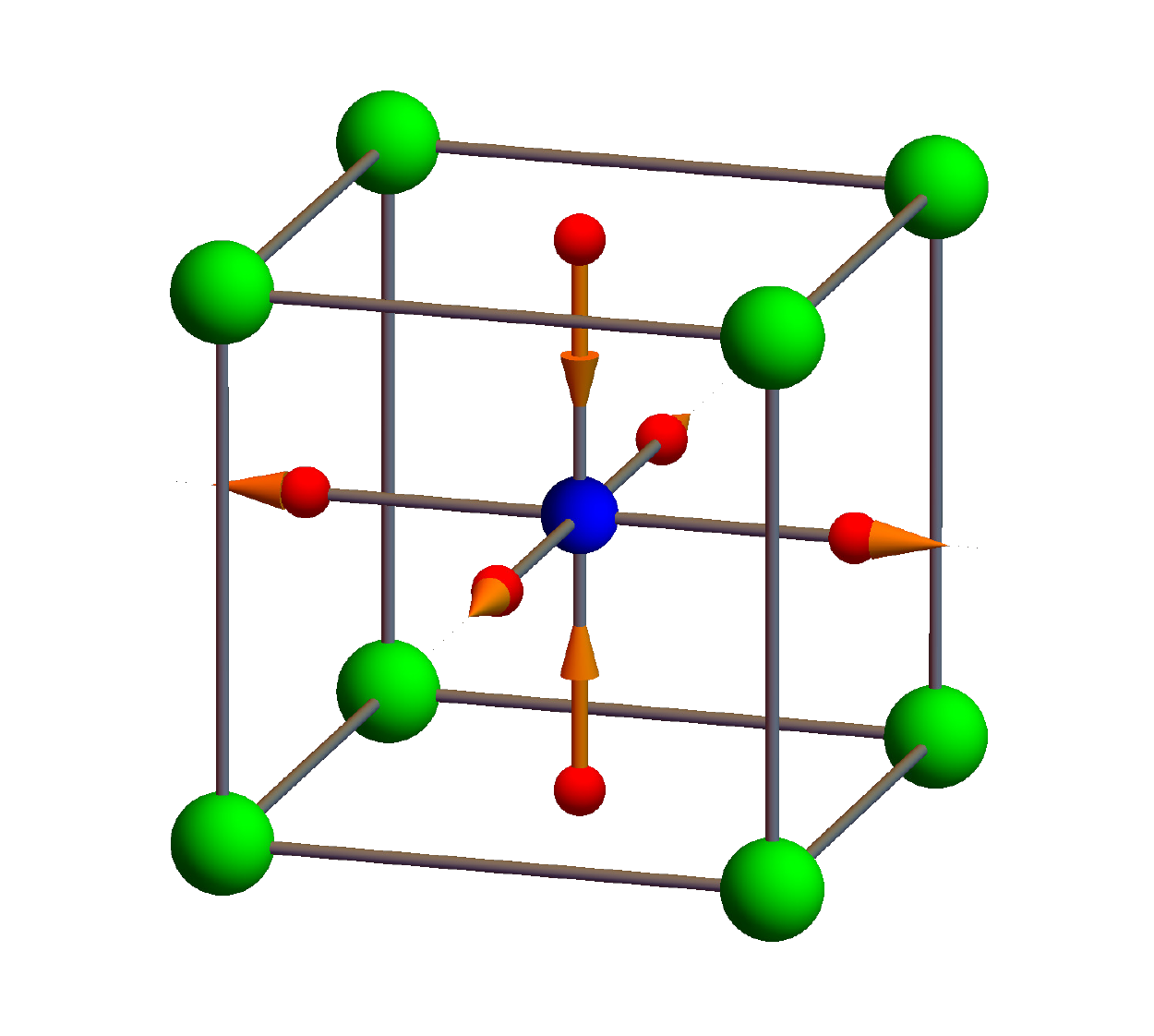}
}
&
&
\multicolumn{1}{c}{
\includegraphics[height=2.5cm]{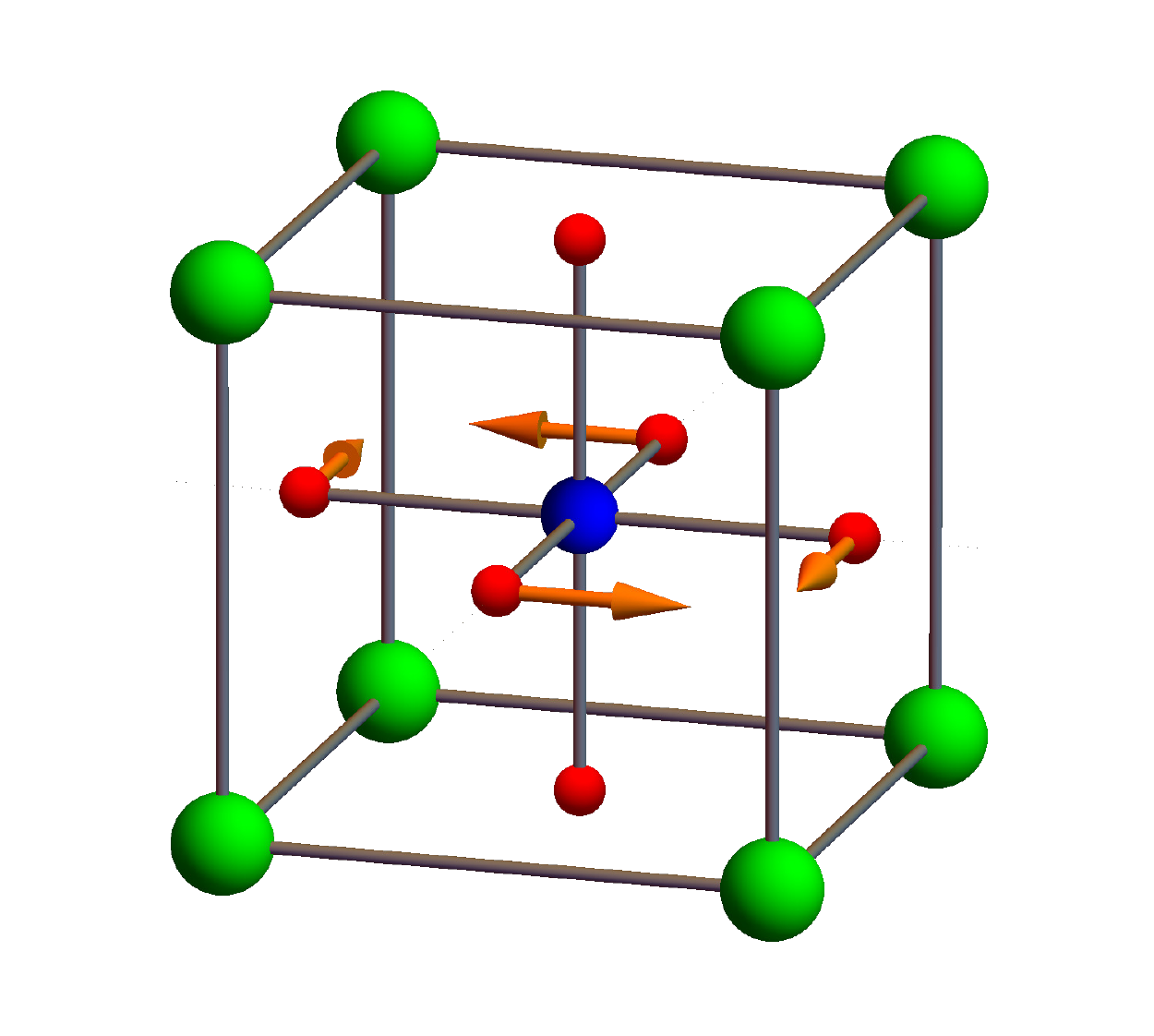}
}
\end{tabular}
\end{center}
\caption{
The APES along the $e_g\theta$ distortion for Ba$_2$NaOsO$_6$ (a), JT distortions for Os (b) and Mo (c) compounds. 
In (a) the red squares and gray filled circles indicate the APES with and without pseudo JT coupling, respectively, and the blue open circles the APES of linear JT model with pseudo JT coupling. 
In (b,c), the orange arrows show the direction of the displacements of oxygen atoms at the minima of APES. 
}
\label{Fig:JTDist}
\end{figure}

\begin{table}[tb]
\begin{ruledtabular}
\caption{The position of one of the minima (a.u.) and energy (meV). 
$\delta l_\text{max}$ indicates the largest displacement of oxygen atom under JT deformation, and $U_\text{min}$ and $\Delta U$ the JT stabilization energy and the energy barrier between the minima and saddle points, respectively [see Fig. \ref{Fig:JTDist}]. 
}
\label{Table:SJT}
\begin{tabular}{cccc}
    & Ba$_2$LiOsO$_6$ & Ba$_2$NaOsO$_6$ & Ba$_2$YMoO$_6$ \\
\hline
$Q_{E\theta}$           & $-12.29$  & $-10.24$  &   1.42  \\
$Q_{T_2\zeta}$          &    0.00   &    0.00   &  11.46  \\
$\delta l_\text{max}$   &   0.022   &   0.018   &  0.018  \\
$U_\text{min}$          & $-32.22$  & $-20.31$  & $-4.84$ \\
$\Delta U$              &  10.86    &    6.27   &   0.67  \\
\end{tabular}
\end{ruledtabular}
\end{table}

\section{Jahn-Teller effect}
\subsection{Static Jahn-Teller deformation}
\label{Sec:SJTE}
The derived parameters show that the energy scales for the multiplet splitting and vibrational frequencies $\omega_\Lambda$ are comparable, particularly in the case of Ba$_2$YMoO$_6$. 
This makes relevant the pseudo JT effect between $\Gamma_7$ and $\Gamma_8$ multiplets, along with JT effect in each of them. 
Indeed, the adiabatic potential energy surfaces (APES) along the JT distortion with and without pseudo JT coupling show non-negligible differences: 
the APES of $\Gamma_8 \otimes e_g$ JT model (gray circles) is modified (red squares) even in the case of Ba$_2A$OsO$_6$ with large $\lambda_\text{SO}$ [Fig. \ref{Fig:JTDist}(a)].
Thus, for the adequate description of $d^1$ site, the consideration of the full $(\Gamma_7 \oplus \Gamma_8) \otimes (e_g \oplus t_{2g})$ JT coupling is essential. 
Fig. \ref{Fig:JTDist}(a) also shows the unexpectedly strong effect of nonlinear vibronic coupling: the positions of the minima and saddle point of the APES within linear model (blue circles) are inverted by nonlinear coupling (red squares).

The global minima and saddle points of the APES were investigated as in simpler case of $\Gamma_8 \otimes (e_g \oplus t_{2g})$ JT problem \cite{Liehr1963} (for details, see \cite{SM}).
The results are summarized in Table \ref{Table:SJT}.
The static JT distortions for Ba$_2A$OsO$_6$ develop only along the $e_g$ mode [Fig. \ref{Fig:JTDist}(b)]. 
The JT stabilization energies $|U_\text{min}|$ are 32.2 and 20.3 meV 
\footnote{
The present static JT stabilization energies are larger by 1.5-3 times for Ba$_2A$OsO$_6$ and 10 times smaller for Ba$_2$YMoO$_6$ than the previous {\it ab initio} values obtained by the ``$z$-axis-only compression'' (10, 15, and 40 meV, respectively) \cite{Xu2016}.
There are two reasons for the discrepancy.
(i) According to Fig. 2b in Ref. \cite{Xu2016}, the ``$z$-axis-only compression'' is a linear combination of $a_{1g}$ and $e_g\theta$ modes.
Since the experimental crystal structure is not fully relaxed in the sense of {\it ab initio} treatment, the contribution from the $a_{1g}$ mode is included in the stabilization energy after the $e_g$ mode.
(ii) Relatively large numerical noise might modify the calculated vibronic coupling parameters due to the large deviation of the {\it ab initio} energy from their fitting curve (Fig. 2c in Ref. \cite{Xu2016}).
}
and the energy barriers $\Delta U$ between the minima and the saddle points in the bottom of the APES are only 10.9 and 6.3 meV for $A=$ Li and Na, respectively [see for $A =$ Na Fig. \ref{Fig:JTDist}(a)].
On the contrary, in Ba$_2$YMoO$_6$ the $t_{2g}$ distortion is dominant [Fig. \ref{Fig:JTDist}(c)]. 
The stabilization energy is only 4.8 meV \cite{Note2} and the energy barrier at the trigonal point ($Q_{T_2\xi} = Q_{T_2\eta} = Q_{T_2\zeta}$) is about 0.7 meV.

In all materials, the largest shifts $\delta l_\text{max}$ of oxygen atom by the static JT deformation were obtained about 0.02 {\AA} which is larger than the experimental resolution
\footnote{
The resolution of neutron scattering measurement of Ba$_2$YMoO$_6$ at 2.7 K is 0.2 \% of the lattice constant ($\approx 0.0167$ {\AA}) and that of the room-temperature x-ray scattering data is half of it \cite{Aharen2010}.
In NMR measurement, the distortion of Ba$_2$NaOsO$_6$ is expected between 0.002-0.066 {\AA} \cite{Lu2017}. 
}, 
which at a glance seems to be contradictory to the absence of the symmetry lowering in the structural data.
However, because of the small warping of the trough $\Delta U$, the dynamical Jahn-Teller effect \cite{Bersuker1989, Kaplan1995}, which causes the delocalization of the nuclear wave function over the trough, has to be fully taken into account.

\begin{figure}[tb]
\includegraphics[height=7.5cm]{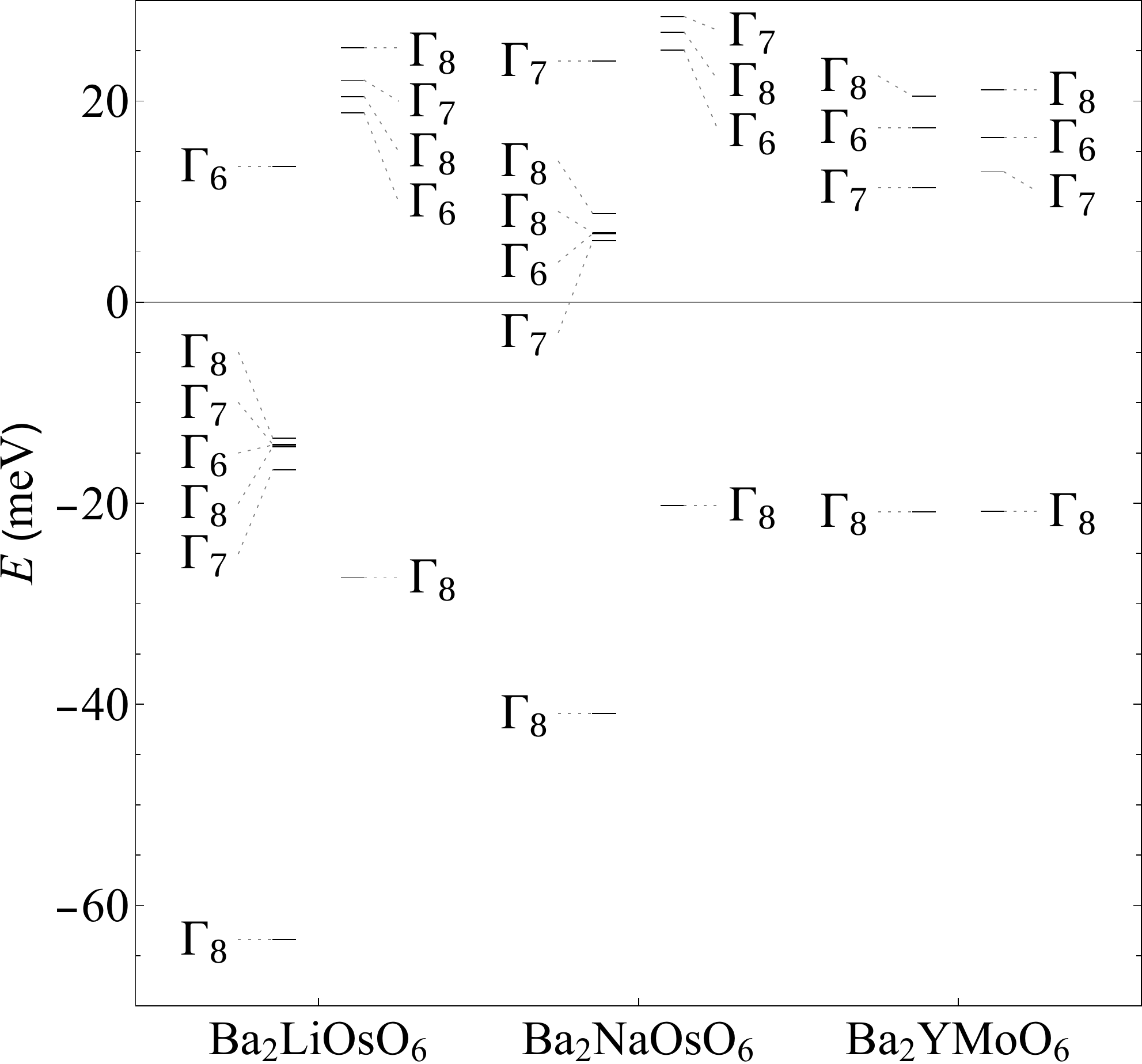}
\caption{Low-energy vibronic levels in double perovskites (meV).
Left and right columns for each compound are the vibronic levels without and with nonlinear vibronic coupling.
The ground energy without vibronic coupling is taken as zero of energy. 
}
\label{Fig:E}
\end{figure}

\subsection{Spin-orbital-lattice entangled states}
\label{Sec:SOLentanglement}
The vibronic eigenstates of the $(\Gamma_7 \oplus \Gamma_8) \otimes (e_g \oplus t_{2g})$ JT system have spin-orbital-lattice entangled form:
\begin{eqnarray}
 |\Psi_{\alpha\Lambda M}\rangle &=& \sum_{\Gamma = \Gamma_7, \Gamma_8} \sum_{N} |\Gamma N\rangle \otimes |\psi_{\Gamma N,\alpha \Lambda M}\rangle,
\label{Eq:Psi}
\end{eqnarray}
where, $\alpha$ is the principal quantum number, $|\Gamma N\rangle$ is the multiplet state, and $|\psi_{\Gamma N,\alpha \Lambda M}\rangle$ is the nuclear part.
The latter is expanded into the eigenstates of harmonic oscillators \cite{Kahn1972, Kahn1975, Iwahara2017}.
The vibronic states (\ref{Eq:Psi}) were obtained by numerical diagonalization of the JT Hamiltonian (see Appendix \ref{A:Lanczos}).

Fig. \ref{Fig:E} shows the vibronic levels without and with nonlinear vibronic coupling for each system. 
In Ba$_2A$OsO$_6$, the nonlinear vibronic coupling significantly destabilizes the linear vibronic states, which is explained by the rise of the minima of APES and reduced magnitude of distortion at the minima [Fig. \ref{Fig:JTDist}(a)]
\footnote{The latter leads to the mixing of the excited linear vibronic states into the ground one. The ground $\Gamma_8$ states mainly consist of the ground linear $\Gamma_8$ vibronic states (87.2 \% for Ba$_2$LiOsO$_6$ and  93.5 \% for Ba$_2$NaOsO$_6$) and the excited ones involving 2-4 vibrational excitations at 46.9 and 78.1 meV, respectively (9.8 \% and 5.2 \%).}.
The resultant dynamical JT stabilizations are several times larger than the exchange interactions measured by Curie-Weiss constants $\Theta$ ($|\Theta| =$ 3.5 meV for $A =$ Li and 0.9-2.8 meV for $A =$ Na \cite{Stitzer2002, Erickson2007}). 
Moreover, a rather weak intersite elastic interaction is expected because OsO$_6$ octahedrons have no common ligand atoms. 
The absence of clear static cooperative JT effect (or orbital ordering) \cite{Erickson2007} is explained by its destruction by the unquenched JT dynamics, as is also the case in fullerides \cite{Iwahara2013}.
The presence of the JT dynamics means that the phase of the materials should be described in terms of the spin-orbital-lattice entangled states instead of spin or spin-orbit coupled states. 
Thus, for example, the low-temperature ordered phase \cite{Stitzer2002} is not simple magnetic one but that of spin-orbital-lattice entangled states. 
The impact of the difference on the physical properties will be discussed in Sec. \ref{Sec:Multiferroicity}.

In the case of Ba$_2$YMoO$_6$, the energy gain by the JT dynamics amounts to as much as three times the static JT energy (as usual in weak JT regime \cite{Bersuker1989}).
The dynamical JT stabilization is comparable in magnitude to the reported $\Theta$'s ($|\Theta|$ ranges from 7.8 to 18.9 meV \cite{Cussen2006, Aharen2010, deVries2010, deVries2013, Qu2013}. see Table \ref{Table:Meff}), implying non-negligible contribution of the JT dynamics to the low-energy states as discussed above.
The first excited states arise at ca 30 meV above the ground one (Fig. \ref{Fig:E}), which should be put in correspondence to the excitation at ca 28 meV observed in inelastic neutron scattering measurements \cite{Carlo2011}. 
The presence of the JT dynamics does not contradict the temperature evolution of the infrared spectra which was attributed to the classical JT distortion at low temperature \cite{Qu2013}. 
Indeed, a similar temperature dependence of infrared spectra of fullerides was explained on the basis of dynamical JT effect \cite{Matsuda2018}.

\begin{figure}[tb]
\begin{tabular}{l}
(a) \\
\includegraphics[height=5cm]{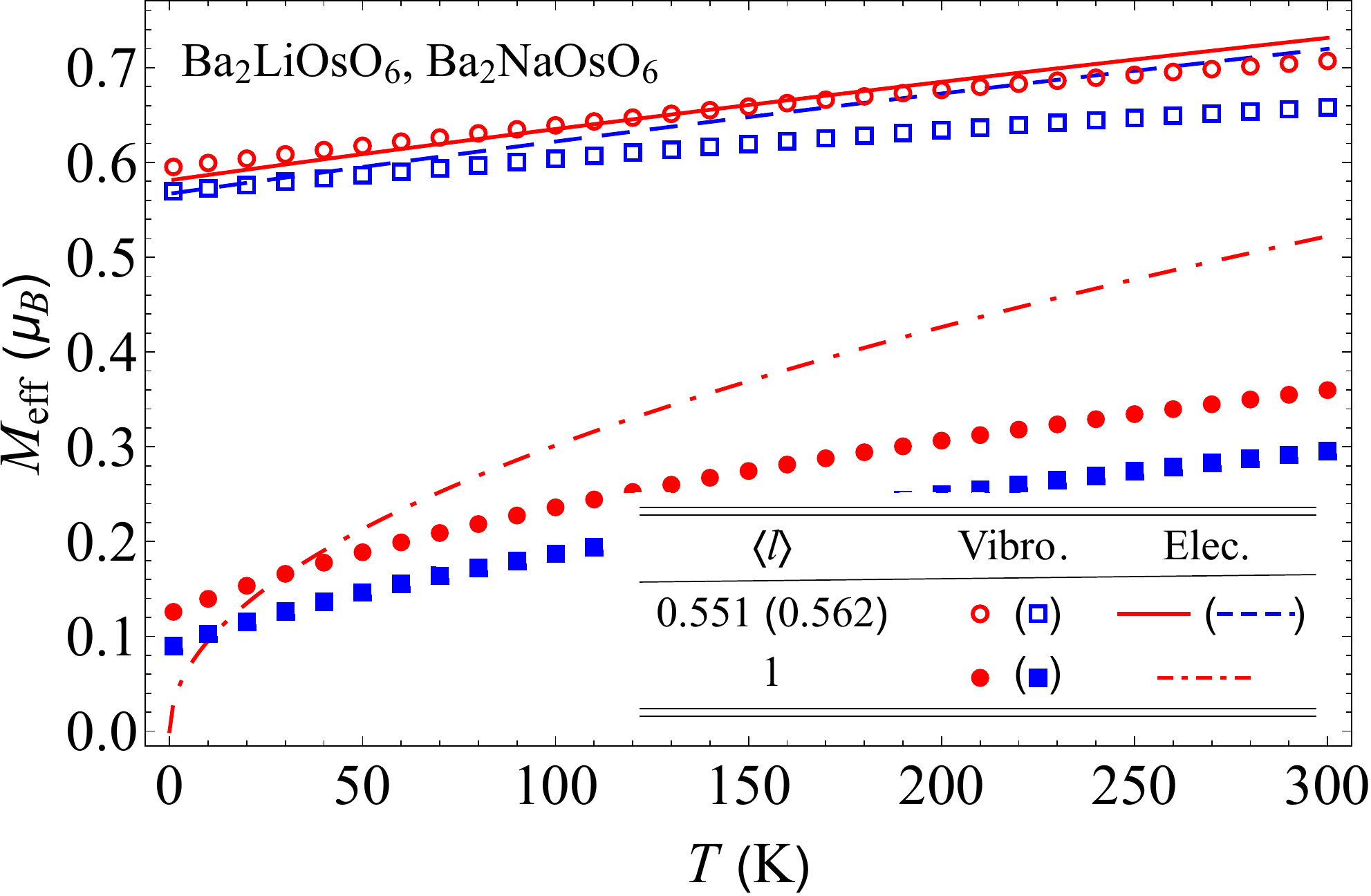}
\\
(b) \\
\includegraphics[height=5cm]{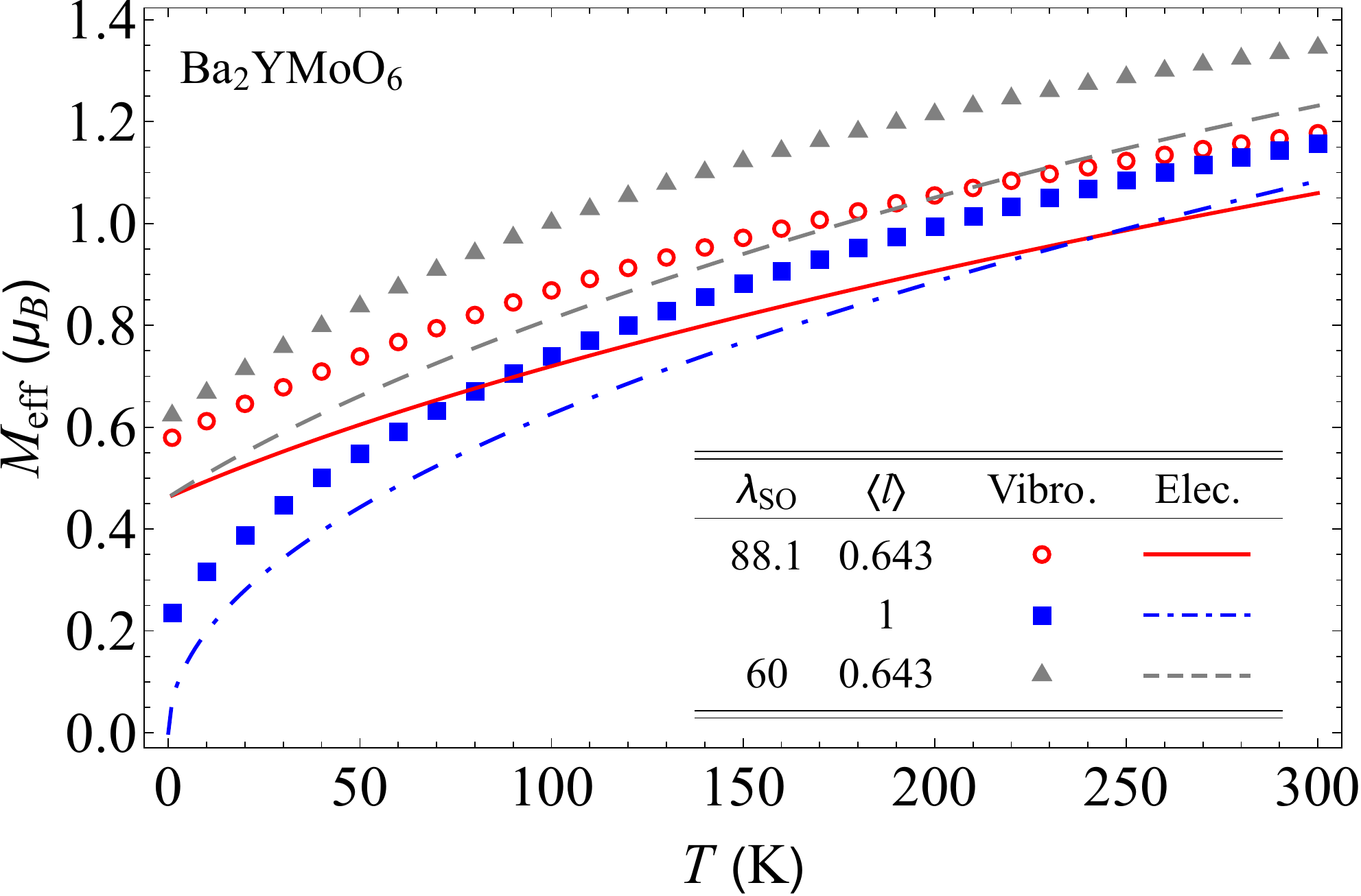}
\end{tabular}
\caption{Effective magnetic moments $M_\text{eff}$ as function of temperature for (a) Os and (b) Mo compounds.
The meaning of the symbols or lines is given in corresponding insets.
``Vibro.'' and ``Elec.'' stand for $M_\text{eff}$ calculated for dynamical JT states and for pure electronic multiplet states, respectively,
and $\langle l \rangle$ (also $\lambda_\text{SO}$ for Mo system) is the value used for the simulation. 
(a) The red circles, solid and dot dashed lines are for Ba$_2$LiOsO$_6$, and the blue squares and dashed line are for Ba$_2$NaOsO$_6$.
}
\label{Fig:Meff}
\end{figure}

\begin{table*}[tb]
\caption{Theoretical and experimental effective magnetic moments $M_\text{eff}$ ($\mu_\text{B}$) and Curie-Weiss temperature $\Theta$ (meV).
$T$ (K) indicates the temperature for the simulation or measurements.}
\label{Table:Meff}
\begin{ruledtabular}
\begin{tabular}{cccccc}
     & $M_\text{eff}$ & $\Theta$ & $T$ & Method & Ref. \\
\hline
\multicolumn{6}{c}{--- High $T$ ---} \\
Ba$_2$LiOsO$_6$ & 0.707 & - & 300 & Theor. (Vibro.) & Present \\ 
                & 0.733 & $-3.5$ & 150-300 & $\chi$ & \cite{Stitzer2002} \\ 
Ba$_2$NaOsO$_6$ & 0.658 & - &     300 & Theor. (Vibro.) & Present \\
                & 0.677 & $-2.8$ & 150-300 & $\chi$ & \cite{Stitzer2002} \\
                & 0.596-0.647 & $-0.9$ - $-1.3$ &  75-200 & $\chi$ & \cite{Erickson2007} \\
Ba$_2$YMoO$_6$  & 1.351 & - & 300 & Theor. (Vibro.) & Present \\
                & 1.231 & - & 300 & Theor. (Elec.) & Present \\
                & 1.34 &  $-7.8$ & 220-300 & $\chi$ & \cite{Cussen2006} \\
                & 1.41 & $-13.8$ & 150-300 & $\chi$ & \cite{deVries2010} \\
                & 1.72 & $-18.9$ & 150-300 & $\chi$ & \cite{Aharen2010} \\
                & 1.44 & $-12.3$ & 160-390 & $\chi$ & \cite{deVries2013} \\
                & 1.52 & $-14.8$ & 150-300 & $\chi$ & \cite{Qu2013}\\
\multicolumn{6}{c}{--- Low $T$ ---} \\
Ba$_2$LiOsO$_6$ & 0.595 & - & 0 & Theor. (Vibro.) & Present \\ 
Ba$_2$NaOsO$_6$ & 0.569 & - & 0 & Theor. (Vibro.) & Present \\
                & $\approx$ 0.6   & - & $\alt 10$ & NMR & \cite{Lu2017} \\
Ba$_2$YMoO$_6$  & 0.624 & - & 0 & Theor. (Vibro.) & Present \\
                & 0.462 & - & 0 & Theor. (Elec.) & Present \\
                & 0.53 & - & $<$ 40 & NMR &  \cite{Aharen2010} \\
                & 0.59 & - & $<$ 25 & $\chi$ & \cite{deVries2013} \\
                & 0.57 & - & $<$ 25 & $\chi$ &  \cite{Qu2013}\\
\end{tabular}
\end{ruledtabular}
\end{table*}

\section{Magnetic properties}
In the present systems, the vibronic states (\ref{Eq:Psi}) inherit the paramagnetic properties from the spin-orbit multiplets.
Particularly, because of the entanglement, the lattice degrees of freedom becomes also relevant to the magnetism. 
Below, two aspects are investigated. 

\subsection{Effective magnetic moment}
\label{Sec:Meff}
The effective magnetic moment $M_\text{eff}$ derived from the magnetic susceptibility $\chi$ at high temperature ($T > |\Theta|$) is expected to be close to that of a single $d^1$ site because the influence of intersite interactions in this case can be neglected.
The temperature dependence of $M_\text{eff}$ was calculated with (points) and without (lines) dynamical JT effect (Fig. \ref{Fig:Meff}).
At $T = 0$ K, $M_\text{eff}$ arises from the $\Gamma_8$ vibronic states only, and as temperature rises it grows due to Van Vleck's second order contribution \cite{Kotani1949, Kotani1960}. 
The vibronic coupling influences $M_\text{eff}$ in two ways:
(i) JT coupling to $\Gamma_8$ multiplet modifies (often reduces) the matrix element of the electronic operator \cite{Child1961, Ham1968, Iwahara2017} and 
(ii) pseudo JT coupling mixes the $\Gamma_7$ and $\Gamma_8$ multiplets.
In the present case, the admixture of the $\Gamma_7$ multiplet with large magnetic moment leads to the enhancement of $M_\text{eff}$.

In the case of Ba$_2A$OsO$_6$, due to the strong $\hat{H}_\text{SO}$, the Van Vleck contribution is small and the temperature dependence of $M_\text{eff}$ is weak [Fig. \ref{Fig:Meff}(a)]. 
The theoretical values with DFT $\langle l \rangle$ are in good agreement with the experimental data both at low and high temperature:
$M_\text{eff} \approx 0.6 \mu_\text{B}$ at $T\approx 0$ K \cite{Lu2017} and 0.60-0.68$\mu_\text{B}$ at high-$T$ \cite{Stitzer2002, Erickson2007} for Ba$_2$NaOsO$_6$ and 0.73$\mu_\text{B}$ at high-$T$ for Ba$_2$LiOsO$_6$ \cite{Stitzer2002}.
As widely accepted \cite{Aharen2010, Gangopadhyay2015, Xu2016, Ahn2017}, the $d$-$p$ hybridization of Os and O atoms enlarges electronic $M_\text{eff}$ (compare the data for the DFT-derived $\langle l \rangle$ and the atomic value).
The JT dynamics slightly quenches the variation of $M_\text{eff}$ [Fig. \ref{Fig:Meff}(a)]. 

On the other hand, in Ba$_2$YMoO$_6$, the pseudo JT coupling plays a crucial role to enhance $M_\text{eff}$ [Fig. \ref{Fig:Meff}(b)].
At $T=0$ K, $M_\text{eff}$ amounts to 0.6$\mu_\text{B}$, which is in line with NMR 0.53$\mu_\text{B}$ \cite{Aharen2010} and low-$T$ susceptibility 0.57-0.59$\mu_\text{B}$ 
\footnote{
NMR probes local properties whereas the magnetic susceptibility reflects the macroscopic properties. 
Thus, the effect of the surroundings might be included in the low-$T$ values from $\chi$.
}
\cite{deVries2013, Qu2013}, and it rapidly grows with $T$.
Taking into account the covalency effect on both $\lambda_\text{SO}$ and $\langle l \rangle$ (gray triangles), $M_\text{eff}$ at $T \approx 300$ K reaches the experimental values (1.3-1.5$\mu_\text{B}$ \cite{Cussen2006, deVries2010, deVries2013, Qu2013}).

\begin{figure}[tb]
\begin{center}
\begin{tabular}{ll}
(a) & (b) \\
\includegraphics[height=3.8cm]{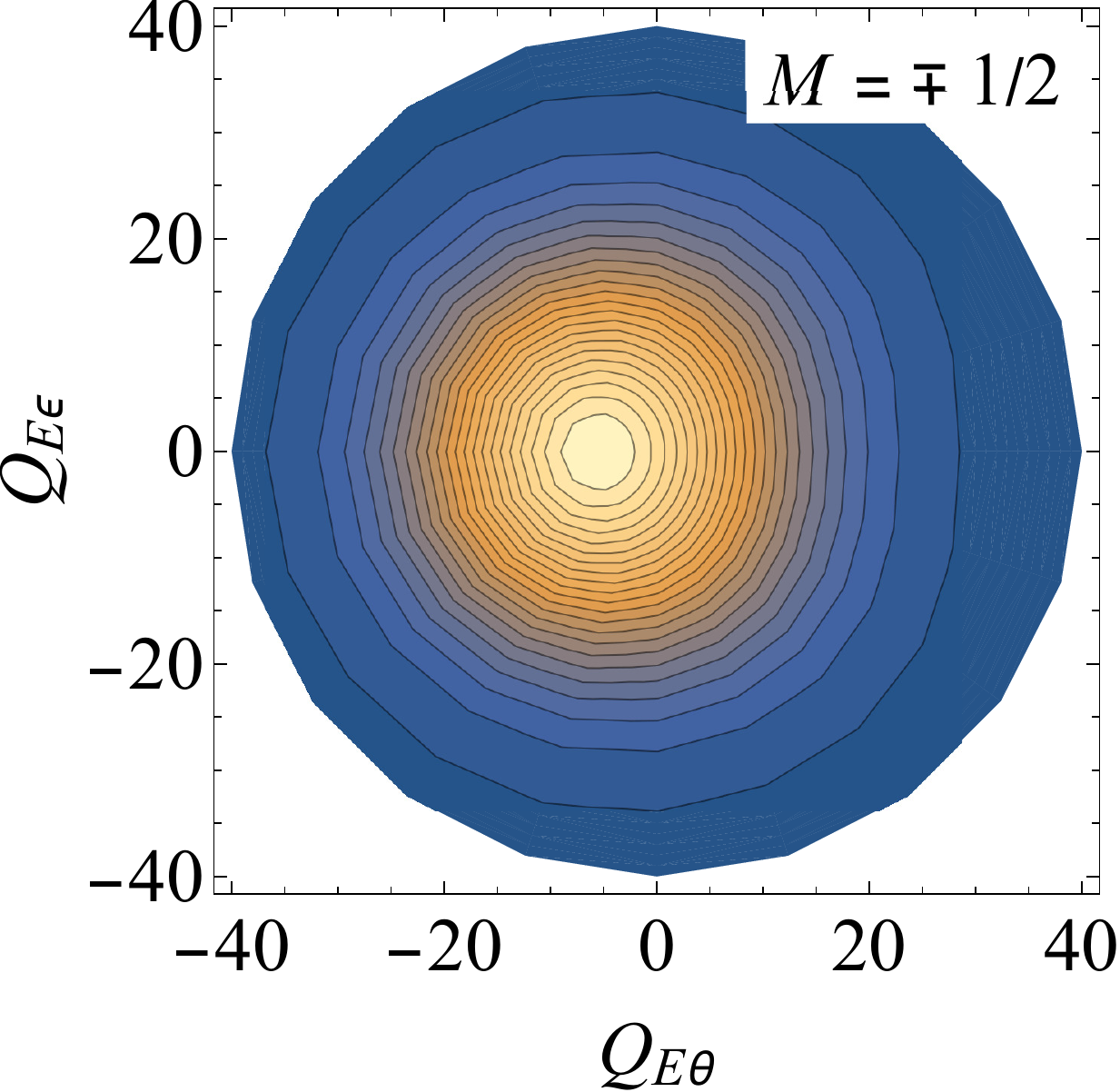}
&
\includegraphics[height=3.8cm]{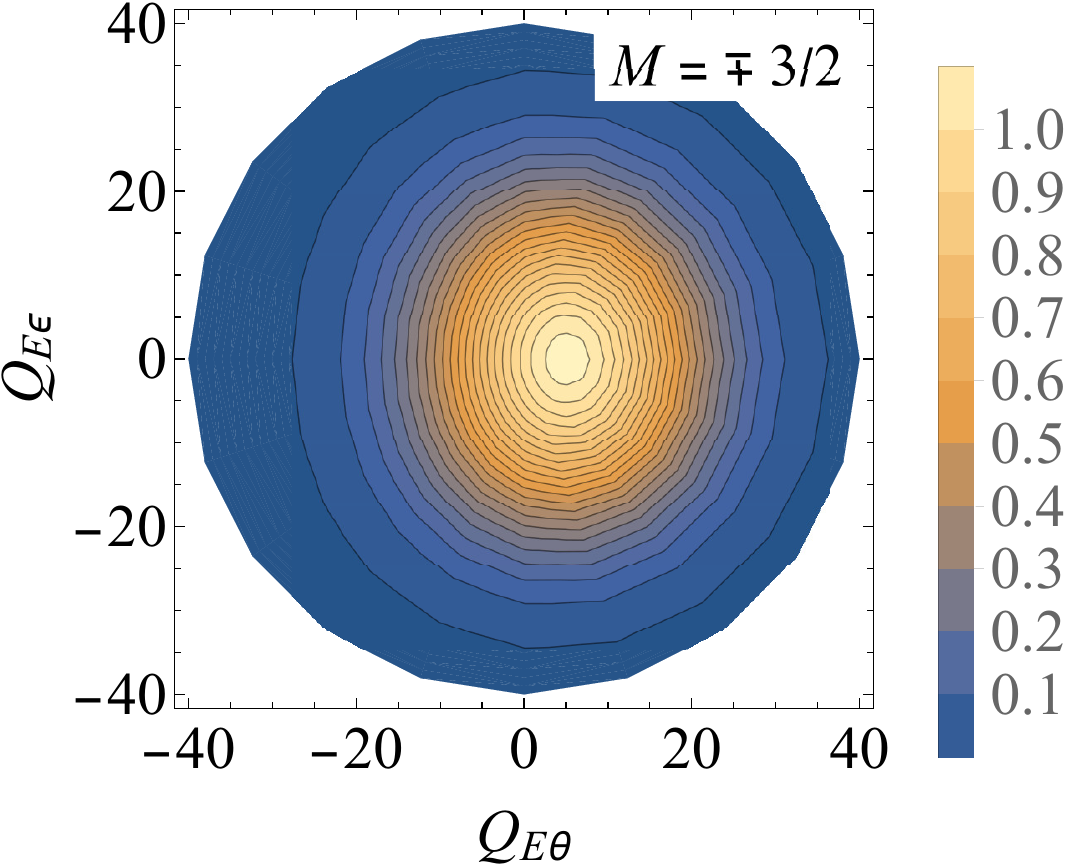}
\\
(c) \\
\multicolumn{2}{c}{
\includegraphics[height=4.5cm]{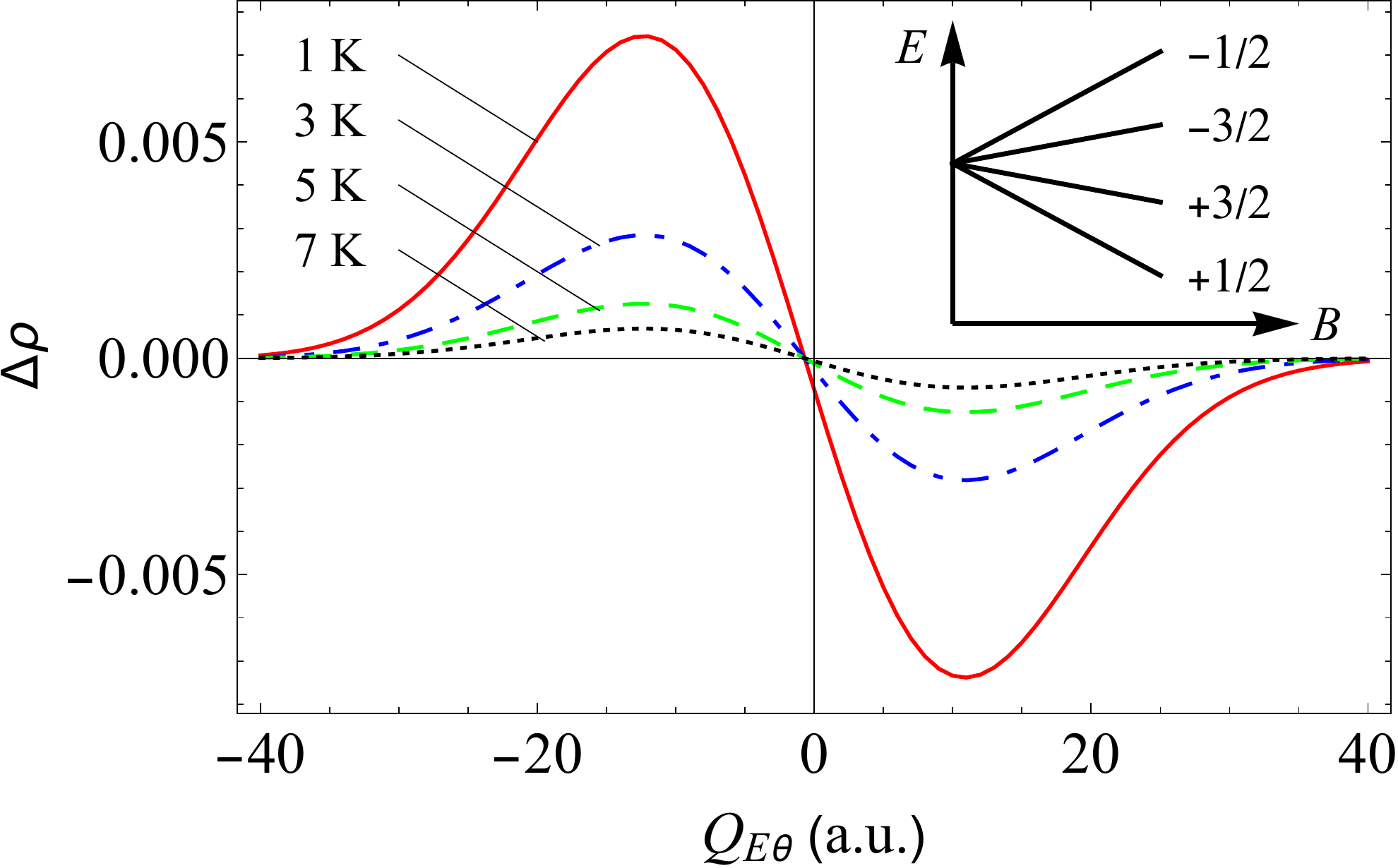}
}
\end{tabular}
\end{center}
\caption{
(a), (b) $\rho_{\Gamma_8 M}(Q_\theta, Q_\epsilon)$ for $M = \mp \frac{1}{2}$ and $\mp \frac{3}{2}$, respectively, and (c) the temperature evolution of the distribution of vibronic states for Ba$_2$NaOsO$_6$. 
In (a), (b), the contour lines are drawn from 0 with increment of 1/20. 
In (c), the distribution with respect to averaged one $\Delta \rho$ is shown under applied field of $|\bm{B}| = 15$ T.
The inset shows the Zeeman splitting of the $\Gamma_8$ vibronic state. 
For the description of $\Delta \rho$, see Appendix \ref{A:rho}.
}
\label{Fig:DJTDist}
\end{figure}

\subsection{Spin-orbital-lattice entanglement driven magneto-elastic response}
\label{Sec:Multiferroicity}
A peculiarity of the present systems is that 
the Zeeman splitting is accompanied by the variation of the $Q_{\Lambda\lambda}$ distribution in the ground $\Gamma_8$ vibronic state, 
\begin{eqnarray}
\rho_{\Gamma_8 M}(\bm{Q}) = 
\langle \Psi_{\Gamma_8M} | \bm{Q} \rangle \langle \bm{Q} |\Psi_{\Gamma_8M}\rangle,
\end{eqnarray}
where, $\bm{Q}$ is a set of normal coordinates. 
Under $\bm{B}\parallel[001]$ the ground $\Gamma_8$ level split as in the inset of Fig. \ref{Fig:DJTDist}(c).
In the case of Os compounds, slight localization at the minima in the APES is observed for $|\Psi_{\Gamma_8, \mp \frac{1}{2}}\rangle$, and around the saddle point for $|\Psi_{\Gamma_8, \mp \frac{3}{2}}\rangle$ [see for the case of Ba$_2$NaOsO$_6$ Fig. \ref{Fig:DJTDist}(a), (b)].
Thus, with the increase of temperature, the center of the distribution $\rho$ shifts from the minima of APES to the symmetric point [Fig. \ref{Fig:JTDist}(c). See for the definition of $\rho$ Appendix \ref{A:rho}].

The temperature evolution of $\rho$ must be related to the observation of the ``broken local point symmetry'' in Ba$_2$NaOsO$_6$ \cite{Lu2017, Liu2017}. 
The expectation value of the JT distortion is reduced by the JT dynamics, and it is consistent with the expected small JT deformation in the NMR study as well as x-ray data \cite{Erickson2007}.
Besides the applied field, the exchange interaction between $d^1$ centers enhances the Zeeman splitting in the presence of magnetic order, which would cause the reduction of the magnetic entropy to $k_\text{B} \ln 2$ \cite{Erickson2007}. 
Thus, the ordering in Ba$_2$NaOsO$_6$ \cite{Lu2017} is not a conventional orbital ordering with classical (static) JT distortions but an ordering of spin-orbital-lattice entangled states. 

Contrary to the Os compounds, in Ba$_2$YMoO$_6$ no magnetic order develops down to 50 mK \cite{deVries2013}.
Despite the stronger exchange interaction \cite{Cussen2006, Aharen2010, deVries2010, deVries2013, Qu2013} than in Ba$_2A$OsO$_6$, the absence of the ordering hinders the large Zeeman splitting of vibronic levels, and hence, the dynamical JT effect develops as supported by neutron diffraction data \cite{Aharen2010} and sustains the magnetic entropy of $k_\text{B} \ln 4$ observed by muon spin resonance \cite{deVries2010}.

\section{Conclusion}
\label{Sec:Conclusion}
The local spin-orbital-lattice entangled states of three cubic $d^1$ double perovskites were derived based on the first principle approach. 
The gain of the energy of the ground coupled states is larger than (Ba$_2A$OsO$_6$) or comparable to (Ba$_2$YMoO$_6$) 
the corresponding Curie-Weiss constants, suggesting the presence of dynamical JT effect in these materials. 
Due to the mixing with spin degrees of freedom, the vibronic states respond strongly to the magnetic field. 
Thus, the first excited vibronic level at $\approx 30$ meV in Ba$_2$YMoO$_6$ suggests its relevance to the magnetic excitations measured in inelastic neutron scattering.
In this compound, the vibronic coupling involving both $\Gamma_8$ and $\Gamma_7$ multiplets gives rise to strong temperature dependence of the effective magnetic moment. 
In Ba$_2A$OsO$_6$, the entanglement gives rise to magneto-elastic response where a small static component of the dynamical JT deformation accompanies the Zeeman splitting, which explains the ``breaking of local point symmetry''.

The relevance of the spin-orbital-lattice entanglement is expected in other cubic $d^1$ \cite{Cussen2006, Coomer2013, Marjerrison2016} and $d^2$ double perovskites \cite{Yamamura2006, Aharen2010d2, Thompson2014, Marjerrison2016d2, Feng2016}, and also in other types of cubic crystals such as $5d^1$ Ta chlorides persisting cubic symmetry down to low temperature \cite{Ishikawa2018}.
For the complete understanding of the unconventional magnetic phases of the family of cubic double perovskites containing heavy transition metal, concomitant treatment of the vibronic and magnetic interactions is found to be crucial. 
The ordering of the spin-orbital-lattice entangled states would be a new direction towards unconventional multifunctional materials. 

\section*{Acknowledgement}
N.I. is supported by Japan Society for the Promotion of Science Overseas Research Fellowship.
V. V. acknowledges the postdoctoral fellowship of the Fonds Wetenschappelijk Onderzoek-Vlaanderen (FWO, Flemish Science Foundation).

\appendix

\section{Derivation of the model Hamiltonian}
\label{A:H}
Here, the transformation of the model Hamiltonian for $t_{2g}^1$ ion in an octahedral environment into the one in the basis of spin-orbit coupled states is shown in detail. 
In this section, the operators for the spin-orbital decoupled and coupled states by lower and upper cases, respectively,
the subscript $g$ of the representations is omitted in the equations for simplicity, and the coordinate axes of the $d^1$ system are chosen to correspond to $C_4$ axes.

\begin{figure}[tb]
\begin{center}
\begin{tabular}{lll}
(a) $e_g \theta$ & (b) $e_g \epsilon$ & (c) $t_{2g} \zeta$ \\
\includegraphics[width=2.5cm]{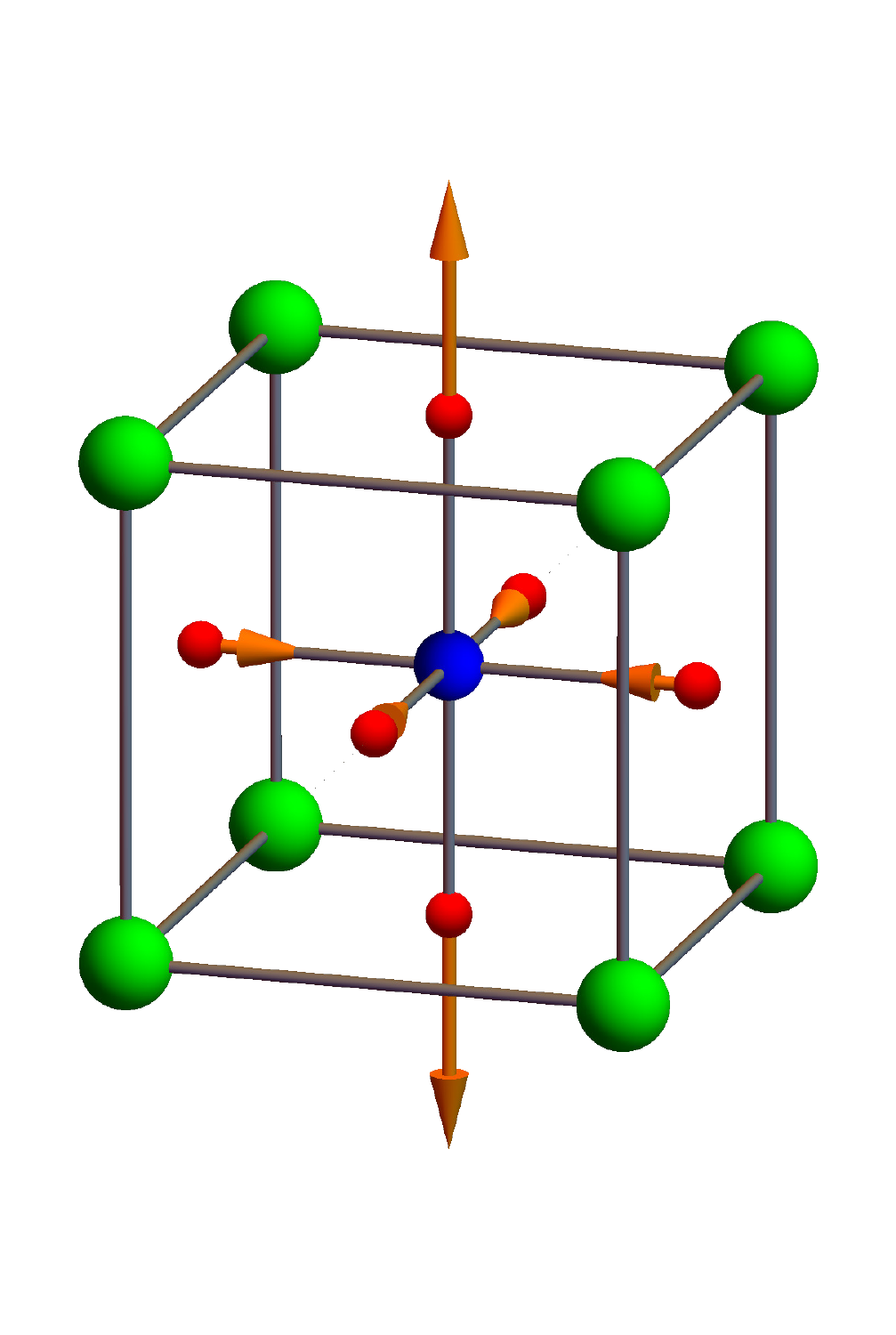}
&
\includegraphics[width=2.5cm]{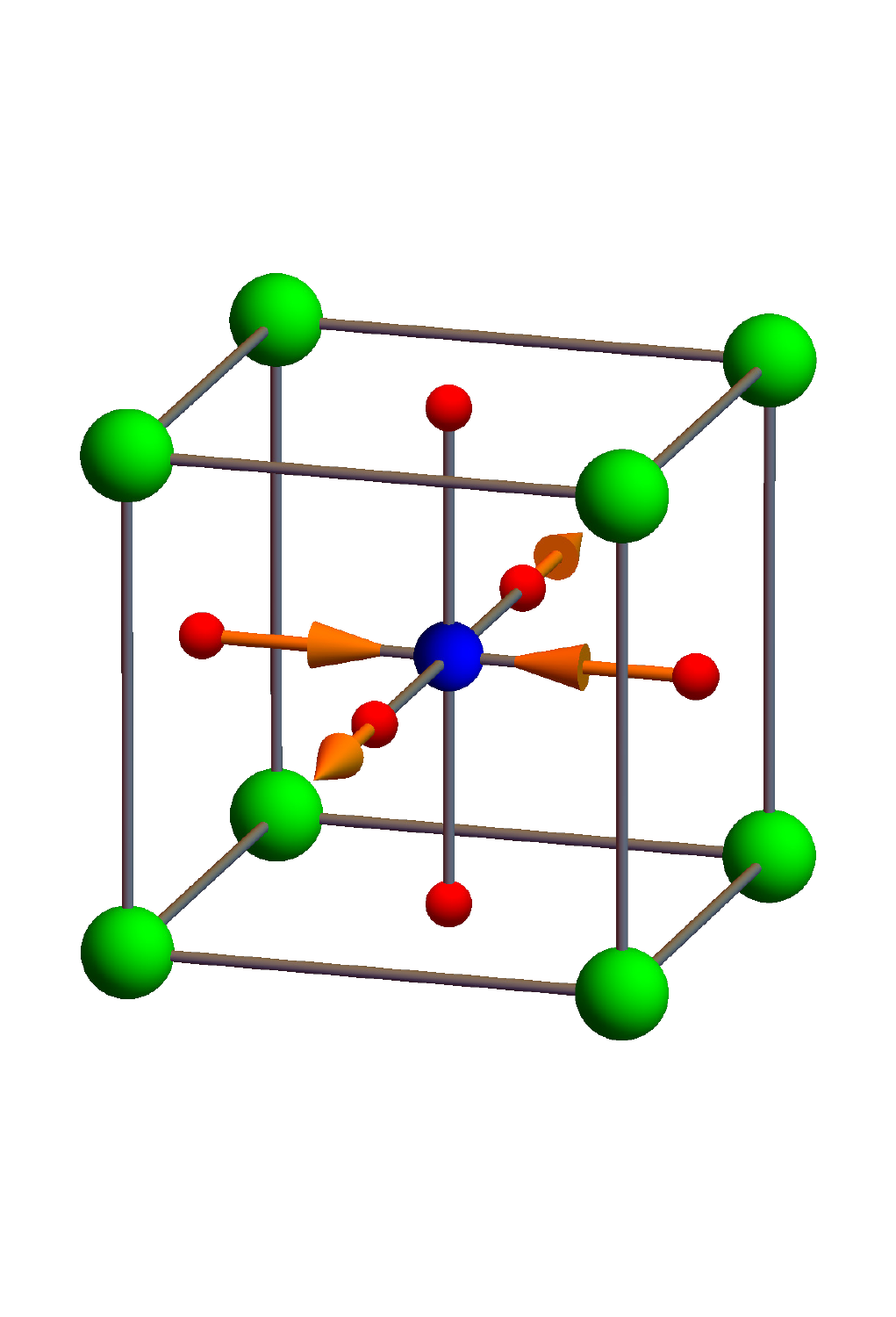}
&
\includegraphics[width=2.5cm]{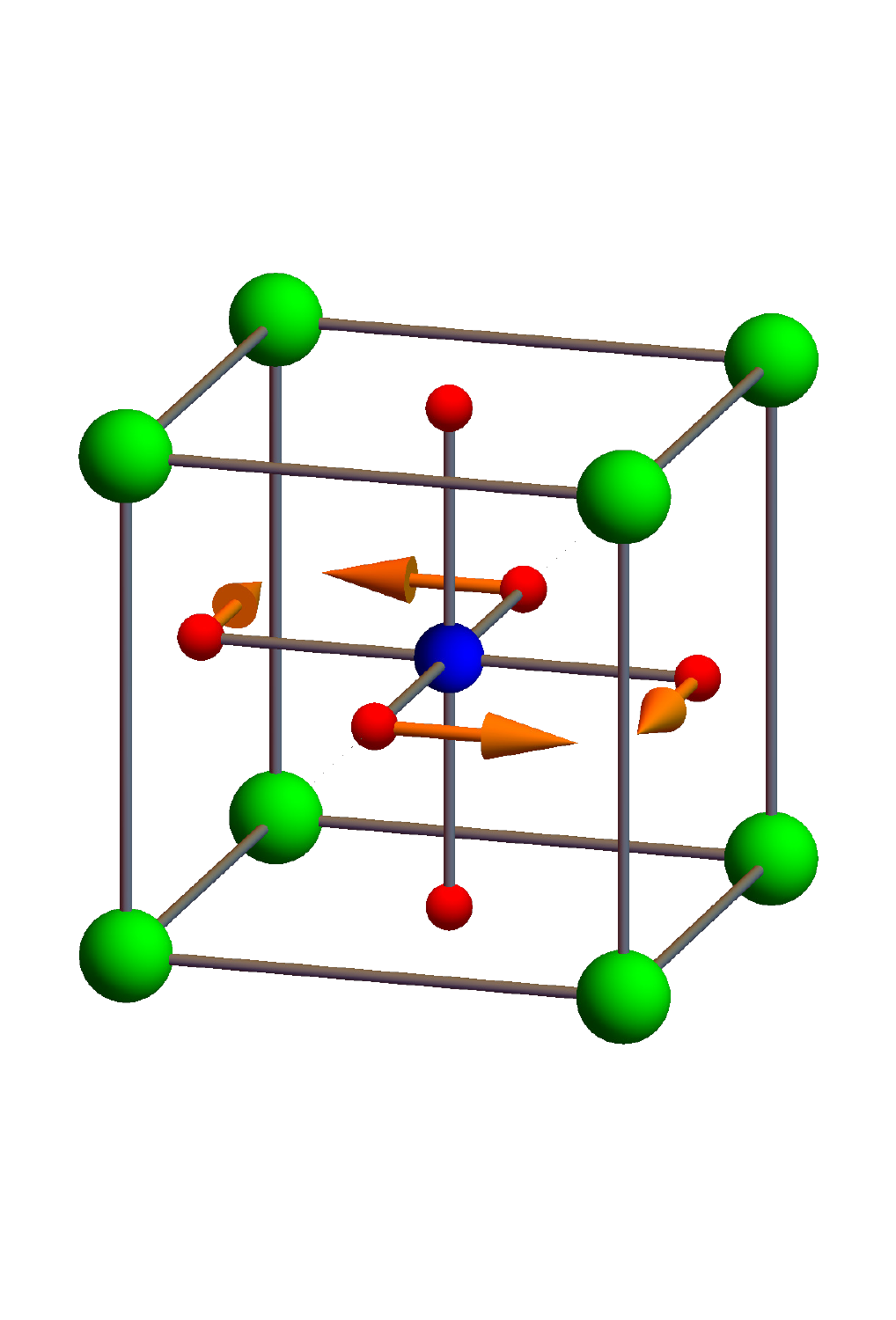}
\end{tabular}
\end{center}
\caption{
Symmetrized mass-weighted normal vibrations. 
(a) $E_g \theta$, 
(b) $E_g \epsilon$, 
(c) $T_{2g} \zeta$.
}
\label{Fig:mode}
\end{figure}

According to the selection rule for the $t_{2g}$ orbitals,
\begin{eqnarray}
t_{2g} \otimes t_{2g} = a_{1g} \oplus e_g \oplus \{t_{1g}\} \oplus t_{2g},
\label{Eq:selection}
\end{eqnarray}
the $t_{2g}$ orbitals have unquenched orbital angular momenta (time-odd $t_{1g}$ operator)
and also couple to the $a_{1g}$, $e_g$ and $t_{2g}$ vibrational modes (Fig. \ref{Fig:mode}). 
In Eq. (\ref{Eq:selection}), the curly bracket indicates that the representation is antisymmetric.

\subsection{Electronic states}
The presence of the unquenched orbital angular momenta indicates the spin-orbit coupling acts on $t_{2g}^1$ configurations in the first order of perturbation.
Projecting the orbital angular momentum operator for the $d$ orbitals into the space of $t_{2g}$ orbitals, we obtain 
\begin{eqnarray}
 \hat{\bm{l}} &=& \langle l \rangle \tilde{\bm{l}},
\end{eqnarray}
where, the reduction of the orbital angular momentum by the covalency effect is included in $\langle l \rangle$,
the components of $\tilde{\bm{l}}$ are written as follows \cite{Kotani1960, Sugano1970}:
\begin{eqnarray}
 \tilde{l}_x &=& 
 \begin{pmatrix}
  0 & 0 & 0 \\
  0 & 0 & i \\
  0 &-i & 0 
 \end{pmatrix},
\quad
 \tilde{l}_y =
 \begin{pmatrix}
  0 & 0 &-i \\
  0 & 0 & 0 \\
  i & 0 & 0 
 \end{pmatrix},
\nonumber\\
\quad
 \tilde{l}_z &=&
 \begin{pmatrix}
  0 & i & 0 \\
 -i & 0 & 0 \\
  0 & 0 & 0 
 \end{pmatrix},
\end{eqnarray}
in the order of the electronic basis $|t_{2} \xi\rangle$, $|t_{2} \eta\rangle$, $|t_{2} \zeta\rangle$.
$\xi, \eta, \zeta$ indicate the basis of $t_{2g}$ representation which transform as $yz$, $zx$, $xy$, respectively, under symmetry operation of $O_h$ group. 

Due to the spin-orbit coupling, 
$\hat{h}_\text{SO} = \lambda_\text{SO} \tilde{\bm{l}} \cdot \hat{\bm{s}}$, 
$t_{2g}^1$ configurations split into spin-orbit multiplets \cite{Koster1963}:
\begin{eqnarray}
 t_{2} \otimes \Gamma_6 = \Gamma_7 \oplus \Gamma_8.
\end{eqnarray}
$\Gamma_6$ is the irreducible representation of electron spin state. 
Since each representation appears only once in the right hand side, the spin-orbit coupled states are determined by using Clebsch-Gordan coefficients as \cite{Koster1963}
\begin{eqnarray}
 \left|\Gamma_7, -\frac{1}{2} \right\rangle  &=& -\frac{i}{\sqrt{3}} |t_{2}\xi,\uparrow\rangle - \frac{1}{\sqrt{3}} |t_{2}\eta,\uparrow\rangle + \frac{i}{\sqrt{3}} |t_{2}\zeta, \downarrow \rangle,
\nonumber\\
 \left|\Gamma_7, +\frac{1}{2} \right\rangle  &=& -\frac{i}{\sqrt{3}} |t_{2}\xi,\downarrow\rangle + \frac{1}{\sqrt{3}} |t_{2}\eta,\downarrow\rangle - \frac{i}{\sqrt{3}} |t_{2}\zeta, \uparrow \rangle,
\nonumber\\
 \left|\Gamma_8, -\frac{3}{2} \right\rangle  &=& -\frac{i}{\sqrt{6}} |t_{2}\xi, \downarrow\rangle + \frac{1}{\sqrt{6}} |t_{2}\eta, \downarrow\rangle + i \sqrt{\frac{2}{3}} |t_{2}\zeta,\uparrow\rangle,
\nonumber\\
 \left|\Gamma_8, -\frac{1}{2} \right\rangle  &=& \frac{i}{\sqrt{2}} |t_{2}\xi,\uparrow\rangle - \frac{1}{\sqrt{2}} |t_{2}\eta,\uparrow\rangle,
\nonumber\\
 \left|\Gamma_8, +\frac{1}{2} \right\rangle  &=& -\frac{i}{\sqrt{2}} |t_{2}\xi,\downarrow\rangle - \frac{1}{\sqrt{2}} |t_{2}\eta,\downarrow\rangle,
\nonumber\\
 \left|\Gamma_8, +\frac{3}{2} \right\rangle  &=& \frac{i}{\sqrt{6}} |t_{2}\xi, \uparrow\rangle + \frac{1}{\sqrt{6}} |t_{2}\eta, \uparrow\rangle + i\sqrt{\frac{2}{3}} |t_{2}\zeta, \downarrow\rangle.
\nonumber\\
\label{Eq:Gamma78}
\end{eqnarray}

With the use of the spin-orbit coupled basis, Eq. (\ref{Eq:Gamma78}), 
the magnetic moment (\ref{Eq:mu_t2}) in Zeeman Hamiltonian, 
the matrix forms of the pseudo orbital and spin angular momentum operators are given as
\begin{eqnarray}
 \hat{L}_x &=& 
 \begin{pmatrix}
  0 & -\frac{2}{3} & \frac{1}{3\sqrt{2}} & 0 & \frac{1}{\sqrt{6}} & 0 \\
 -\frac{2}{3} & 0 & 0 & -\frac{1}{\sqrt{6}} & 0 & -\frac{1}{3\sqrt{2}} \\
 \frac{1}{3\sqrt{2}} & 0 & 0 & \frac{1}{\sqrt{3}} & 0 & -\frac{2}{3} \\
 0 & -\frac{1}{\sqrt{6}} & \frac{1}{\sqrt{3}} & 0 & 0 & 0 \\
 \frac{1}{\sqrt{6}} & 0 & 0 & 0 & 0 & \frac{1}{\sqrt{3}} \\
 0 & -\frac{1}{3\sqrt{2}} & -\frac{2}{3} & 0 & \frac{1}{\sqrt{3}} & 0 \\
 \end{pmatrix},
\nonumber\\
 \hat{L}_y &=& 
 \begin{pmatrix}
  0 & -\frac{2i}{3} & \frac{i}{3\sqrt{2}} & 0 & -\frac{i}{\sqrt{6}} & 0 \\
 \frac{2i}{3} & 0 & 0 & -\frac{i}{\sqrt{6}} & 0 & \frac{i}{3\sqrt{2}} \\
 -\frac{i}{3\sqrt{2}} & 0 & 0 & \frac{i}{\sqrt{3}} & 0 & \frac{2i}{3} \\
 0 & \frac{i}{\sqrt{6}} & -\frac{i}{\sqrt{3}} & 0 & 0 & 0 \\
 \frac{i}{\sqrt{6}} & 0 & 0 & 0 & 0 & \frac{i}{\sqrt{3}} \\
 0 & -\frac{i}{3\sqrt{2}} & -\frac{2i}{3} & 0 & -\frac{i}{\sqrt{3}} & 0 \\
 \end{pmatrix},
\nonumber\\
 \hat{L}_z &=& 
 \begin{pmatrix}
  \frac{2}{3} & 0 & 0 & 0 & 0 & -\frac{\sqrt{2}}{3} \\
  0 & -\frac{2}{3} & -\frac{\sqrt{2}}{3} & 0 & 0 & 0 \\
  0 & -\frac{\sqrt{2}}{3} & -\frac{1}{3} & 0 & 0 & 0 \\
  0 & 0 & 0 & -1 & 0 & 0 \\
  0 & 0 & 0 & 0 & 1 & 0 \\
  -\frac{\sqrt{2}}{3} & 0 & 0 & 0 & 0 & \frac{1}{3} 
 \end{pmatrix},
\end{eqnarray}
and 
\begin{eqnarray}
 \hat{S}_x &=& 
 \begin{pmatrix}
  0 & -\frac{1}{6} & \frac{1}{3\sqrt{2}} & 0 & \frac{1}{\sqrt{6}} & 0 \\
 -\frac{1}{6} & 0 & 0 & -\frac{1}{\sqrt{6}} & 0 & -\frac{1}{3\sqrt{2}} \\
 \frac{1}{3\sqrt{2}} & 0 & 0 & -\frac{1}{2\sqrt{3}} & 0 & \frac{1}{3} \\
  0 & -\frac{1}{\sqrt{6}} & -\frac{1}{2\sqrt{3}} & 0 & 0 & 0 \\
  \frac{1}{\sqrt{6}} & 0 & 0 & 0 & 0 & -\frac{1}{2\sqrt{3}} \\
  0 & -\frac{1}{3\sqrt{2}} & \frac{1}{3} & 0 & -\frac{1}{2\sqrt{3}} & 0 \\
 \end{pmatrix},
\nonumber\\
 \hat{S}_y &=& 
 \begin{pmatrix}
  0 & -\frac{i}{6} & \frac{i}{3\sqrt{2}} & 0 & -\frac{i}{\sqrt{6}} & 0 \\
  \frac{i}{6} & 0 & 0 & -\frac{i}{\sqrt{6}} & 0 & \frac{i}{3\sqrt{2}} \\
 -\frac{i}{3\sqrt{2}} & 0 & 0 & -\frac{i}{2\sqrt{3}} & 0 & -\frac{i}{3} \\
 0 & \frac{i}{\sqrt{6}} & \frac{i}{2\sqrt{3}} & 0 & 0 & 0 \\
 \frac{i}{\sqrt{6}} & 0 & 0 & 0 & 0 & -\frac{i}{2\sqrt{3}} \\
 0 & -\frac{i}{3\sqrt{2}} & \frac{i}{3} & 0 & \frac{i}{2\sqrt{3}} & 0 \\
 \end{pmatrix},
\nonumber\\
 \hat{S}_z &=& 
 \begin{pmatrix}
  \frac{1}{6} & 0 & 0 & 0 & 0 & -\frac{\sqrt{2}}{3} \\
  0 & -\frac{1}{6} & -\frac{\sqrt{2}}{3} & 0 & 0 & 0 \\
  0 & -\frac{\sqrt{2}}{3} & \frac{1}{6} & 0 & 0 & 0 \\
  0 & 0 & 0 & \frac{1}{2} & 0 & 0 \\
  0 & 0 & 0 & 0 & -\frac{1}{2} & 0 \\
 -\frac{\sqrt{2}}{3} & 0 & 0 & 0 & 0 & -\frac{1}{6} 
 \end{pmatrix},
\end{eqnarray}
respectively.
The basis is in the same order as Eq. (\ref{Eq:Gamma78}).
The spin-orbit, $\hat{h}_\text{SO} = \lambda_\text{SO} \tilde{\bm{l}} \cdot \hat{\bm{s}}$, and Zeeman, $\hat{h}_\text{Zee} = -\hat{\bm{m}} \cdot \bm{B}$, Hamiltonian matrices in the coupled basis are obtained by simply replacing the $\tilde{\bm{l}}$ and $\hat{\bm{s}}$ by $\hat{\bm{L}}$ and $\hat{\bm{S}}$, respectively.

\subsection{Vibronic coupling}
\label{Sec:Vibronic}
The $t_{2g}$ orbital couples to $a_{1g}$, $e_g$ and $t_{2g}$ vibrations (\ref{Eq:selection}). 
We take the $O_h$ structure which is fully relaxed with respect to the $a_g$ normal mode as reference structure. 
The totally symmetric part contains harmonic and anharmonic potentials:
\begin{eqnarray}
 \hat{h}_{A_1} &=& 
    \sum_{\gamma = \theta, \epsilon} \frac{\omega_E^2}{2} \hat{Q}_{E\gamma}^2 
  + \sum_{\gamma = \xi, \eta, \zeta} \frac{\omega_{T_2}^2}{2} \hat{Q}_{T_2\gamma}^2 
\nonumber\\
 &+&\sum_{\gamma = \theta, \epsilon} \frac{v_{A_1}^{EEE}}{3!} \{ \hat{Q}_{E} \otimes \hat{Q}_E \otimes \hat{Q}_E \}_{A_1}
\nonumber\\
 &+&\sum_{\gamma = \theta, \epsilon} \frac{v_{A_1}^{EEEE}}{4!} \{ \hat{Q}_{E} \otimes \hat{Q}_E \otimes \hat{Q}_E \otimes \hat{Q}_E \}_{A_1}.
\label{Eq:hA1}
\end{eqnarray}
Here, $\omega_\Gamma$ is the frequency for $\Gamma$ mode ($\Gamma = E, T_2$). 
The symmetrized products are shown below. 
The basis of $e_g$ representation expressed by $\theta$ and $\epsilon$ transform as $(-x^2-y^2+2z^2)/\sqrt{6}$ and $(x^2-y^2)/\sqrt{2}$, respectively, under symmetry operations.  

The vibronic couplings with the $e_g$ and $t_{2g}$ modes induce the Jahn-Teller effect.
The linear term is given by 
\begin{eqnarray}
 \hat{h}_\text{LJT} &=& \sum_{\gamma = \theta, \epsilon} v_E \hat{Q}_{E\gamma} \hat{\tau}_{E\gamma}
 + \sum_{\gamma = \xi, \eta, \zeta} v_{T_2} \hat{Q}_{T_2\gamma} \hat{\tau}_{T_2\gamma}. 
\label{Eq:hLJT}
\end{eqnarray}
Here, $v_\Gamma$ are the linear orbital vibronic coupling parameters, and $\hat{\bm{\tau}}$ are matrices of Clebsch-Gordan coefficients:
\begin{eqnarray}
 \hat{\tau}_{E\theta} &=& 
\begin{pmatrix}
 -\frac{1}{2} & 0 & 0 \\
 0 & -\frac{1}{2} & 0 \\
 0 & 0 & 1
\end{pmatrix},
\quad
 \hat{\tau}_{E\epsilon} = 
\begin{pmatrix}
 \frac{\sqrt{3}}{2} & 0 & 0 \\
 0 & -\frac{\sqrt{3}}{2} & 0 \\
 0 & 0 & 0 
\end{pmatrix},
\nonumber\\
 \hat{\tau}_{T_2\xi} &=& 
 \begin{pmatrix}
   0 & 0 & 0 \\
   0 & 0 & \frac{1}{\sqrt{2}} \\
   0 & \frac{1}{\sqrt{2}} & 0 \\
 \end{pmatrix},
\quad
 \hat{\tau}_{T_2\eta} = 
 \begin{pmatrix}
   0 & 0 & \frac{1}{\sqrt{2}} \\
   0 & 0 & 0 \\
   \frac{1}{\sqrt{2}} & 0 & 0 \\
 \end{pmatrix},
\nonumber\\
 \hat{\tau}_{T_2\zeta} &=& 
 \begin{pmatrix}
   0 & \frac{1}{\sqrt{2}} & 0 \\
   \frac{1}{\sqrt{2}} & 0 & 0 \\
   0 & 0 & 0 \\
 \end{pmatrix}.
\label{Eq:tau}
\end{eqnarray}
The phase factors of the the Jahn-Teller active modes are chosen as shown in Fig. \ref{Fig:mode}.

Transforming the electronic basis from the spin-orbital decoupled states into the coupled states (\ref{Eq:Gamma78}), $\hat{\bm{\tau}}$'s in $\hat{h}_\text{LJT}$ become
\begin{eqnarray}
 \hat{T}_{E\theta} &=& 
 \begin{pmatrix}
   0 & 0 & 0 & 0 & 0 & \frac{1}{\sqrt{2}}\\
   0 & 0 & -\frac{1}{\sqrt{2}} & 0 & 0 & 0 \\
   0 & -\frac{1}{\sqrt{2}} & \frac{1}{2} & 0 & 0 & 0\\
   0 & 0 & 0 & -\frac{1}{2} & 0 & 0 \\
   0 & 0 & 0 & 0 & -\frac{1}{2} & 0 \\
   \frac{1}{\sqrt{2}} & 0 & 0 & 0 & 0 & \frac{1}{2}
 \end{pmatrix},
\nonumber\\
 \hat{T}_{E\epsilon} &=&
 \begin{pmatrix}
  0 & 0 & 0 & -\frac{1}{\sqrt{2}} & 0 & 0 \\
  0 & 0 & 0 & 0 & \frac{1}{\sqrt{2}} & 0 \\
  0 & 0 & 0 & 0 & \frac{1}{2} & 0 \\
  -\frac{1}{\sqrt{2}} & 0 & 0 & 0 & 0 & \frac{1}{2} \\
  0 & \frac{1}{\sqrt{2}} & \frac{1}{2} & 0 & 0 & 0 \\
  0 & 0 & 0 & \frac{1}{2} & 0 & 0 \\
 \end{pmatrix},
\nonumber\\
 \hat{T}_{T_2\xi} &=& 
 \begin{pmatrix}
  0 & 0 & -\frac{i}{2} & 0 & \frac{i}{2\sqrt{3}} & 0 \\
  0 & 0 & 0 & -\frac{i}{2\sqrt{3}} & 0 & \frac{i}{2} \\
  \frac{i}{2} & 0 & 0 & \frac{i}{\sqrt{6}} & 0 & 0 \\
  0 & \frac{i}{2\sqrt{3}} & -\frac{i}{\sqrt{6}} & 0 & 0 & 0 \\
  -\frac{i}{2\sqrt{3}} & 0 & 0 & 0 & 0 & -\frac{i}{\sqrt{6}} \\
  0 & -\frac{i}{2} & 0 & 0 & \frac{i}{\sqrt{6}} & 0 \\
 \end{pmatrix},
\nonumber\\
 \hat{T}_{T_2\eta} &=& 
 \begin{pmatrix}
  0 & 0 & -\frac{1}{2} & 0 & -\frac{1}{2\sqrt{3}} & 0 \\
  0 & 0 & 0 & -\frac{1}{2\sqrt{3}} & 0 & -\frac{1}{2} \\
 -\frac{1}{2} & 0 & 0 & \frac{1}{\sqrt{6}} & 0 & 0 \\
  0 & -\frac{1}{2\sqrt{3}} & \frac{1}{\sqrt{6}} & 0 & 0 & 0 \\
  -\frac{1}{2\sqrt{3}} & 0 & 0 & 0 & 0 & -\frac{1}{\sqrt{6}} \\
  0 & -\frac{1}{2} & 0 & 0 & -\frac{1}{\sqrt{6}} & 0 \\
 \end{pmatrix},
\nonumber\\
 \hat{T}_{T_2\zeta} &=& 
 \begin{pmatrix}
  0 & 0 & 0 & -\frac{i}{\sqrt{3}} & 0 & 0 \\
  0 & 0 & 0 & 0 & -\frac{i}{\sqrt{3}} & 0 \\
  0 & 0 & 0 & 0 & -\frac{i}{\sqrt{6}} & 0 \\
  \frac{i}{\sqrt{3}} & 0 & 0 & 0 & 0 & -\frac{i}{\sqrt{6}} \\
  0 & \frac{i}{\sqrt{3}} & \frac{i}{\sqrt{6}} & 0 & 0 & 0 \\
  0 & 0 & 0 & \frac{i}{\sqrt{6}} & 0 & 0 \\
 \end{pmatrix}.
\end{eqnarray}
Replacing $\hat{\bm{\tau}}$ in $\hat{h}_\text{LJT}$ with $\hat{\bm{T}}$, we obtain $\hat{H}_\text{LJT}$.

The non-linear vibronic Hamiltonian is derived in the same manner:
\begin{eqnarray}
 \hat{h}_\text{NLJT} &=& 
  \sum_{\gamma = \theta, \epsilon} \frac{v_E^{EE}}{2!} \{ \hat{Q}_E \otimes \hat{Q}_E \}_{E\gamma} \hat{\tau}_{E\gamma} 
\nonumber\\
 &+& \sum_{\gamma = \theta, \epsilon} \frac{v_E^{T_2T_2}}{2!} \{ \hat{Q}_{T_2} \otimes \hat{Q}_{T_2} \}_{E\gamma} \hat{\tau}_{E\gamma} 
\nonumber\\
 &+& \sum_{\gamma = \xi, \eta, \zeta} \frac{v_{T_2}^{T_2T_2}}{2!} \{ \hat{Q}_{T_2} \otimes \hat{Q}_{T_2} \}_{T_2\gamma} \hat{\tau}_{T_2\gamma} 
\nonumber\\
 &+& \sum_{\gamma = \xi, \eta, \zeta} \frac{v_{T_2}^{ET_2}}{2!} \{ \hat{Q}_{E} \otimes \hat{Q}_{T_2} \}_{T_2\gamma} \hat{\tau}_{T_2\gamma} 
\nonumber\\
 &+& \sum_{\gamma = \theta, \epsilon} \frac{v_E^{EEE}}{3!} \{ \hat{Q}_{E} \otimes \hat{Q}_{E} \otimes \hat{Q}_E \}_{E\gamma} \hat{\tau}_{E\gamma} 
\nonumber\\
 &+& \sum_{n=1,2} \sum_{\gamma = \theta, \epsilon} \frac{v_{nE}^{EEEE}}{4!} \{ \hat{Q}_E \otimes \hat{Q}_{E} \otimes \hat{Q}_{E} \otimes \hat{Q}_E \}_{nE\gamma} \hat{\tau}_{E\gamma}.
\nonumber\\
\label{Eq:hNLJT}
\end{eqnarray}
Here, only the terms treated in this work are written.
By the same transformation as $\hat{h}_\text{LJT}$, we obtain $\hat{H}_\text{NLJT}$.

The symmetrized products of the coordinates appearing in Eqs. (\ref{Eq:hA1}) and (\ref{Eq:hNLJT}) are calculated as follows. 
The symmetrized quadratic coordinates are in general calculated as 
\begin{eqnarray}
 \{ \hat{Q}_{\Lambda_1} \otimes \hat{Q}_{\Lambda_2} \}_{\Lambda \lambda} &=& 
 \sum_{\lambda_1 \lambda_2} \hat{Q}_{\Lambda_1 \lambda_1} \hat{Q}_{\Lambda_2 \lambda_2} 
\nonumber\\
 &\times&
 \langle \Lambda_1 \lambda_1 \Lambda_2 \lambda_2| (\Lambda_1 \Lambda_2) \Lambda \lambda \rangle.
\label{Eq:QQ}
\end{eqnarray}
Here, $\langle \Lambda_1 \lambda_1 \Lambda_2 \lambda_2|(\Lambda_1\Lambda_2) \Lambda \lambda \rangle$ is Clebsch-Gordan coefficient tabulated in Ref. \cite{Koster1963}.
The explicit form of the symmetrized products of our interests are 
\begin{eqnarray}
 \{ \hat{Q}_E \otimes \hat{Q}_E \}_A &=& \frac{1}{\sqrt{2}}\left( \hat{Q}_{E\theta}^2 + \hat{Q}_{E\epsilon}^2 \right),
\nonumber\\
 \{ \hat{Q}_{T_2} \otimes \hat{Q}_{T_2} \}_A &=& \frac{1}{\sqrt{3}} \left( \hat{Q}_{T_2\xi}^2 + \hat{Q}_{T_2 \eta}^2 + \hat{Q}_{T_2\zeta}^2 \right),
\nonumber\\
 \{ \hat{Q}_E \otimes \hat{Q}_E \}_{E\theta} &=& -\frac{1}{\sqrt{2}}\left( \hat{Q}_{E\theta}^2 - \hat{Q}_{E\epsilon}^2 \right),
\nonumber\\
 \{ \hat{Q}_E \otimes \hat{Q}_E \}_{E\epsilon} &=& \sqrt{2} \hat{Q}_{E\theta} \hat{Q}_{E\epsilon},
\nonumber\\
 \{ \hat{Q}_{T_2} \otimes \hat{Q}_{T_2} \}_{E\theta} &=& -\frac{1}{\sqrt{6}} \left( \hat{Q}_{T_2\xi}^2 + \hat{Q}_{T_2 \eta}^2 - 2\hat{Q}_{T_2\zeta}^2 \right),
\nonumber\\
 \{ \hat{Q}_{T_2} \otimes \hat{Q}_{T_2} \}_{E\epsilon} &=& \frac{1}{\sqrt{2}} \left( \hat{Q}_{T_2\xi}^2 - \hat{Q}_{T_2 \eta}^2 \right),
\nonumber\\
 \{ \hat{Q}_{T_2} \otimes \hat{Q}_{T_2} \}_{T_2\xi} &=& \sqrt{2} \hat{Q}_{T_2\eta} \hat{Q}_{T_2\zeta},
\nonumber\\
 \{ \hat{Q}_{T_2} \otimes \hat{Q}_{T_2} \}_{T_2\eta} &=& \sqrt{2} \hat{Q}_{T_2\zeta} \hat{Q}_{T_2\xi},
\nonumber\\
 \{ \hat{Q}_{T_2} \otimes \hat{Q}_{T_2} \}_{T_2\zeta} &=& \sqrt{2} \hat{Q}_{T_2\xi} \hat{Q}_{T_2\eta},
\nonumber\\
 \{ \hat{Q}_{E} \otimes \hat{Q}_{T_2} \}_{T_2\xi} &=& \left( -\frac{1}{2} \hat{Q}_{E\theta} + \frac{\sqrt{3}}{2} \hat{Q}_{E \epsilon} \right) \hat{Q}_{T_2\xi},
\nonumber\\
 \{ \hat{Q}_{E} \otimes \hat{Q}_{T_2} \}_{T_2\eta} &=& \left( -\frac{1}{2} \hat{Q}_{E\theta} - \frac{\sqrt{3}}{2} \hat{Q}_{E \epsilon} \right) \hat{Q}_{T_2\eta},
\nonumber\\
 \{ \hat{Q}_{E} \otimes \hat{Q}_{T_2} \}_{T_2\zeta} &=& \hat{Q}_{E\theta} \hat{Q}_{T_2\zeta},
\end{eqnarray}
for quadratic terms. 
The cubic products can be calculated by using Eq. (\ref{Eq:QQ}) twice. 
The cubic terms of our interest are calculated as follows:
\begin{eqnarray}
 \{ \hat{Q}_E \otimes \{ \hat{Q}_E \otimes \hat{Q}_E \}_E \}_{A_1} &=& \frac{1}{2} \left(-\hat{Q}_{E\theta}^3 + 3\hat{Q}_{E\theta} \hat{Q}_{E\epsilon}^2 \right),
\nonumber\\
 \{ \hat{Q}_E \otimes \{ \hat{Q}_E \otimes \hat{Q}_E \}_E \}_{E\theta} &=& \frac{1}{2} \hat{Q}_{E\theta} \left(\hat{Q}_{E\theta}^2 + \hat{Q}_{E\epsilon}^2 \right),
\nonumber\\
 \{ \hat{Q}_E \otimes \{ \hat{Q}_E \otimes \hat{Q}_E \}_E \}_{E\theta} &=& \frac{1}{2} \hat{Q}_{E\epsilon} \left(\hat{Q}_{E\theta}^2 + \hat{Q}_{E\epsilon}^2 \right).
\end{eqnarray}
The fourth order terms for the $E$ mode are calculated as 
\begin{widetext}
\begin{eqnarray}
 \{ \hat{Q}_E \otimes \hat{Q}_E \otimes \hat{Q}_E \otimes \hat{Q}_E \}_{A_1} &=& 
 \{ \hat{Q}_E \otimes \hat{Q}_E \}_{A_1}^2 = \frac{1}{2} \left( \hat{Q}_{E\theta}^2 + \hat{Q}_{E\epsilon}^2 \right)^2, 
\nonumber\\
 \{ \hat{Q}_E \otimes \hat{Q}_E \otimes \hat{Q}_E \otimes \hat{Q}_E \}_{1E\theta} &=& 
 \{ \hat{Q}_E \otimes \hat{Q}_E \}_{A_1} \{ \hat{Q}_E \otimes \hat{Q}_E \}_{E\theta} = 
 -\frac{1}{2} \left( \hat{Q}_{E\theta}^2 - \hat{Q}_{E\epsilon}^2 \right) \left( \hat{Q}_{E\theta}^2 + \hat{Q}_{E\epsilon}^2 \right),
\nonumber\\
 \{ \hat{Q}_E \otimes \hat{Q}_E \otimes \hat{Q}_E \otimes \hat{Q}_E \}_{1E\epsilon} &=& 
 \{ \hat{Q}_E \otimes \hat{Q}_E \}_{A_1} \{ \hat{Q}_E \otimes \hat{Q}_E \}_{E\epsilon} = 
 \hat{Q}_{E\theta} \hat{Q}_{E\epsilon} \left( \hat{Q}_{E\theta}^2 + \hat{Q}_{E\epsilon}^2 \right),
\nonumber\\
 \{ \hat{Q}_E \otimes \hat{Q}_E \otimes \hat{Q}_E \otimes \hat{Q}_E \}_{2E\theta} &=& 
 \{ \{ \hat{Q}_E \otimes \hat{Q}_E \}_{E} \otimes \{ \hat{Q}_E \otimes \hat{Q}_E \}_{E} \}_{E\theta} = 
 -\frac{1}{2\sqrt{2}} \left( \hat{Q}_{E\theta}^4 + \hat{Q}_{E\epsilon}^4 - 6\hat{Q}_{E\theta}^2 \hat{Q}_{E\epsilon}^2 \right),
\nonumber\\
 \{ \hat{Q}_E \otimes \hat{Q}_E \otimes \hat{Q}_E \otimes \hat{Q}_E \}_{2E\epsilon} &=& 
 \{ \{ \hat{Q}_E \otimes \hat{Q}_E \}_{E} \otimes \{ \hat{Q}_E \otimes \hat{Q}_E \}_{E} \}_{E\epsilon} = 
 -\sqrt{2} \left( \hat{Q}_{E\theta}^2 - \hat{Q}_{E\epsilon}^2 \right) \hat{Q}_{E\theta} \hat{Q}_{E\epsilon}.
\end{eqnarray}
\end{widetext}

\section{Computational details}
\subsection{{\it Ab initio} Method}
\label{A:Abinitio}
The electronic and vibronic coupling parameters were derived from the cluster calculations with post Hartree-Fock (HF) methods and $\langle l \rangle$'s were extracted with density functional theory (DFT) calculations.
The clusters were generated from the experimental crystal structures \cite{Aharen2010, Stitzer2002} retaining the $O_h$ symmetry.
In the post HF calculations, the $d^1$ metal ion and the nearest six oxygen atoms were treated {\it ab initio} with ANO-RCC-VQZP basis functions and the surrounding 280 atoms were replaced by \textit{ab initio} embedding model potential (AIMP) \cite{Seijo1999}.
The atomic bielectronic integrals were calculated using Cholesky decomposition with threshold $1 \times 10^{-9}$ $E_h$.
Inversion symmetry was employed in all calculations.

All \textit{ab initio} calculations were carried out with {\tt Molcas} 8.0 program \cite{Molcas8} and were of complete-active-space self-consistent-field (CASSCF)/extended multi-state complete active space second-order perturbation theory (XMS-CASPT2) \cite{Granovsky, Shiozaki}/spin-orbit restricted-active-space state-interaction (SO-RASSI) type. 
The active space of all CASSCF calculations included seven electrons in six orbitals.
Three orbitals are $4d$ or $5d$ orbitals and other three orbitals are of ligand $\pi$-type.
Three roots were optimized at the CASSCF level, and then XMS-CASPT2 were done on the 3 roots from CASSCF.
In XMS-CASPT2 calculations, IPEA shift was set to 0, while IMAG shift was set to 0.1.
SO-RASSI calculations mixed the roots obtained from XMS-CASPT2 by spin-orbit coupling.
The scalar relativistic effects were included in the basis set.

\subsection{DFT Method}
\label{A:DFT}
The clusters for the DFT calculations contain 89 atoms which are treated explicitly with def2-TZVP basis set and def2/J auxiliary basis sets.
DFT calculations were done with hybrid functional B3LYP with RIJCOSX approximation.
The basis function contains the scalar relativistic effects.
For the DFT calculations, {\tt ORCA} 4.0.0.2 \cite{orca} was used.
For the SCF, condition ``TightSCF'' is used.  The grid for density was ``Grid5''.

\subsection{Calculations of electronic and vibronic coupling parameters}
The spin-orbit coupling parameters $\lambda_\text{SO}$ were obtained from the {\it ab initio} multiplet levels,
$E_{\Gamma_7} - E_{\Gamma_8} = \frac{3}{2} \lambda_\text{SO}$. 
The expectation values of orbital angular momentum $\langle l \rangle$ were calculated by using {\it ab initio} or DFT wave functions at $O_h$ structure.
In the latter case, the orbital angular momentum matrices in the atomic orbital basis were calculated using {\tt Molpro} 2012.1 \cite{molpro}.

The frequencies and vibronic parameters were derived by fitting the {\it ab initio} ${}^2T_{2g}$ adiabatic potential energy surface (APES) to the $t_{2g} \otimes (e_g \oplus t_{2g})$ model vibronic Hamiltonian \cite{SM}.
The step of deformation is $\Delta Q = 0.5$ a.u. 
The unit of $k$-th order vibronic coupling parameter is $E_h/(m_e a_0^2)^{k/2}$, where, $E_h$ is Hartree, $m_e$ is electron mass and $a_0$ is Bohr radius.

\subsection{Numerical diagonalization of vibronic Hamiltonian}
\label{A:Lanczos}
The $(\Gamma_7 \oplus \Gamma_8) \otimes (e_g \oplus t_{2g})$ JT Hamiltonians for the $d^1$ systems were numerically diagonalized with the derived parameters.
The nuclear part of the vibronic state [Eq. \ref{Eq:Psi}] is expressed as
\begin{eqnarray}
 |\psi_{\Gamma M,\alpha\Lambda \lambda}\rangle &=& \sum_{n_\theta, n_\epsilon, n_\xi, n_\eta, n_\zeta} |\bm{n}\rangle \psi_{\Gamma M \bm{n}, \alpha\Lambda\lambda},
\label{Eq:psi}
\end{eqnarray}
where, $|\bm{n}\rangle = |n_\theta, n_\epsilon, n_\xi, n_\eta, n_\zeta\rangle$ ($n_\gamma \ge 0$) are the eigenstates of $\hat{H}_0$, and 
the coefficient $\psi_{\Gamma M \bm{n}, \alpha\Lambda\lambda}$ is defined by $\langle \bm{n}|\psi_{\Gamma M,\alpha \Lambda \lambda}\rangle$.
The vibronic basis $|\Gamma \gamma\rangle \otimes |\bm{n}\rangle$ are truncated as
\begin{eqnarray}
 0 \le \sum_\gamma n_\gamma \le 9.
\end{eqnarray}

The diagonalization of the dynamical JT Hamiltonian matrix was done in two steps. 
First, the linear JT Hamiltonian matrices,
\begin{eqnarray}
 \hat{H}^{(1)} &=& \hat{H}_\text{SO} + \sum_{\Gamma \gamma} \frac{\hat{P}_{\Gamma \gamma}^2}{2} + \hat{H}_{A_1} + \hat{H}_\text{LJT},
\label{Eq:H1}
\end{eqnarray}
were diagonalized using Lapack (ZHEEV).
Then, using the lowest 1000 linear vibronic states as the basis, the nonlinear JT Hamiltonian matrices,
\begin{eqnarray}
 \hat{H}^{(2)} &=& \hat{H}^{(1)} + \hat{H}_\text{NLJT},
\label{Eq:H2}
\end{eqnarray}
were calculated and diagonalized.

\subsection{Effective magnetic moment}
The effective magnetic moments were calculated within pure electronic (Elec.) and vibronic (Vibro.) models.
The model Hamiltonians for these two cases are $\hat{H}_\text{SO} + \hat{H}_\text{Zee}$ and $\hat{H}^{(2)} + \hat{H}_\text{Zee}$, respectively, 
where, $\hat{H}^{(2)}$ corresponds to Eq. (\ref{Eq:H2}).
The magnetic field $\bm{B}$ was applied along the $z$ axis, $\bm{B} = (0,0,B)$.
The magnetic moments were calculated by 
\begin{eqnarray}
M_\text{eff} = \sqrt{3k_\text{B}T \chi(T)},
\end{eqnarray}
where $\chi(T)$ is the magnetic susceptibility,
$\chi(T) = \beta^{-1} {\partial^2 \ln Z(T,B) }/{\partial B^2}|_{B \rightarrow 0}$,
$\beta = 1/(k_\text{B}T)$, and $Z(T,B)$ is the distribution function. 
In both cases, the Van Vleck's contribution is directly included in the energy levels.

\subsection{Distribution of vibronic states under Zeeman splitting}
\label{A:rho}
At low temperature such that only the ground $\Gamma_8$ vibronic levels are occupied, the spatial distribution of the vibronic state is calculated as 
\begin{eqnarray}
 \rho(\bm{Q}, \bm{B}, T) &=& 
 \frac{\sum_{M} \rho_{\Gamma_8 M}(\bm{Q}) e^{-E_{\Gamma_8M}(\bm{B})\beta}}{\sum_{M} e^{-E_{\Gamma_8M}(\bm{B})\beta}},
\end{eqnarray}
where, the sum is over the ground $\Gamma_8$ vibronic states, and $E_{\Gamma_8M}(\bm{B})$ is the Zeeman split vibronic level. 
The difference of the distribution $\Delta \rho$ in Fig. \ref{Fig:JTDist}(c) is defined by 
\begin{eqnarray}
 \Delta \rho(\bm{Q}, \bm{B}, T) &=& 
 \rho(\bm{Q}, \bm{B}, T) - \bar{\rho}(\bm{Q}),
\end{eqnarray}
where, $\bar{\rho}$ is the averaged density over the ground vibronic states $M$, $\bar{\rho} = \sum_M \rho_{\Gamma_8M}/4$. 

%

%

\end{document}


\title{
Supplemental Materials \\
for \\
``Spin-orbital-lattice entangled states in cubic $d^1$ double perovskites''
}
\author{Naoya Iwahara}
\email[]{naoya.iwahara@gmail.com}
\author{Veacheslav Vieru}
\author{Liviu F. Chibotaru}
\email[]{liviu.chibotaru@gmail.com}
\affiliation{Theory of Nanomaterials Group, 
University of Leuven, 
Celestijnenlaan 200F, B-3001 Leuven, Belgium}
\date{\today}

\begin{abstract}
Supplemental Materials contain followings:
\begin{enumerate}
 \item Fitting of the APES to the model JT Hamiltonian
 \item Analysis of the APES and Jahn-Teller deformed coordinates
 \item The low-energy vibronic levels and expectation values of magnetic moments
\end{enumerate}
\end{abstract}

\maketitle

\section{Fitting of the APES}
\begin{figure*}[tb]
\begin{center}
\begin{tabular}{ccc}
\multicolumn{1}{l}{(a)} & \multicolumn{1}{l}{(b)} & \multicolumn{1}{l}{(c)} \\
\includegraphics[width=5.5cm]{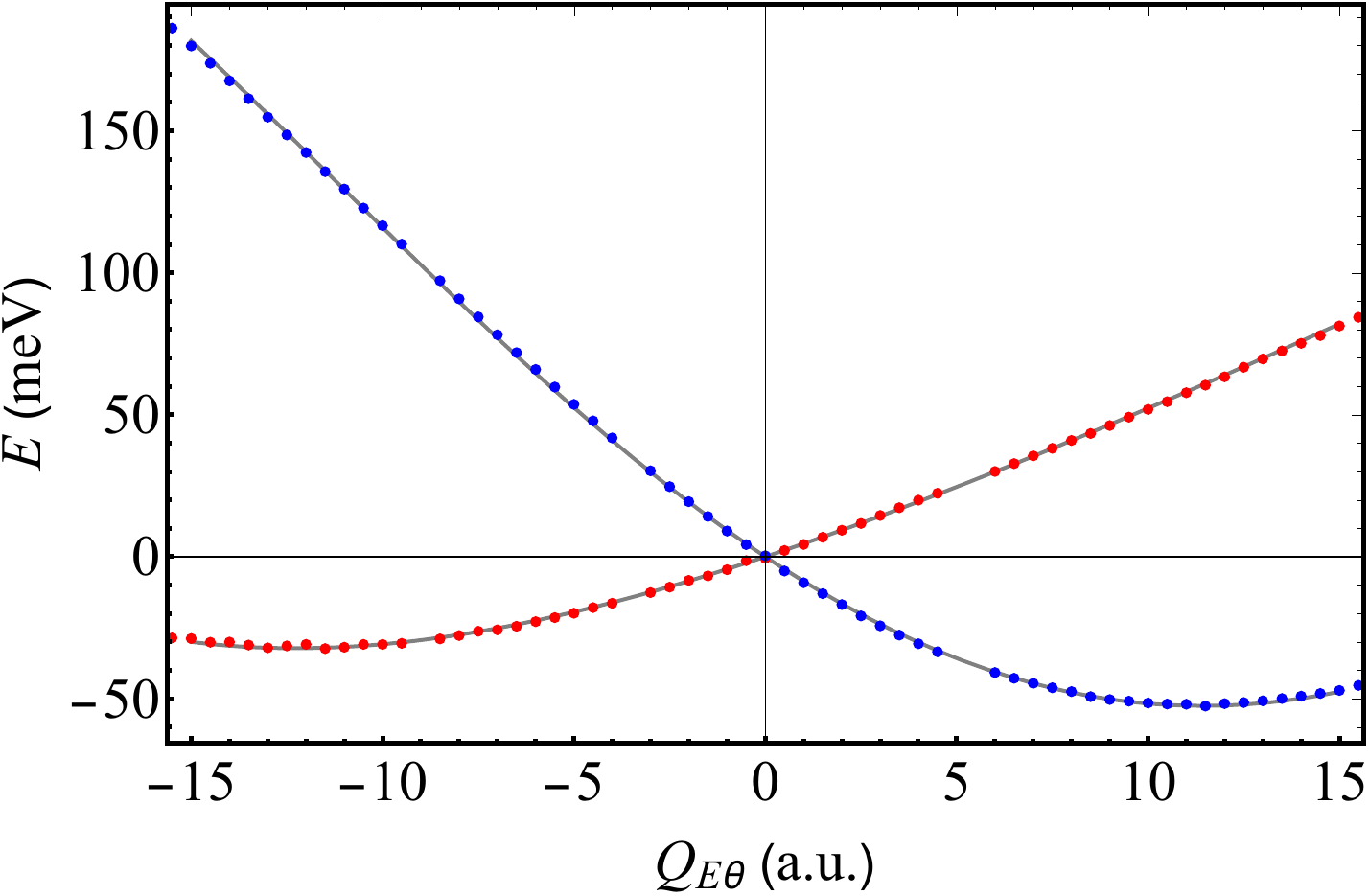}
&
\includegraphics[width=5.5cm]{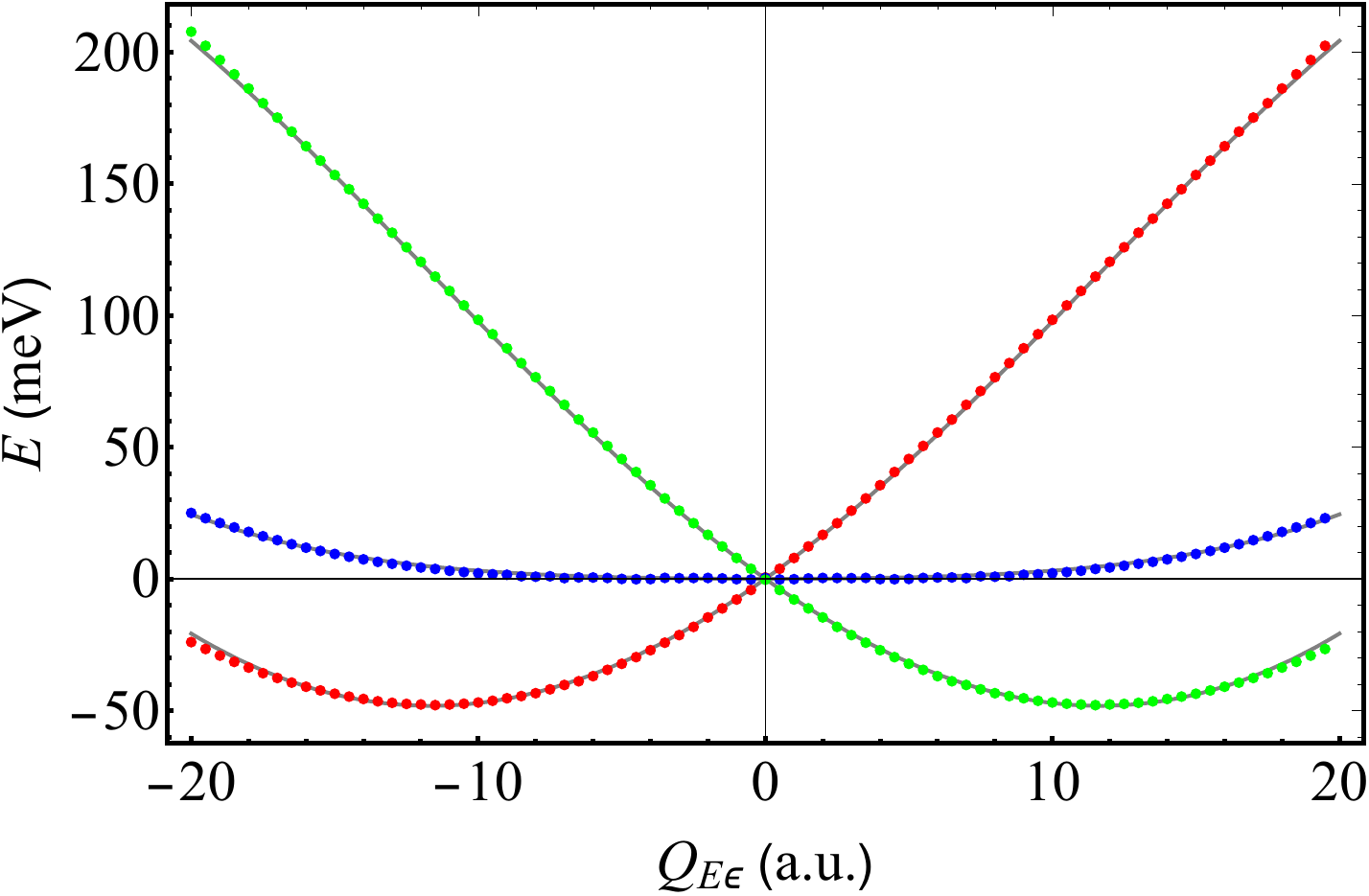}
&
\includegraphics[width=5.5cm]{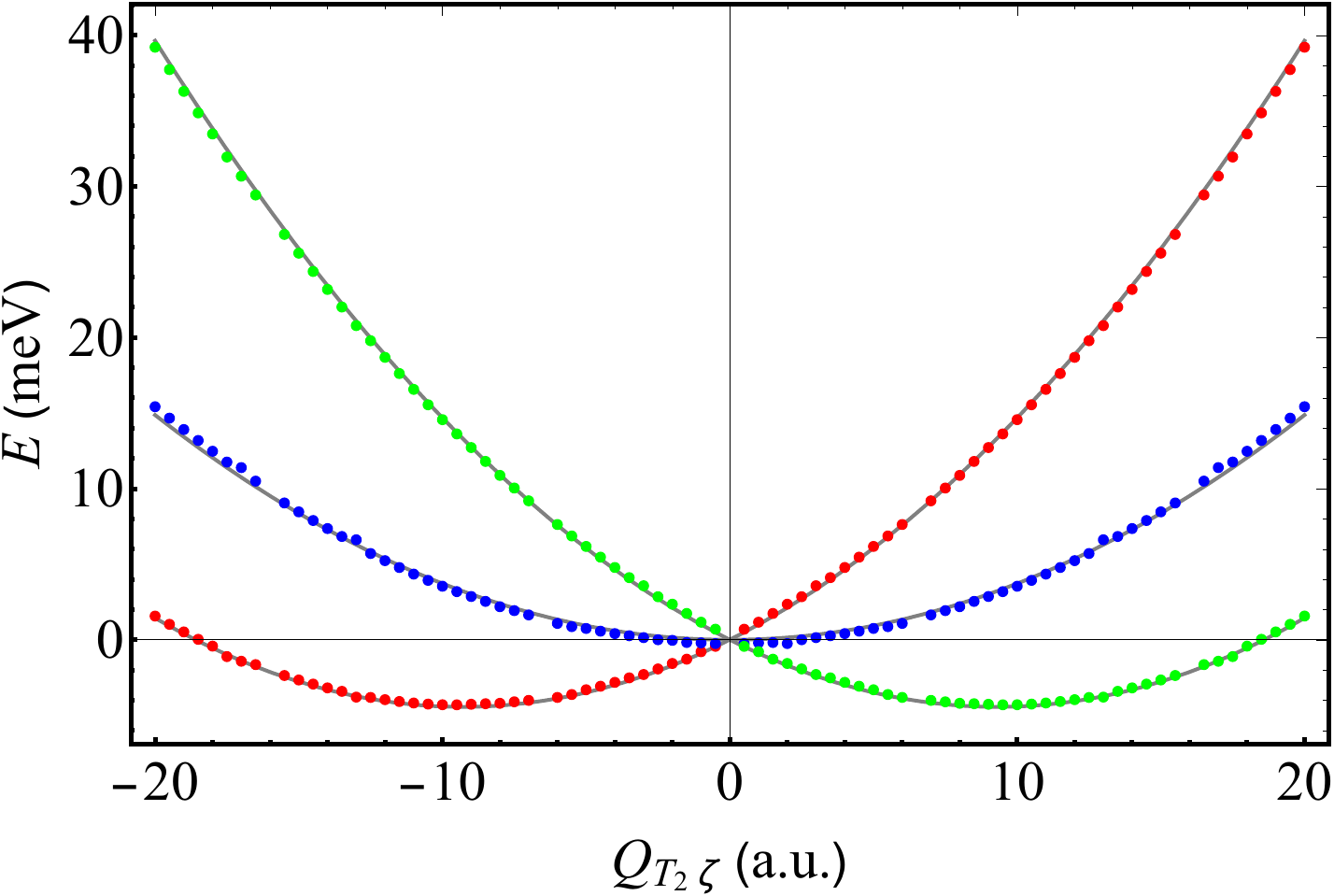}
\\
\multicolumn{1}{l}{(d)} & \multicolumn{1}{l}{(e)} \\
\includegraphics[width=5.5cm]{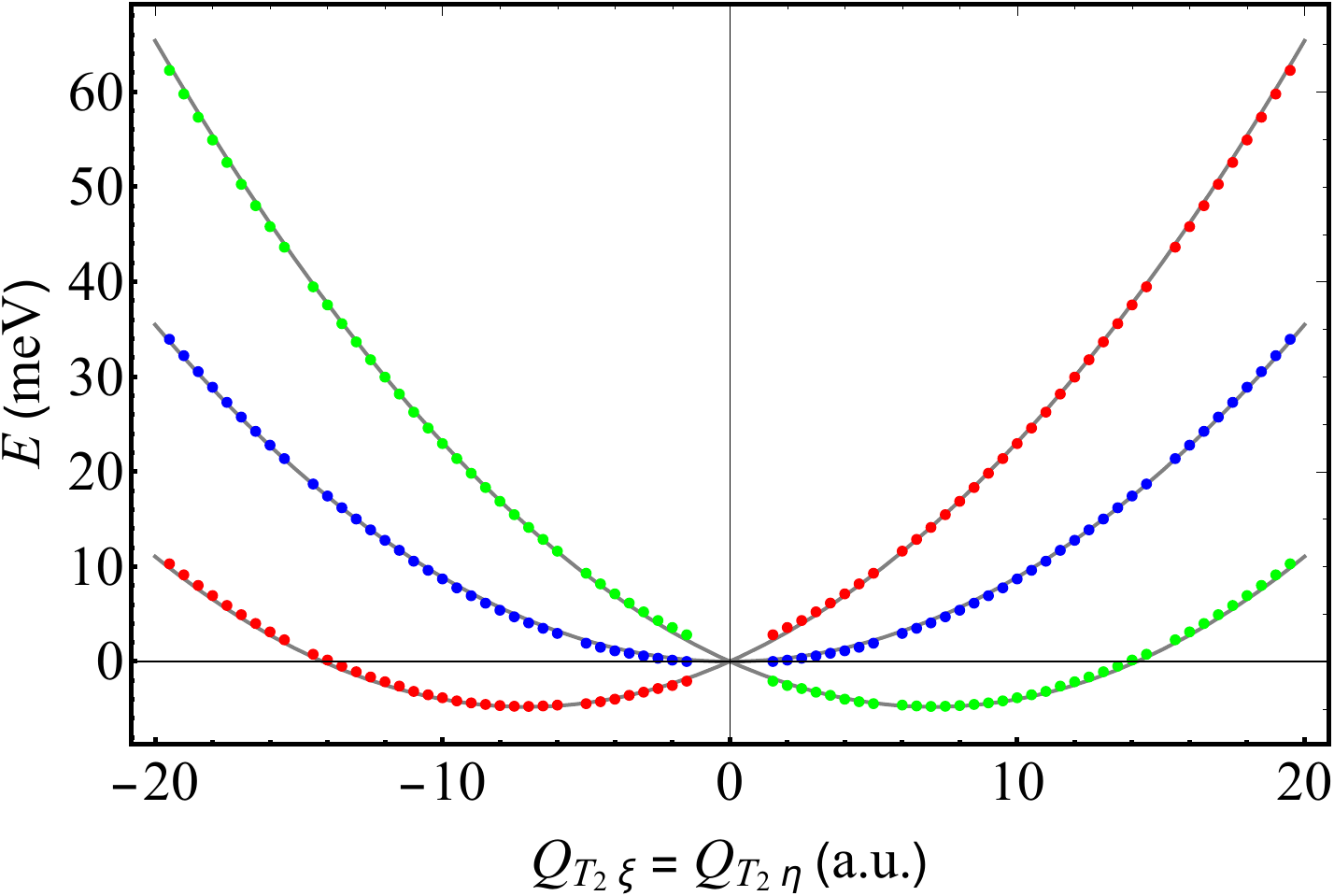}
&
\includegraphics[width=5.5cm]{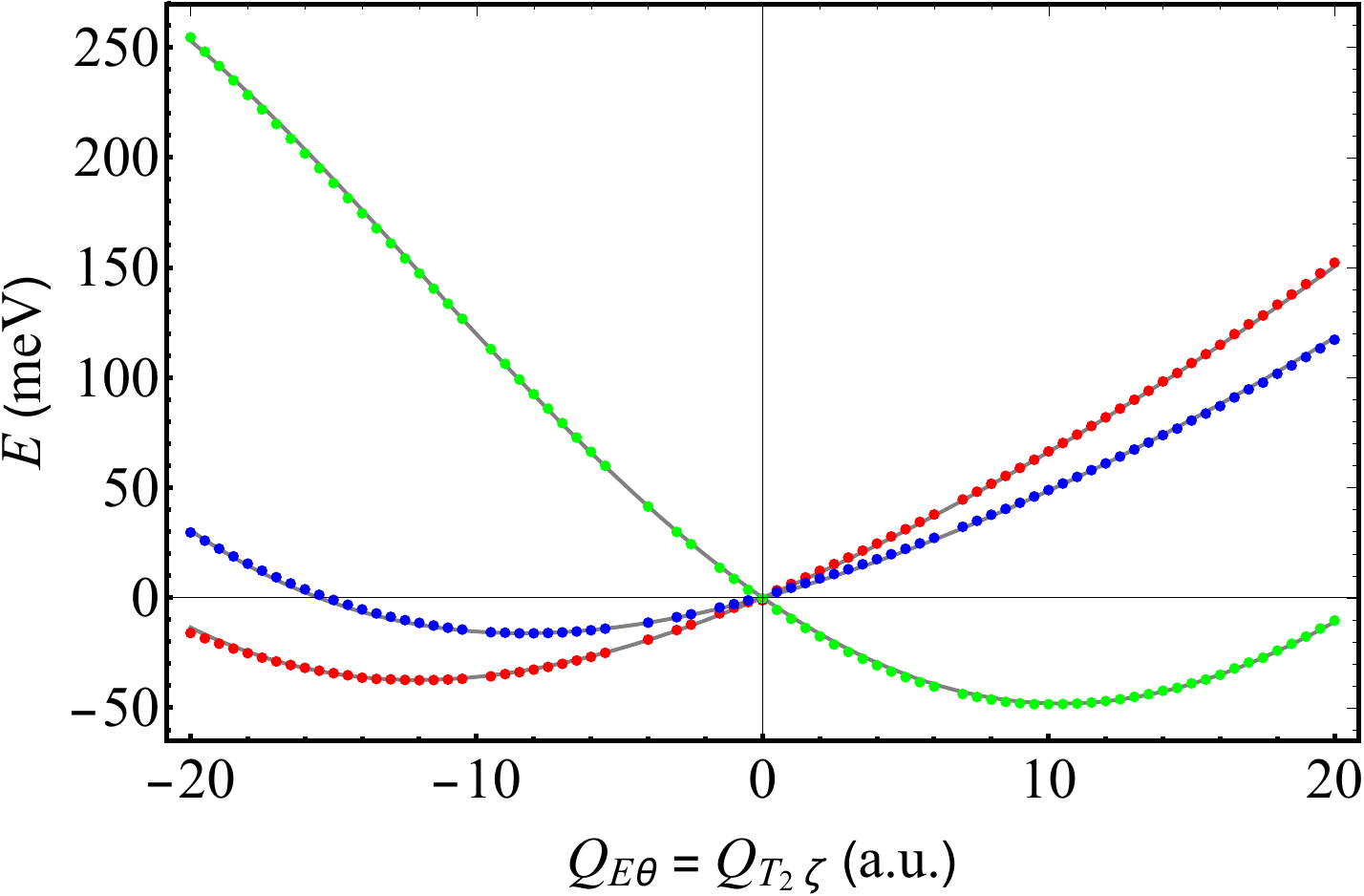}
\end{tabular}
\end{center}
\caption{
APES of Ba$_2$LiOsO$_6$ with (a) $E_{g}\theta$, (b) $E_g \epsilon$, (c) $T_{2g}\zeta$, (d) $T_{2g}\xi + T_{2g}\eta$, and (e) $T_{2g}\zeta + E_{g}\theta$ distortions of octahedron.
Points show the {\it ab initio} calculated data and the continous lines show the calculations using the JT Hamiltonian. 
}
\label{Fig:APES_BLOO}
\end{figure*}

\begin{figure*}[tb]
\begin{center}
\begin{tabular}{ccc}
\multicolumn{1}{l}{(a)} & \multicolumn{1}{l}{(b)} & \multicolumn{1}{l}{(c)} \\
\includegraphics[width=5.5cm]{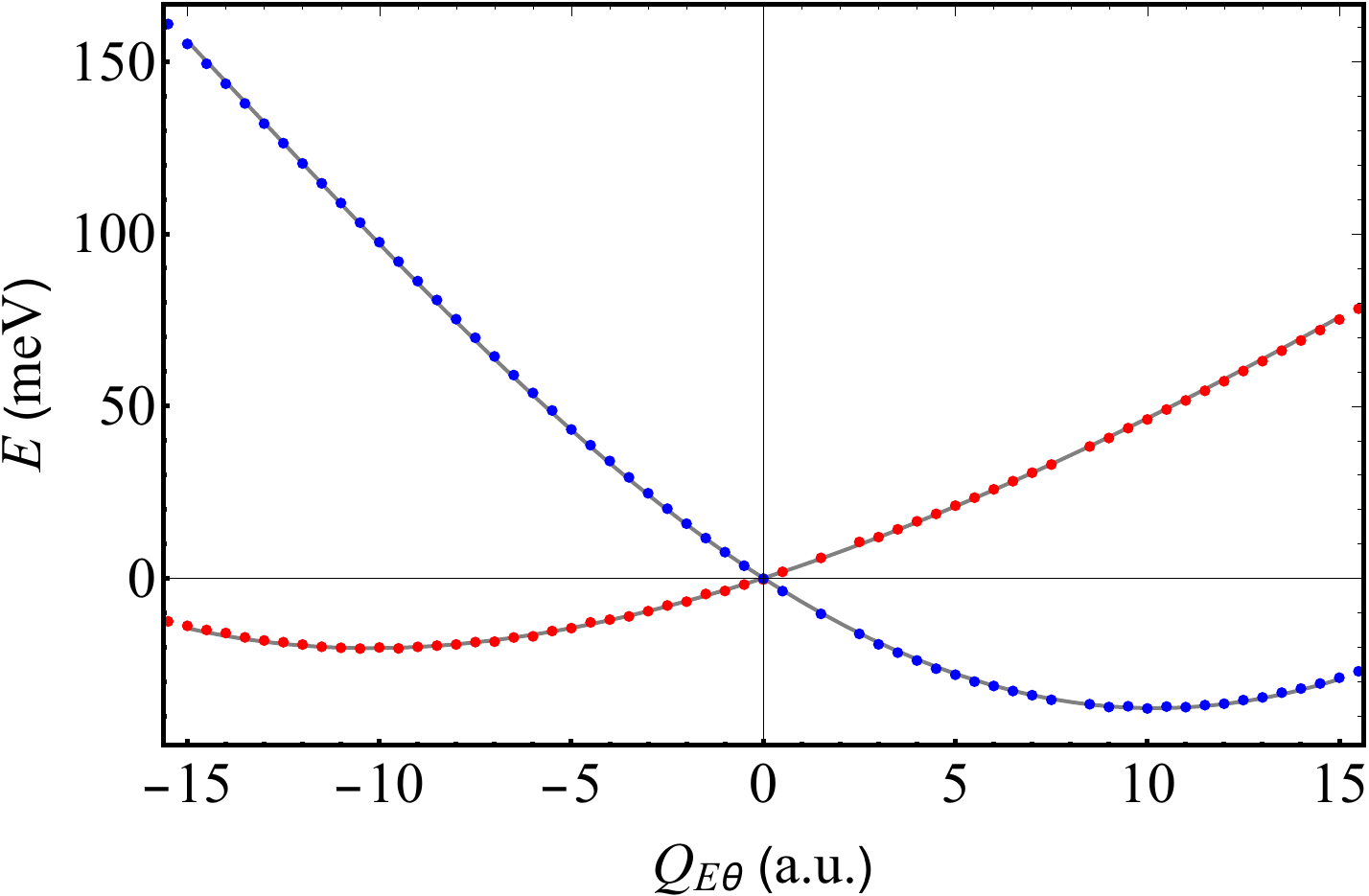}
&
\includegraphics[width=5.5cm]{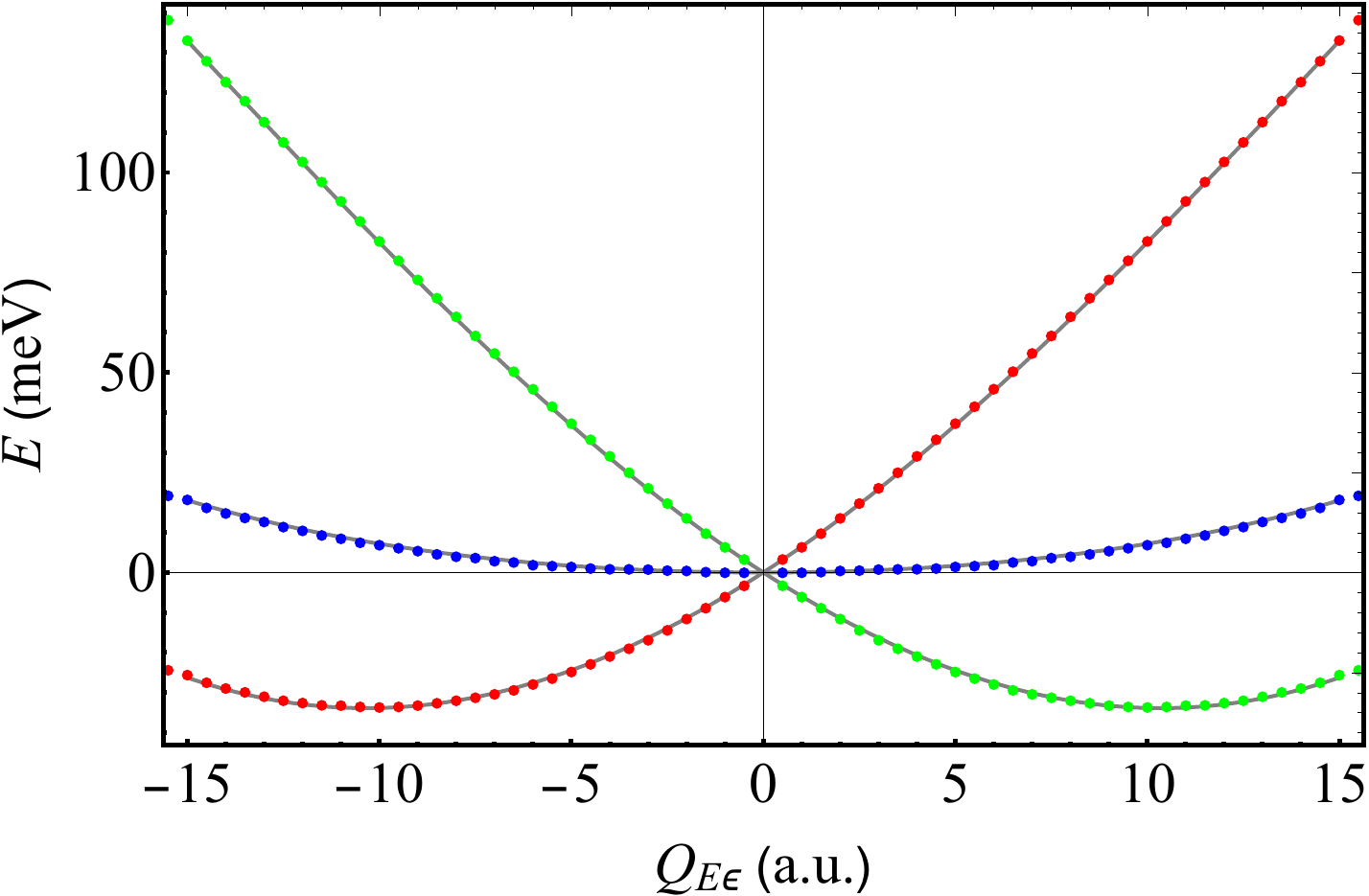}
&
\includegraphics[width=5.5cm]{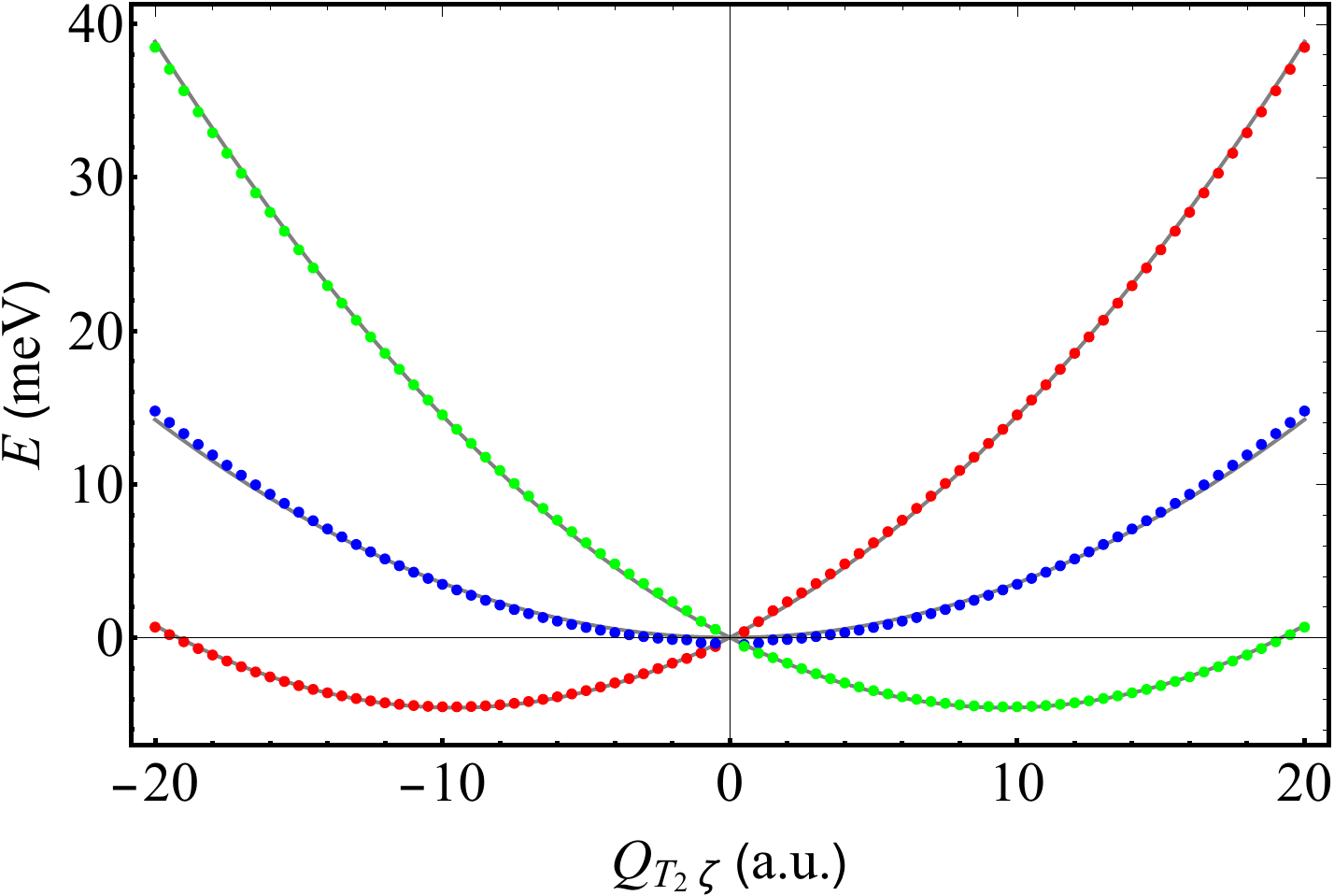}
\\
\multicolumn{1}{l}{(d)} & \multicolumn{1}{l}{(e)} \\
\includegraphics[width=5.5cm]{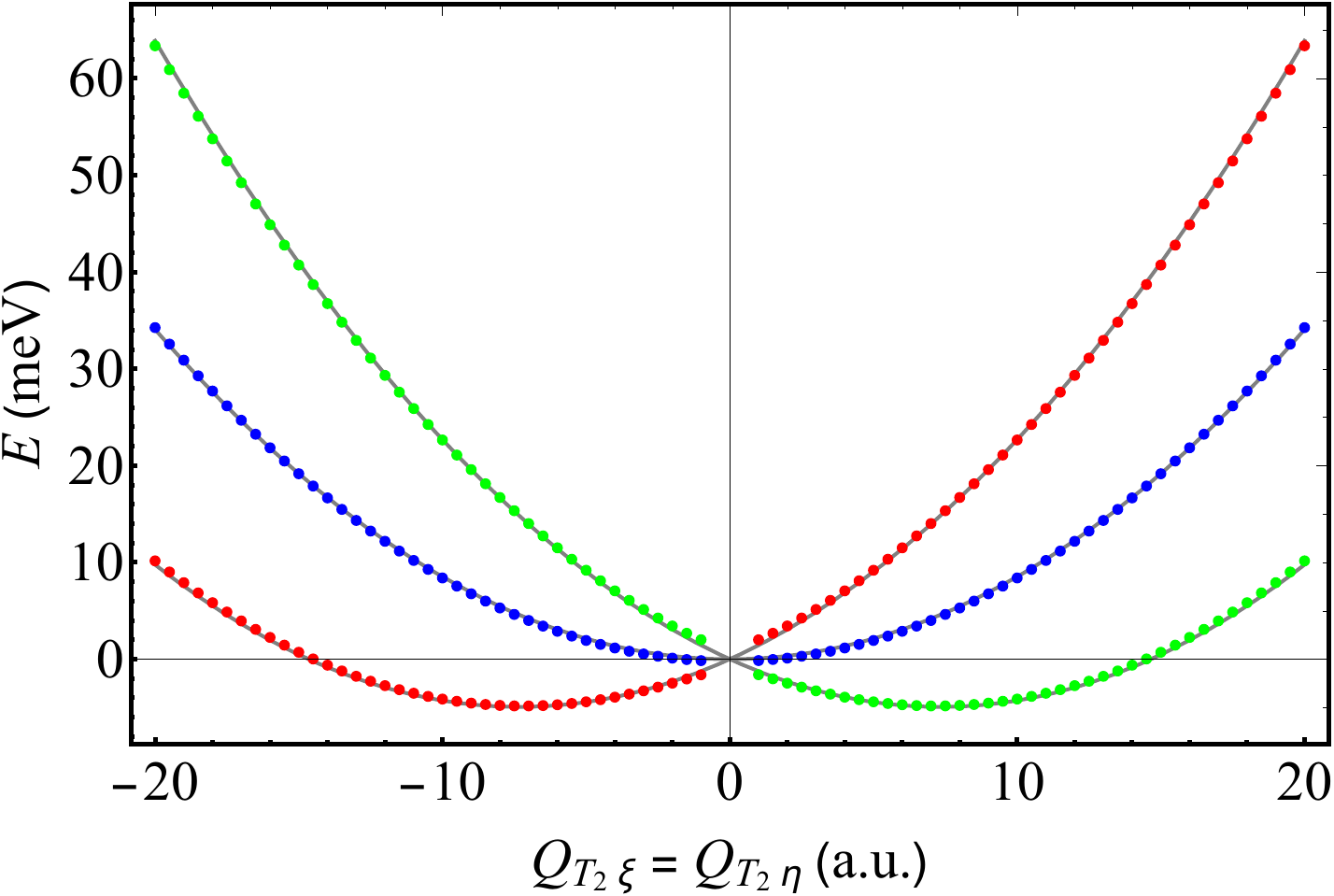}
&
\includegraphics[width=5.5cm]{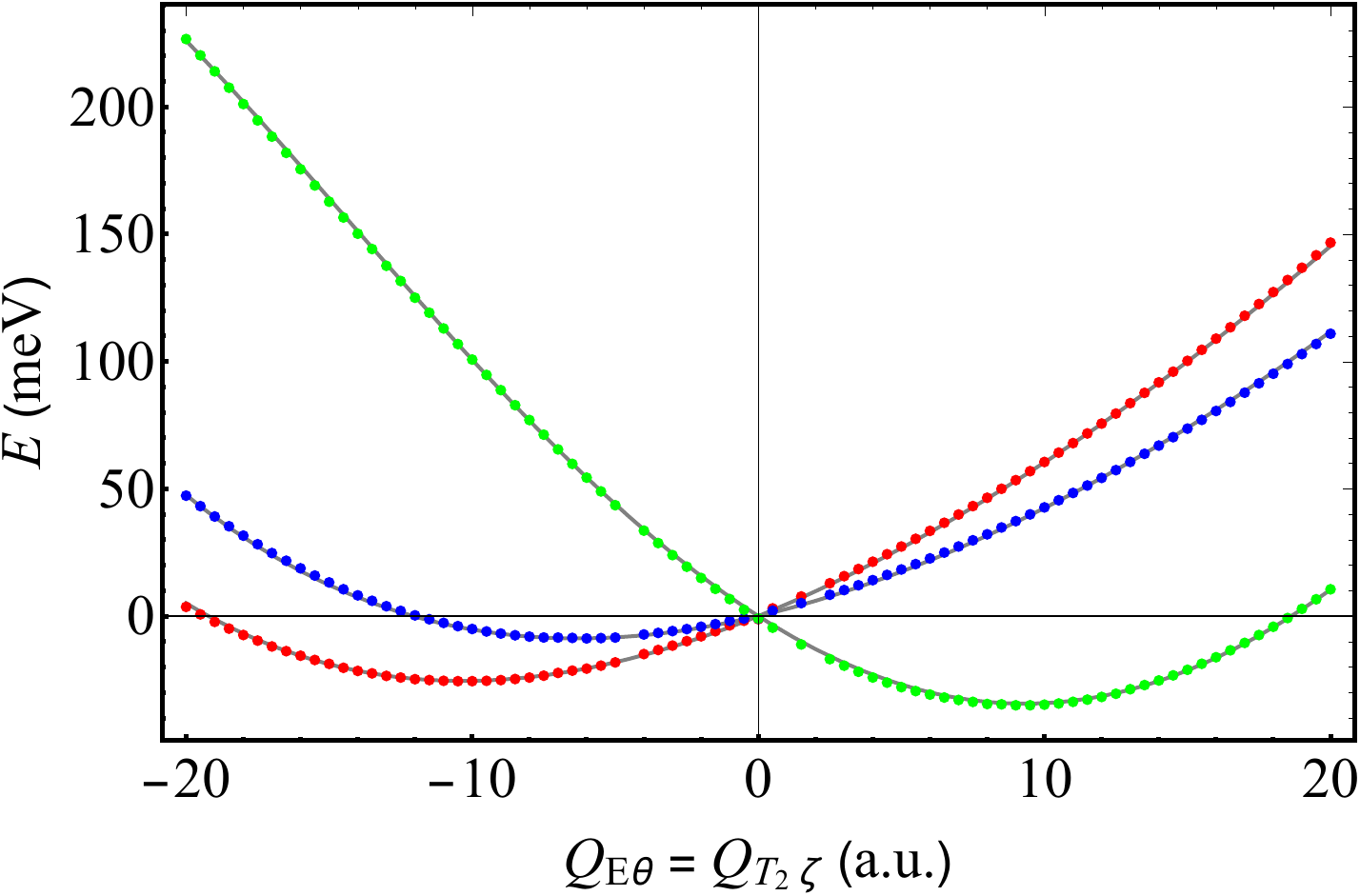}
\end{tabular}
\end{center}
\caption{
APES of Ba$_2$NaOsO$_6$ with (a) $E_{g}\theta$, (b) $E_g \epsilon$, (c) $T_{2g}\zeta$, (d) $T_{2g}\xi + T_{2g}\eta$, and (e) $T_{2g}\zeta + E_{g}\theta$ distortions of octahedron.
Points show the {\it ab initio} calculated data and the continous lines show the calculations using the JT Hamiltonian. 
}
\label{Fig:APES_BNOO}
\end{figure*}

\begin{figure*}[tb]
\begin{center}
\begin{tabular}{ccc}
\multicolumn{1}{l}{(a)} & \multicolumn{1}{l}{(b)} & \multicolumn{1}{l}{(c)} \\
\includegraphics[width=5.5cm]{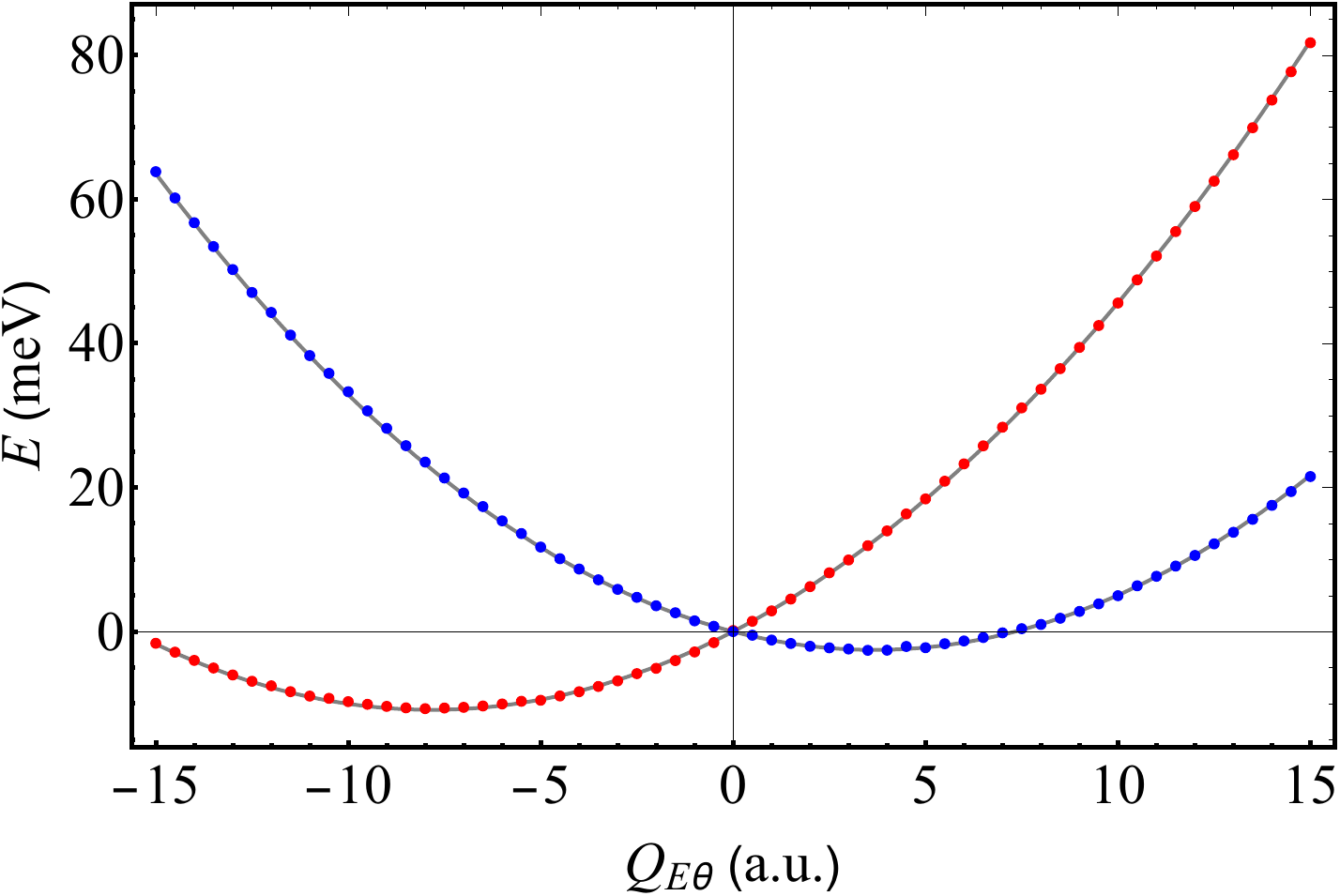}
&
\includegraphics[width=5.5cm]{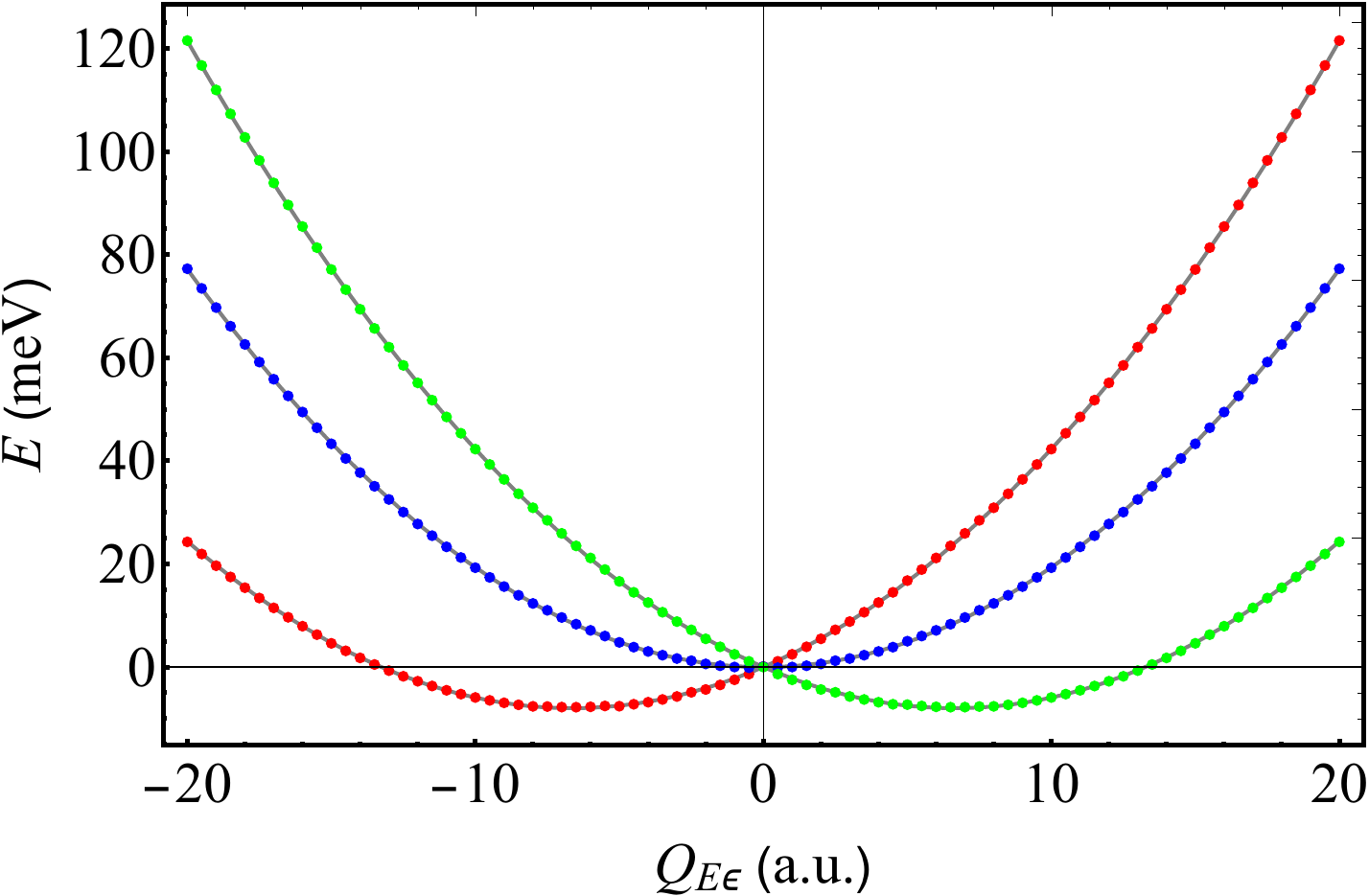}
&
\includegraphics[width=5.5cm]{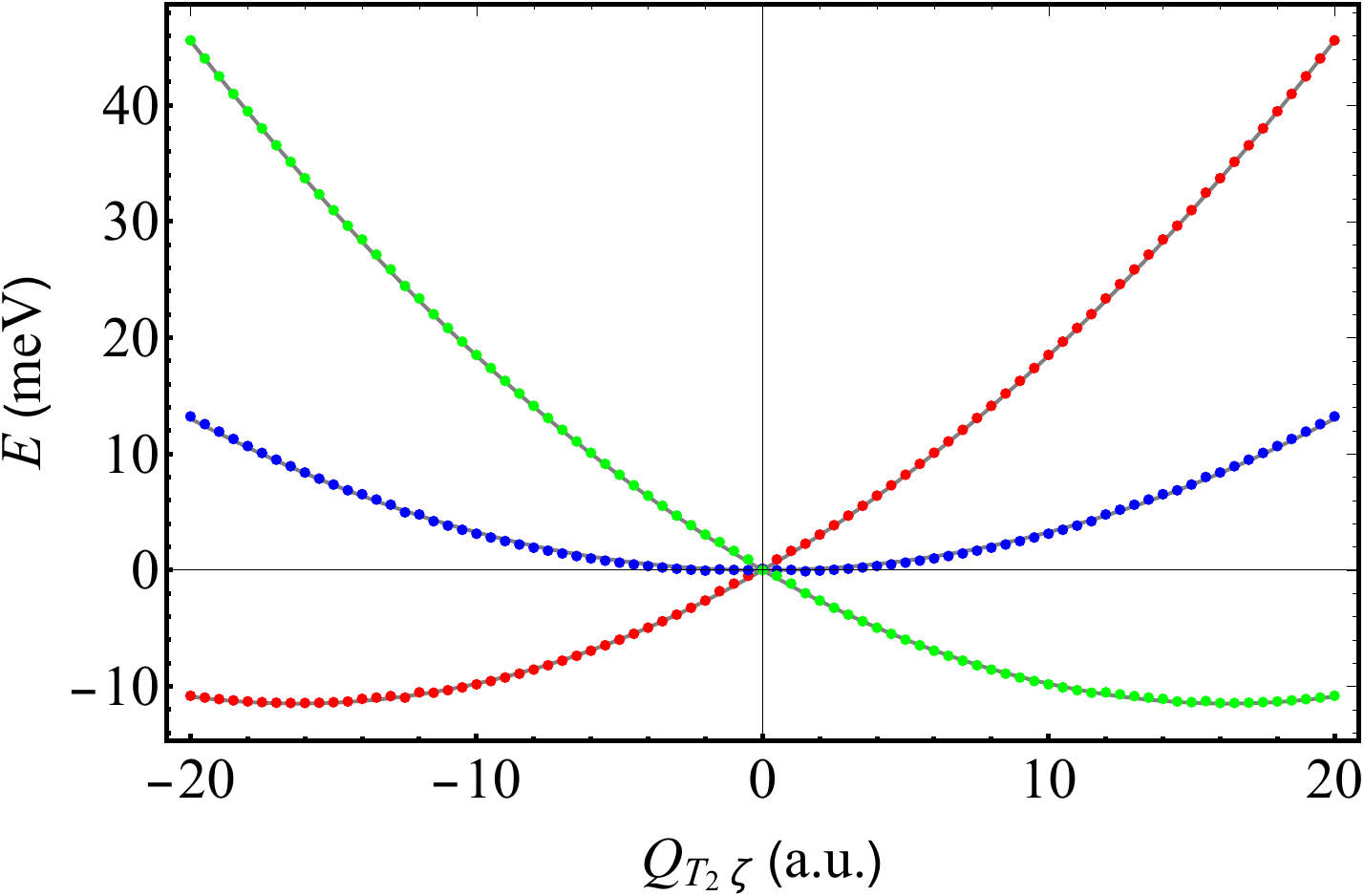}
\\
\multicolumn{1}{l}{(d)} & \multicolumn{1}{l}{(e)} \\
\includegraphics[width=5.5cm]{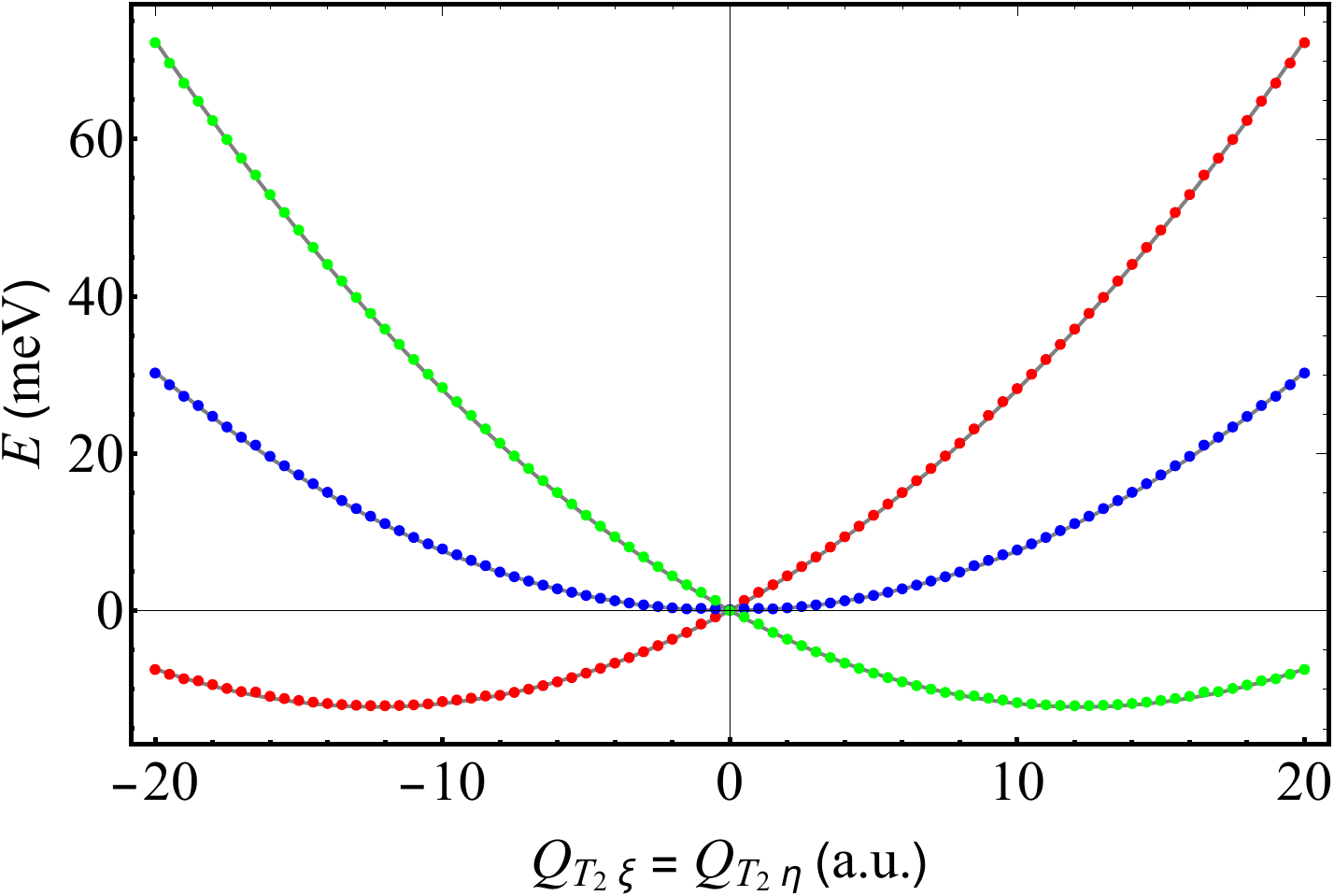}
&
\includegraphics[width=5.5cm]{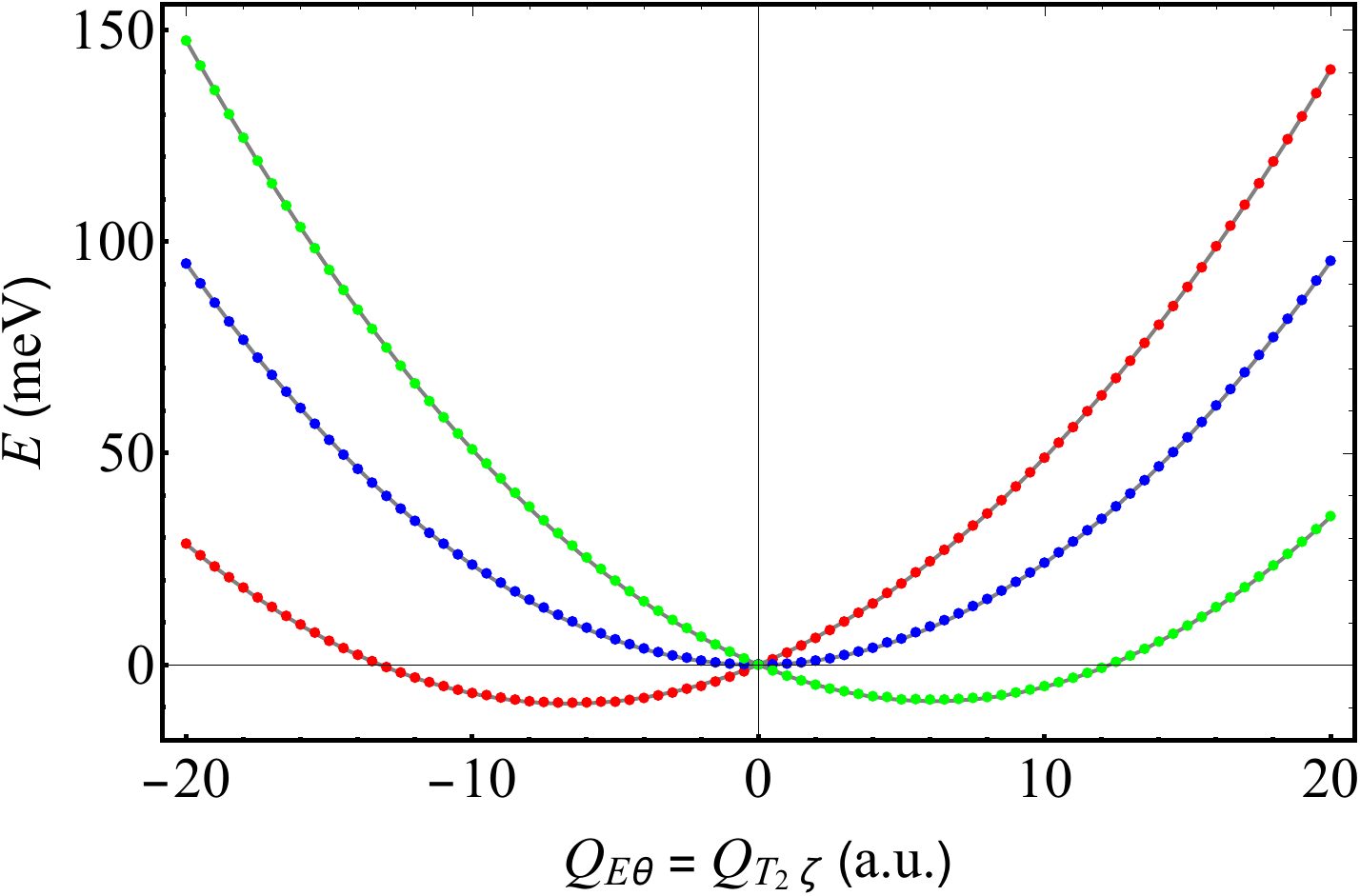}
\end{tabular}
\end{center}
\caption{
APES of Ba$_2$YMoO$_6$ with (a) $E_{g}\theta$, (b) $E_g \epsilon$, (c) $T_{2g}\zeta$, (d) $T_{2g}\xi + T_{2g}\eta$, and (e) $T_{2g}\zeta + E_{g}\theta$ distortions of octahedron.
Points show the {\it ab initio} calculated data and the continous lines show the calculations using the JT Hamiltonian. 
}
\label{Fig:APES_BYMO}
\end{figure*}

The fittings of the {\it ab initio} ${}^2T_{2g}$ APES to the model $t_{2g} \otimes (e_g \oplus t_{2g})$ JT Hamiltonian are shown in Figs. \ref{Fig:APES_BLOO}, \ref{Fig:APES_BNOO}, and \ref{Fig:APES_BYMO}.

\section{Static Jahn-Teller effect}
\label{Sec:SJT}
Since the APES of $(\Gamma_7 \oplus \Gamma_8) \otimes (e_g \oplus t_{2g})$ JT system is complex, 
we first derive that of $\Gamma_8 \otimes (e_g \oplus t_{2g})$ JT problem.
This is a good starting point to find the global minima of the APES of the former system when the splitting between the two electronic terms are large enough. 
Based on the knowledge on the $\Gamma_8 \otimes (e_g \oplus t_{2g})$ JT system, the global minima of $(\Gamma_7 \oplus \Gamma_8) \otimes (e_g \oplus t_{2g})$ JT system is searched numerically. 

\subsection{APES of $\Gamma_8 \otimes (e_g \oplus t_{2g})$ model}
The expression of the adiabatic potential energy surface (APES) of the $\Gamma_8$ system is shown. 
The main part of the problem was done by Liehr \cite{SMLiehr1963} taking into account up to quadratic vibronic term.
In this work, we include up to forth order for the $E_g$ mode. 
Finding all minima of the APES is not so simple, we only show two types of minima as before \cite{SMLiehr1963}.

The static Jahn-Teller Hamiltonian is written as 
\begin{eqnarray}
 {H}_\text{SJT} &=& \left[\sum_{\Lambda \lambda} \frac{\omega_\Lambda^2}{2}Q_{\Lambda \lambda}^2 + U' \right] {I}
 + {V}_\text{JT},
\\
 {V}_\text{JT} &=& \sum_{\Lambda \lambda} \Theta_{\Lambda \lambda} {C}_{\Lambda \lambda}.
\end{eqnarray}
Here, $\Theta_{\Lambda \lambda}$ is the polynomial of coordinates defined by
\begin{eqnarray}
 \Theta_{E\lambda} &=& 
 \frac{v_E}{2} Q_{E\lambda} 
+ \frac{1}{2!}\frac{v_E^{EE}}{2} \{Q_{E} \otimes Q_{E}\}_{E\lambda}
\nonumber\\
 &+& \frac{1}{2!}\frac{v_E^{T_2T_2}}{2} \{Q_{T_2} \otimes Q_{T_2}\}_{E\lambda}
\nonumber\\
 &+& \frac{1}{3!}\frac{v_E^{EEE}}{2} \{Q_E \otimes Q_E \otimes Q_E \}_{E\lambda}
\nonumber\\
 &+& \sum_{n=1,2} \frac{1}{4!}\frac{v_{nE}^{EEEE}}{2} \{Q_E \otimes Q_E \otimes Q_E \otimes Q_E \}_{nE\lambda},
\nonumber\\
 \Theta_{T_2\lambda} &=&
 \frac{v_{T_2}}{\sqrt{6}} Q_{T_2\lambda} 
+ \frac{1}{2!}\frac{v_{T_2}^{T_2T_2}}{\sqrt{6}} \{Q_{T_2} \otimes Q_{T_2}\}_{T_2\lambda} 
\nonumber\\
 &+& \frac{1}{2!} \frac{v_{T_2}^{ET_2}}{\sqrt{6}} \{Q_{E} \otimes Q_{T_2}\}_{T_2\lambda}.
\end{eqnarray}
$\{Q_{\Lambda_1} \otimes ... \otimes Q_{\Lambda_n}\}_{\Lambda \lambda}$ indicate the symmetrized products of $n$ coordinates. 
$I$ is four dimensional identity matrix, and ${C}$'s are defined as follows:
\begin{eqnarray}
 {C}_{E\theta} &=&
 \begin{pmatrix}
  1 & 0 & 0 & 0 \\
  0 & -1& 0 & 0 \\
  0 & 0 & -1& 0 \\
  0 & 0 & 0 & 1 
 \end{pmatrix},
\quad
 {C}_{E\epsilon} = 
 \begin{pmatrix}
  0 & 0 & 1 & 0 \\
  0 & 0 & 0 & 1 \\
  1 & 0 & 0 & 0 \\
  0 & 1 & 0 & 0 
 \end{pmatrix},
\nonumber\\
 {C}_{T_2\xi} &=&
 \begin{pmatrix}
  0 & i & 0 & 0 \\
 -i & 0 & 0 & 0 \\
  0 & 0 & 0 & -i \\
  0 & 0 & i & 0 
 \end{pmatrix},
\quad
 {C}_{T_2\eta} =
 \begin{pmatrix}
  0 & 1 & 0 & 0 \\
  1 & 0 & 0 & 0 \\
  0 & 0 & 0 &-1 \\
  0 & 0 &-1 & 0 
 \end{pmatrix},
\nonumber\\
 {C}_{T_2\zeta} &=& 
 \begin{pmatrix}
   0 & 0 &-i & 0 \\
   0 & 0 & 0 &-i \\
   i & 0 & 0 & 0 \\
   0 & i & 0 & 0 
 \end{pmatrix}.
\end{eqnarray}
The basis of the matrices is in the increasing order of $M$ of $|\Gamma_8,M\rangle$. 
In this section, the operator sign $~\hat{}~$ is omitted because coordinate representation is used. 

Expressing $\Theta_{\Lambda \lambda}$ by polar coordinates, 
\begin{eqnarray}
 \Theta_{E\theta} &=& X \cos \gamma \cos \phi, 
\nonumber\\
 \Theta_{E\epsilon} &=& X \cos \gamma \sin \phi,
\nonumber\\
 \Theta_{T_2\xi} &=& X \sin \gamma \sin \alpha \cos \beta,
\nonumber\\
 \Theta_{T_2\eta} &=& X \sin \gamma \sin \alpha \sin \beta,
\nonumber\\
 \Theta_{T_2\zeta} &=& X \sin \gamma \cos \alpha,
\end{eqnarray}
the Jahn-Teller term ${V}_\text{JT}$ is diagonalized as follows:
\begin{eqnarray}
 D_R^\dagger D_{T_2}^\dagger D_E^\dagger {V}_{\text{JT}}
 D_E D_{T_2} D_R
 &=& 
 \begin{pmatrix}
  X & 0 & 0 & 0 \\
  0 & X & 0 & 0 \\
  0 & 0 &-X & 0 \\
  0 & 0 & 0 &-X \\
 \end{pmatrix}.
\nonumber\\
\end{eqnarray}
Here,
\begin{eqnarray}
 D_E &=& 
\left(
\begin{array}{cccc}
 \cos \frac{\phi}{2} & 0 & \sin \frac{\phi }{2} & 0 \\
 0 & \sin \frac{\phi }{2} & 0 & -\cos \frac{\phi}{2} \\
 \sin \frac{\phi }{2} & 0 & -\cos \frac{\phi }{2} & 0 \\
 0 & \cos \frac{\phi }{2} & 0 & \sin \frac{\phi }{2} \\
\end{array}
\right),
\nonumber\\
 D_{T_2} &=& 
 \left(
\begin{array}{cccc}
 e^{-\frac{i \beta }{2}} & 0 & 0 & 0 \\
 0 & e^{\frac{i \beta }{2}}  & 0 & 0 \\
 0 & 0 & e^{-\frac{i \beta }{2}} & 0 \\
 0 & 0 & 0 & e^{\frac{i \beta }{2}}  \\
\end{array}
\right)
\nonumber\\
 &\times&
 \left(
\begin{array}{cccc}
 0 & 0 & \sin \frac{\alpha }{2} & -\cos \frac{\alpha }{2} \\
 0 & 0 & \cos \frac{\alpha }{2} &  \sin \frac{\alpha }{2} \\
  \sin \frac{\alpha }{2} & -\cos \frac{\alpha }{2} & 0 & 0 \\
  \cos \frac{\alpha }{2} &  \sin \frac{\alpha }{2} & 0 & 0 \\
\end{array}
\right),
\nonumber\\
 D_{R} &=& 
 \left(
\begin{array}{cccc}
 -i \sin \frac{\gamma }{2} & 0 & i \cos \frac{\gamma }{2} & 0 \\
 0 & i \sin \frac{\gamma }{2} & 0 & -i \cos \frac{\gamma }{2} \\
 \cos \frac{\gamma }{2} & 0 & \sin \frac{\gamma }{2} & 0 \\
 0 & \cos \frac{\gamma }{2} & 0 & \sin \frac{\gamma }{2} \\
\end{array}
\right).
\end{eqnarray}
Thus, the adiabatic potential energy surface (APES) is written as 
\begin{eqnarray}
 U &=& \sum_{\Lambda \lambda} \frac{\omega_\Lambda^2}{2} Q_{\Lambda \lambda}^2 + U' - X. 
\label{Eq:APES}
\end{eqnarray}

Now, we search the extrema of the APES (\ref{Eq:APES}). 
To this end, first we transform the basis of the $T_{2g}$ representation into the trigonal one [Eq. (3.2-7) in Ref. \cite{SMLiehr1963}]:
\begin{eqnarray}
 \left(
  Q_{T_2 \xi}, Q_{T_2 \eta}, Q_{T_2 \zeta}
 \right)
 &=& 
 \left(
  Q'_{A}, Q'_{E\theta}, Q'_{E\epsilon}
 \right)
\nonumber\\
 &\times&
 \begin{pmatrix}
  \frac{1}{\sqrt{3}} &  \frac{1}{\sqrt{3}} & \frac{1}{\sqrt{3}} \\
 -\frac{1}{\sqrt{6}} & -\frac{1}{\sqrt{6}} & \sqrt{\frac{2}{3}} \\
  \frac{1}{\sqrt{2}} & -\frac{1}{\sqrt{2}} & 0 \\
 \end{pmatrix}.
\label{Eq:Q_trig}
\end{eqnarray}
In the new coordinate systems, $(Q'_{A}, Q'_{E\theta}, Q'_{E\epsilon})$, the first one transform totally symmetric under $C_3$ rotation, and
the second and the third ones transform as if they are $E\theta$ and $E\epsilon$ coordinates, respectively. 
Polar coordinates are introduced as
\begin{eqnarray}
Q_{E\theta} &=& \rho_E \cos \theta_E, 
\quad
Q_{E\epsilon} = \rho_E \sin \theta_E, 
\nonumber\\
Q'_{E\theta} &=& Q'_E \cos \theta'_{T_2}, 
\quad
Q'_{E\epsilon} = Q'_E \sin \theta'_{T_2},
\nonumber\\
Q_A' &=& \rho_{T_2}' \cos \gamma_{T_2}', \quad 
Q_E' = \rho_{T_2}' \sin \gamma_{T_2}'.
\label{Eq:Q_polar1}
\end{eqnarray}
Using the new coordinates, Eqs. (\ref{Eq:Q_trig}) and (\ref{Eq:Q_polar1}), 
the elastic term in Eq. (\ref{Eq:APES}) becomes 
\begin{eqnarray}
 U_0 = \frac{\omega_E^2}{2} \rho_E^2 + \frac{\omega_{T_2}^2}{2} \rho_{T_2}^{\prime 2} - \frac{v^{EEE}_{A_1}}{12} \rho_E^3 \cos 3\theta_E + \frac{v^{EEEE}_{A_1}}{48} \rho_E^4, 
\nonumber\\
\end{eqnarray}
and $X^2 = \Theta_{E\theta}^2 + \Theta_{E\epsilon}^2 + \Theta_{T_2\xi}^2 + \Theta_{T_2\eta}^2 + \Theta_{T_2\zeta}^2$
is calculated as 
\begin{widetext}
\begin{eqnarray}
 X^2 &=& 
 \frac{v_E^2}{4} \rho_E^2 
+ \left(\frac{v_E v_E^{EEE}}{24} + \frac{(v_E^{EE})^2}{32}\right) \rho_E^4 
+ \left(\frac{(v_E^{EEE})^2}{576} + \frac{v^{EE}_E v^{EEEE}_{1E}}{192\sqrt{2}}\right) \rho_E^6
+ \left(\frac{(v_{1E}^{EEEE})^2}{9216} + \frac{(v^{EEEE}_{2E})^2}{18432}\right) \rho_E^8
\nonumber\\
 &+& \frac{v_{T_2}^2}{6} \rho_{T_2}^{\prime 2} 
 + \frac{(v_{T_2}^{ET_2})^2}{12} \rho_E^2 \rho_{T_2}^{\prime 2}
 + \frac{1}{24\sqrt{6}}\left( 3v_{T_2} \cos \gamma'_{T_2} + 5 v_{T_2} v_{T_2}^{T_2T_2} \cos 3\gamma'_{T_2} \right) \rho_{T_2}^{\prime 3}
\nonumber\\
 &+&
   \frac{(v_{E}^{T_2T_2})^2}{96} \left(\sin^4 \gamma'_{T_2} + 2\sin^4 2\gamma'_{T_2} \right) \rho_{T_2}^{\prime 4}
 + \frac{(v_{T_2}^{T_2T_2})^2}{144} \left(4 \cos^4 \gamma'_{T_2} + 3\sin^4 \gamma'_{T_2} \right) \rho_{T_2}^{\prime 4}
\nonumber\\
&-& \left\{
 \frac{v_E v_E^{EE}}{4\sqrt{2}} \rho_E^3
 + \frac{1}{96\sqrt{2}} \left[ v_E \left(\sqrt{2} v_{1E}^{EEEE} + v_{2E}^{EEEE} \right) + 2 v_E^{EE} v_E^{EEE} \right] \rho_E^5
 \right.
\nonumber\\
&+&
 \left.
   \frac{1}{1152\sqrt{2}} v_E^{EEE} \left(\sqrt{2} v_{1E}^{EEEE} + v_{2E}^{EEEE} \right) \rho_E^7
 + \frac{v_{T_2} v_{T_2}^{T_2T_2}}{6\sqrt{3}} \sin^3 \gamma'_{T_2} \rho_{T_2}^{\prime 3} 
 \right.
\nonumber\\
&+&
 \left.
   \frac{3(v_E^{T_2T_2})^2 - 2(v_{T_2}^{T_2T_2})^2}{36\sqrt{2}} \cos \gamma'_{T_2} \sin^3 \gamma'_{T_2} \rho_{T_2}^{\prime 4}
 \right\} \cos 3\theta_E
\nonumber\\
&+&
 \left( \frac{v_{E}^{EE} v_{2E}^{EEEE}}{384} \rho_E^6 + \frac{v_{1E}^{EEEE} v_{2E}^{EEEE}}{4608\sqrt{2}} \rho_E^8 \right) \cos 6\theta_E
\nonumber\\
&+&
 \left[ 
   \left( \frac{v_{T_2} v_{T_2}^{ET_2}}{6} + \frac{v_E v_E^{T_2T_2}}{4\sqrt{6}} \right) \rho_E \rho_{T_2}^{\prime 2} 
 + \frac{v_{E}^{T_2T_2} v_E^{EEE}}{48\sqrt{6}} \rho_E^3 \rho_{T_2}^{\prime 2} 
 - \frac{v_E^{T_2T_2} v_{1E}^{EEEE}}{192\sqrt{3}} \sin 2\gamma_T' \rho_E^4 \rho_{T_2}^{\prime 2}
 \right]
 \cos \left(\theta_E + 2\theta_{T_2}'\right) 
\nonumber\\
&+&
   \left( \frac{v_{T_2} v_{T_2}^{ET_2}}{3\sqrt{2}} + \frac{v_E v_E^{T_2T_2}}{4\sqrt{3}} \right) \rho_E \rho_{T_2}^{\prime 2} \sin 2\gamma_T'
 \cos \left( \theta_E - \theta_{T_2}' \right)
\nonumber\\
&+& 
 \left( \frac{(v_{T_2}^{ET_2})^2}{12\sqrt{2}} - \frac{v_E^{EE} v_E^{T_2T_2}}{8\sqrt{6}} \right) \rho_E^2 \rho_{T_2}^{\prime 2} \sin 2\gamma_T'
 \cos \left( 2\theta_E + \theta_{T_2}'\right) 
\nonumber\\
&+&
 \left[
 \left( \frac{(v_{T_2}^{ET_2})^2}{24} - \frac{v_E^{EE} v_E^{T_2T_2}}{16\sqrt{3}} \right) \rho_E^2 \rho_{T_2}^{\prime 2} 
 - \frac{v_E^{T_2T_2} v_{1E}^{EEEE}}{192\sqrt{6}}  \sin^2 \gamma_{T_2}' \rho_E^4 \rho_{T_2}^{\prime 2}
 \right] 
 \cos \left(2\theta_E - 2\theta_{T_2}' \right)
\nonumber\\
&-&
 \frac{v_E^{T_2T_2} v_{2E}^{EEEE}}{384\sqrt{3}} \sin^2 \gamma_T' \cos \left(4\theta_E + 2\theta_{T_2}'\right) \rho_E^2 \rho_{T_2}^{\prime 4} 
- \frac{v_E^{T_2T_2} v_{2E}^{EEEE}}{192\sqrt{6}} \sin 2\gamma_T' \cos \left(4\theta_E - \theta_{T_2}'\right) \rho_E^2 \rho_{T_2}^{\prime 4}.
\label{Eq:X2}
\end{eqnarray}
It is readily found that each term is invariant under the $C_3$ rotation ($n = 0, \pm 1$): 
\begin{eqnarray*}
 \theta_E \rightarrow \theta_E + \frac{2\pi}{3} n, 
\quad
 \theta_T' \rightarrow \theta_T' + \frac{2\pi}{3} n.
\end{eqnarray*}

The derivatives of $U$ with respect to $\theta_E$ and $\theta_{T_2}'$,
\begin{eqnarray}
 \frac{\partial U}{\partial \theta_E} &=& 
 -\frac{v_{A_1}^{EEE}}{4} \rho_E^3 \sin 3\theta_E - \frac{1}{2X^2} \frac{\partial X^2}{\partial \theta_E},
\quad
 \frac{\partial U}{\partial \theta_{T_2}'} = -\frac{1}{2X^2} \frac{\partial X^2}{\partial \theta_{T_2}'},
\end{eqnarray}
become zero for arbitrary $\rho_E$, $\rho_{T_2}'$, and $\gamma_{T_2}'$ when
\begin{eqnarray}
 \theta_E = \frac{\pi}{3} j, \quad \theta'_{T_2} = \frac{\pi}{3} (j - 3k), 
\label{Eq:angle1}
\end{eqnarray}
for integers $j,k$. 
Similarly, the derivative of $U$ with respect to $\gamma_T'$ at Eq. (\ref{Eq:angle1}) is 
\begin{eqnarray}
 2X^2 \frac{\partial U}{\partial \gamma_{T_2}'} &=& \frac{\partial X^2}{\partial \gamma_{T_2}'}
\nonumber\\
 &=&
 \left\{
   \frac{1}{8\sqrt{3}} v_{T_2} v_{T_2}^{T_2T_2} \left[ -\sqrt{2} (3+5\cos 2\gamma'_{T_2}) + 2 (-1)^{j+k} \sin 2\gamma'_{T_2} \right] 
 \right.
\nonumber\\
  &+& 
 \left.
 \frac{1}{288} \left[ 3 (v_E^{T_2T_2})^2 - 2 (v_{T_2}^{T_2T_2})^2 \right] 
    \left[ 9 \cos \gamma'_{T_2} + 7 \cos 3\gamma'_{T_2} + (-1)^{j+k} 4\sqrt{2} \sin 3\gamma'_{T_2} \right] \rho_{T_2}^{\prime} 
 \right\}
 \rho_{T_2}^{\prime 3} 
 \sin \gamma'_{T_2} 
\nonumber\\
 &+& 
  \frac{1}{384 \sqrt{6}} 
 \left\{
  (-1)^j 16\sqrt{6} \left(4 v_{T_2} v_{T_2}^{ET_2} + \sqrt{6} v_E v_E^{T_2T_2} \right) 
  + 4\sqrt{6} \left[(2 v_{T_2}^{ET_2})^2 - \sqrt{3} v_E^{EE} v_E^{T_2T_2} \right] \rho_E
 \right.
\nonumber\\
 &+&
 \left.
   (-1)^j 8 v_E^{EEE} v_E^{T_2T_2} \rho_E^2 
  - \left(2v_{1E}^{EEEE} + \sqrt{2} v_{2E}^{EEEE} \right) \rho_E^3
 \right\}
 \rho_E \rho_{T_2}^{\prime 2}
 \left[ (-1)^{j+k} 2\sqrt{2} \cos 2\gamma'_{T_2} + \sin 2\gamma'_{T_2} \right].
\label{Eq:dX2dgT2}
\end{eqnarray}
There are two solutions that each term becomes zero (sufficient conditions for $\partial U/\partial \gamma_{T_2}' = 0$):
\begin{itemize}
 \item[(a)] $\gamma'_T = l \pi$ or $\sin \gamma'_T = (-1)^l \frac{2}{3}\sqrt{2}$ and $\cos \gamma'_T = -(-1)^{l+j-k}\frac{1}{3}$ ($l$: integers),
 \item[(b)] $\sin \gamma'_T = \pm (-1)^{j-k} \sqrt{\frac{2}{3}}$ and $\cos \gamma'_T = \pm \frac{1}{\sqrt{3}}$.
\end{itemize}
For each case, we derive the Jahn-Teller distortions. 
\begin{itemize}
\item[(a)]
First, we consider the former case, $\gamma'_T = l \pi$. 
Since $\sin \gamma'_T = 0$, the coefficients of the first and the second terms of Eq. (\ref{Eq:dX2dgT2}) become zero. 
The third and fourth lines also becomes zero when either $\rho_E = 0$ or 
\begin{eqnarray}
  (-1)^j 16\sqrt{6} \left(4 v_{T_2} v_{T_2}^{ET_2} + \sqrt{6} v_E v_E^{T_2T_2} \right) 
  &+& 4\sqrt{6} \left[(2 v_{T_2}^{ET_2})^2 - \sqrt{3} v_E^{EE} v_E^{T_2T_2} \right] \rho_E
\nonumber\\
  &+& (-1)^j 8 v_E^{EEE} v_E^{T_2T_2} \rho_E^2 
  - \left(2v_{1E}^{EEEE} + \sqrt{2} v_{2E}^{EEEE} \right) \rho_E^3 = 0.
\label{Eq:casea_rho}
\end{eqnarray}
The latter without cubic and quartic vibronic couplings is fulfilled when 
\begin{eqnarray}
 \rho_E = (-1)^j 4 \frac{4 v_{T_2} v_{T_2}^{ET_2} + \sqrt{6} v_E v_E^{T_2T_2}}{\sqrt{3} v_E^{EE} v_E^{T_2T_2} - (2v_{T_2}^{ET_2})^2}.
\end{eqnarray}
Since the denominator is much smaller than the numerator (first order coupling is usually stronger than the quadratic coupling), this deformation of the $E_g$ mode is far from that for the global minima (for large deformation, the energy becomes large due to elastic term). 
Thus, at least in the present case, the latter possibility can be ruled out. 
When $\rho_E = 0$, the APES becomes 
\begin{eqnarray}
 U &=& \frac{\omega_{T_2}^2}{2} \rho_{T_2}^{\prime 2} 
 - \frac{\rho'_{T_2}}{6} \left|\sqrt{6} v_{T_2} + (-1)^l v_{T_2}^{T_2T_2} \rho'_{T_2} \right|.
\label{Eq:APES1}
\end{eqnarray}
Choosing $l$ so that the signs of $v_{T_2}$ and $(-1)^l v_{T_2}^{T_2T_2}$ become the same, $\rho_{T_2}'$ for the minimum of the APES is obtained as 
\begin{eqnarray}
 \rho_{T_2}' &=& \sqrt{\frac{3}{2}} \frac{v_{T_2}}{3\omega_{T_2}^2 - v_{T_2}^{T_2T_2}},
\end{eqnarray}
and the energy is 
\begin{eqnarray}
 U = - \frac{v_{T_2}^2}{12\omega_{T_2}^2 - 4 (v_{T_2}^{T_2T_2})}.
\end{eqnarray}

Since $\sin \gamma'_{T_2} = 0$ and $\rho_E = 0$, the positions of extrema are described as 
\begin{eqnarray}
 Q_{E\theta} &=& Q_{E\epsilon} = Q'_{E\theta} = Q'_{E\epsilon} = 0, \quad Q'_{A} = (-1)^l \rho_{T_2}' \ne 0. 
\end{eqnarray}
From the definition of the trigonal coordinates, Eq. (\ref{Eq:Q_trig}), this deformation indicates that all $T_{2g}$ distortions equally contribute to the JT distortion, and the symmetry of the deformed structure has trigonal symmetry. 

Now, we consider the second case.
By the similar discussion, the potential energy surface reduces to 
\begin{eqnarray}
 U &=& \frac{\omega_{T_2}^2}{2} \rho_{T_2}^{\prime 2} 
- \frac{\rho'_{T_2}}{6} \left|\sqrt{6}v_{T_2} + (-1)^{j+k+l}v_{T_2}^{T_2T_2} \rho'_{T_2} \right|.
\end{eqnarray}
and $\rho_{T_2}'$ for the minima and the stabilization energy correspond to the first case.
The deformed structure is given by one of the trigonal structures for $(-1)^{j-k}=1$ or 
\begin{eqnarray}
 Q_\xi = Q_\eta = (-1)^l \frac{1}{3\sqrt{3}} \rho'_{T_2}, \quad Q_\zeta = -(-1)^l \frac{5}{3\sqrt{3}} \rho'_{T_2}
\end{eqnarray}
for $(-1)^{j-k} = -1$.

\item[(b)]
All terms of $\partial U/\partial \gamma'_{T_2}$ becomes zero. 
The APES becomes 
\begin{eqnarray}
 U &=& 
 \frac{\omega_E^2}{2} \rho_E^2 - \frac{v_{A_1}^{EEE}}{12} \rho_E^3 \cos 3\theta_E + \frac{v_{A_1}^{EEEE}}{48} \rho_E^4
+ \frac{\omega_{T_2}^2}{2} \rho_{T_2}^{\prime 2} 
- \left\{
   \left(
    \frac{v_E}{2}\rho_E - (-1)^j \frac{v_E^{EE}}{4\sqrt{2}} \rho_E^2 + \frac{v_E^{EEE}}{24} \rho_E^3
   \right.
   \right.
\nonumber\\
&-&
\left.
\left.
    (-1)^j \frac{v_{1E}^{EEEE}}{96} \rho_E^4 - (-1)^j \frac{v_{2E}^{EEEE}}{96\sqrt{2}} \rho_E^4 
   + (-1)^j \frac{v_E^{T_2T_2}}{2\sqrt{6}} \rho_{T_2}^{\prime 2}
   \right)^2
  + 
   \frac{1}{6} \left( v_{T_2} + v_{T_2}^{ET_2} \rho_E \right)^2 \rho_{T_2}^{\prime 2}
  \right\}^\frac{1}{2}.
\label{Eq:APES2}
\end{eqnarray}
$\rho_E$ and $\rho_{T_2}'$ are determined by minimizing the APES. 
One of the $T_2$ coordinates for the minima is given by 
\begin{eqnarray}
 Q_\xi = Q_\eta = 0, \quad Q_\zeta = \pm \rho_{T_2}'.
\end{eqnarray}
Here, $(-1)^{j-k} = 1$.
\end{itemize}
The minima of the $\Gamma_8 \otimes (e_g \oplus t_{2g})$ JT system is found (i) along $Q'_A$ or (ii) in the $Q_\theta-Q_\zeta$ plane. 
\end{widetext}

\begin{figure*}[tbh]
\begin{tabular}{lll}
(a) Ba$_2$LiOsO$_6$ & (b) Ba$_2$NaOsO$_6$ & (c) Ba$_2$YMoO$_6$ \\
\includegraphics[height=4.5cm]{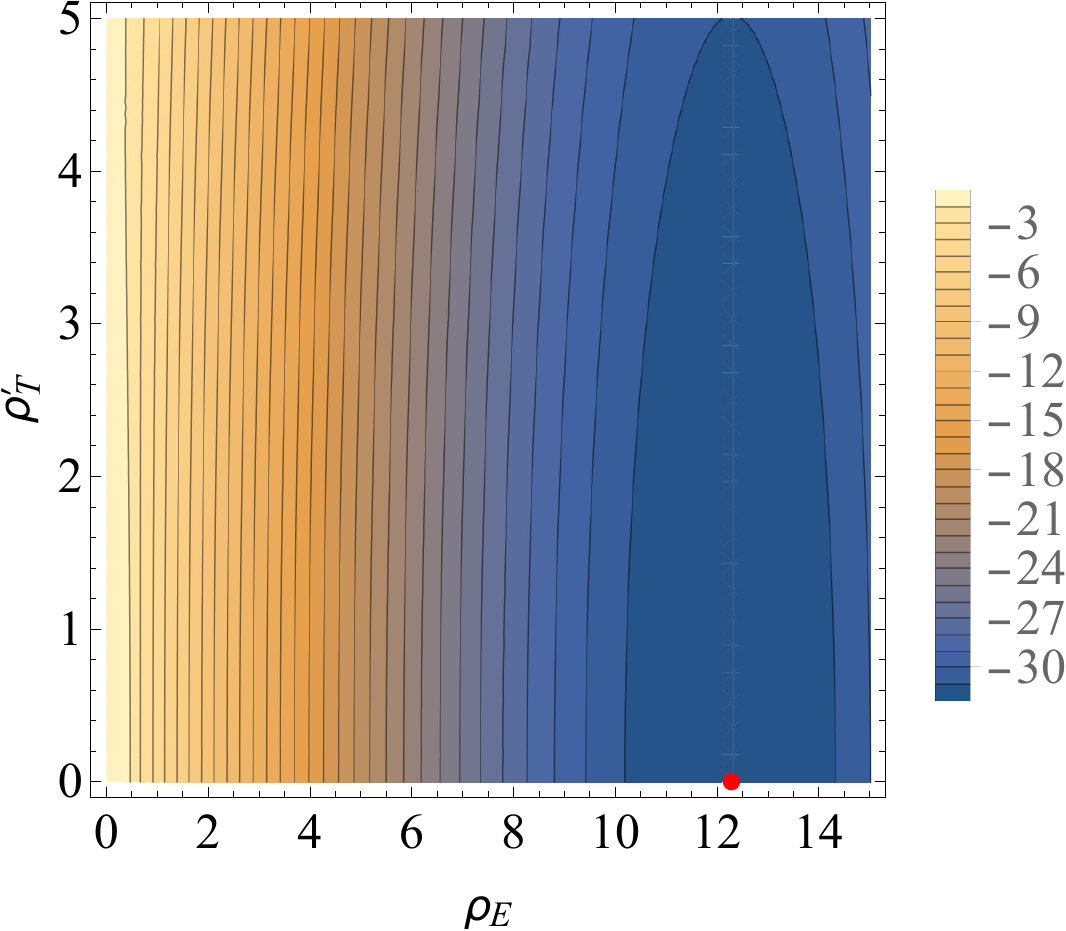}
&
\includegraphics[height=4.5cm]{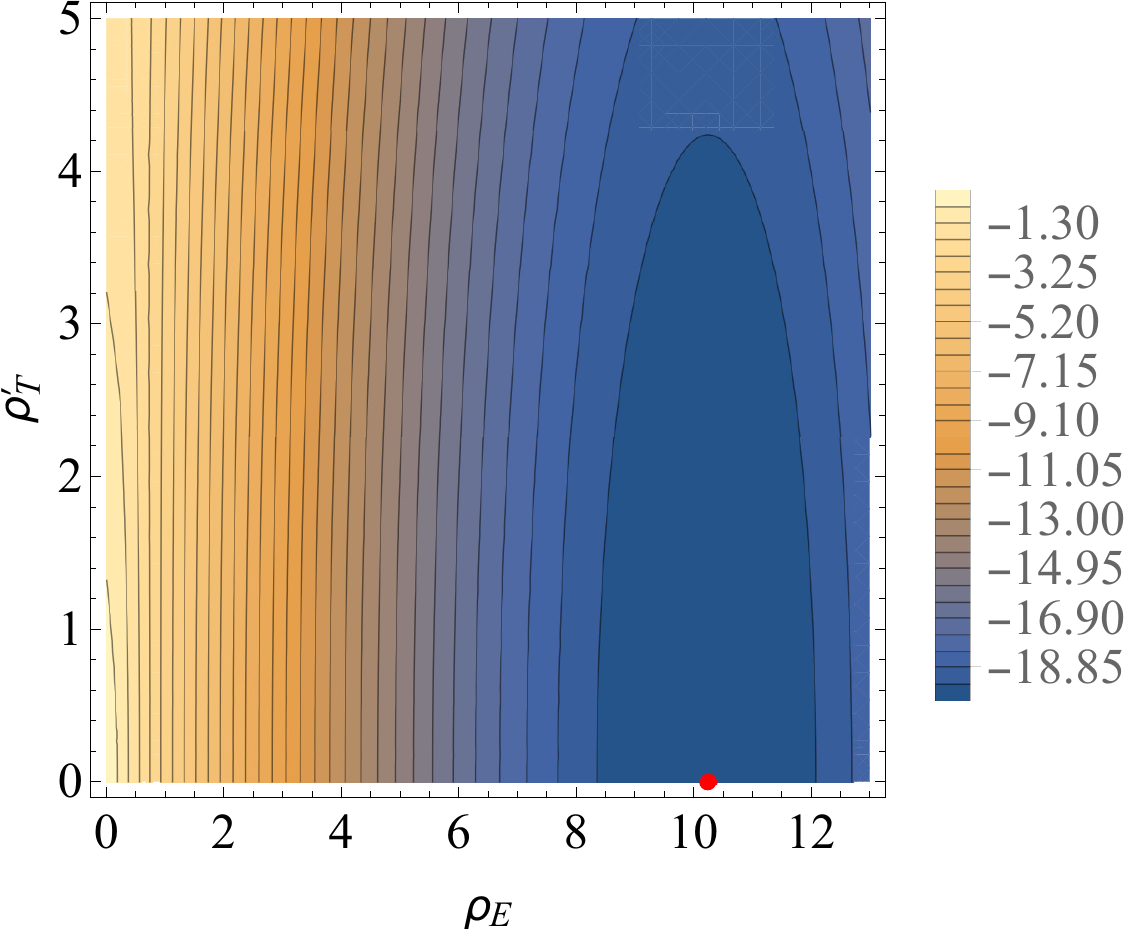}
&
\includegraphics[height=4.5cm]{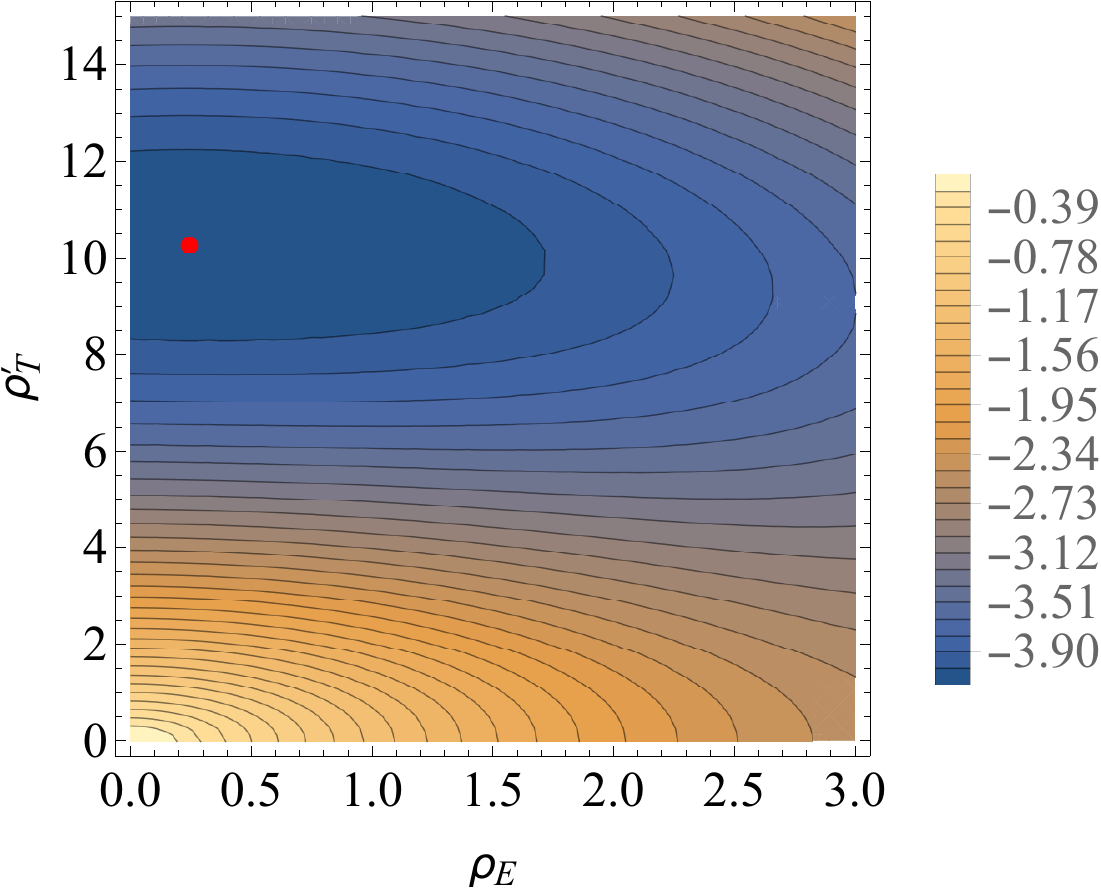}
\\
(d) Ba$_2$LiOsO$_6$ & (e) Ba$_2$NaOsO$_6$ & (f) Ba$_2$YMoO$_6$ \\
\includegraphics[height=4.5cm]{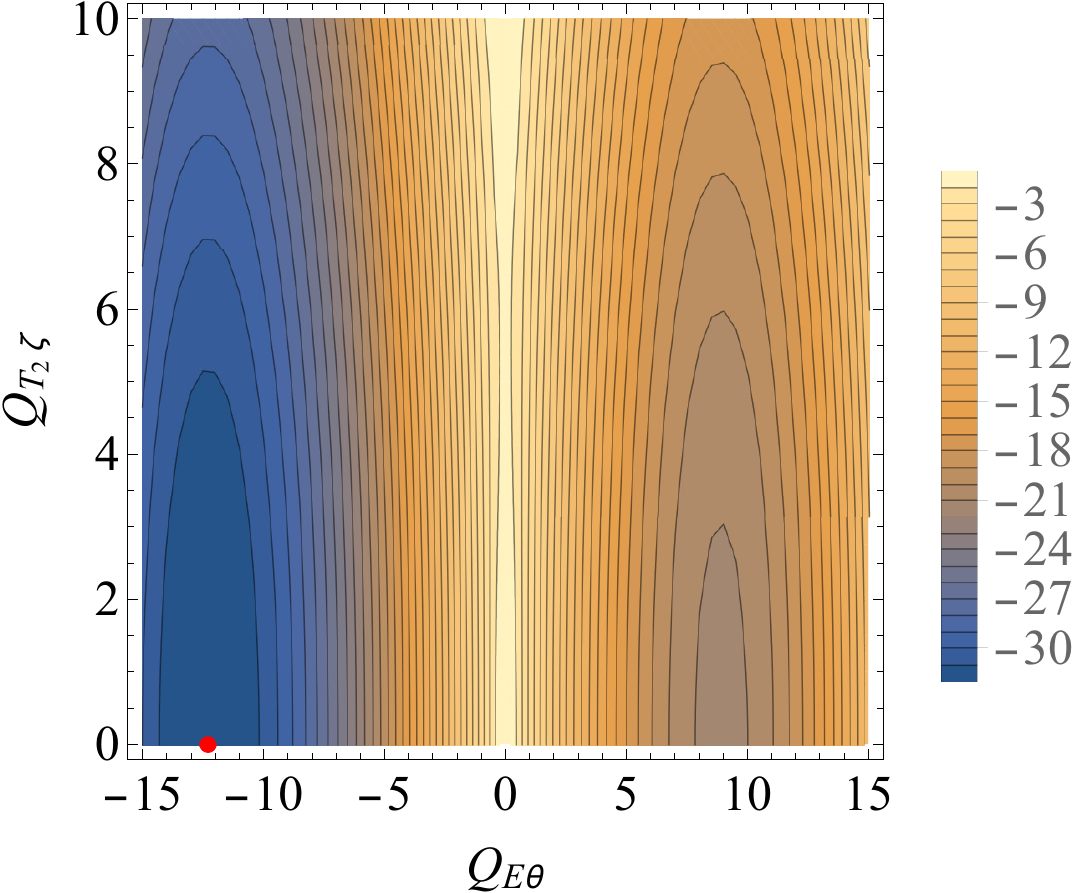}
&
\includegraphics[height=4.5cm]{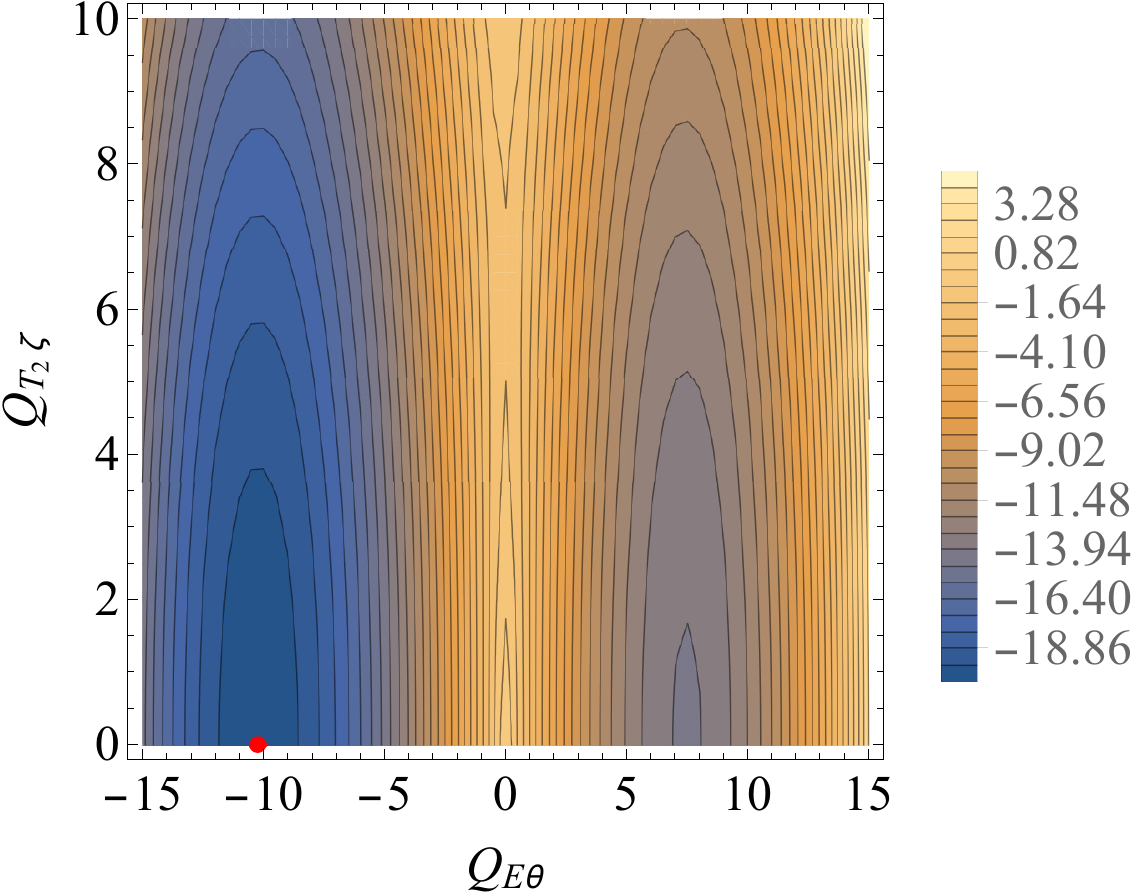}
&
\includegraphics[height=4.5cm]{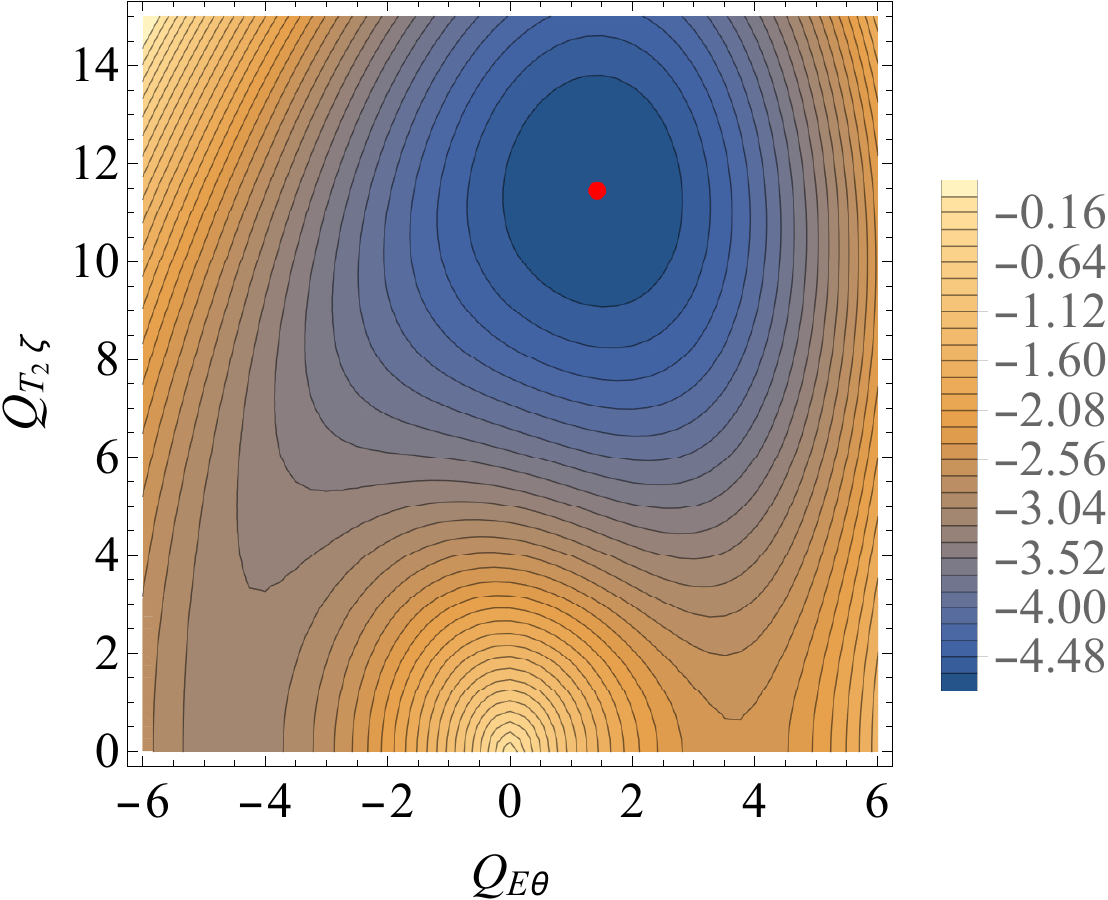}
\end{tabular}
\caption{APES's of $\Gamma_8 \otimes (e_g \oplus t_{2g})$ JT model, Eq. (\ref{Eq:APES2}), for (a) Ba$_2$LiOsO$_6$, (b) Ba$_2$NaOsO$_6$ and (c) Ba$_2$YMoO$_6$, and 
those of $(\Gamma_7 \oplus \Gamma_8) \otimes (e_g \oplus t_{2g})$ JT model for (d) Ba$_2$LiOsO$_6$, (e) Ba$_2$NaOsO$_6$ and (f) Ba$_2$YMoO$_6$. 
The red point correspond to the global minimum. The unit of energy is meV and the unit of coordinate is atomic unit (a.u.).}
\label{Fig:APES}
\end{figure*}




\begin{table}[tbh]
\begin{ruledtabular}
\caption{
Cartesian coordinates of oxygens corresponding to the global minima of the APES (\AA).
The position of metal is the origin.
The bond length between metal and oxygen at high-symmetric structures are 
1.888372, 1.86955, 1.96138 {\AA} for Ba$_2$LiOsO$_6$, Ba$_2$NaOsO$_6$, and Ba$_2$YMoO$_6$, respectively.
} 
\label{Table:coord}
\begin{tabular}{cccccc}
 \multicolumn{3}{c}{Ba$_2$LiOsO$_6$} & \multicolumn{3}{c}{Ba$_2$NaOsO$_6$} \\
 $x$ & $y$ & $z$ & $x$ & $y$ & $z$ \\
\hline
  0        & 1.8993    & 0         & 0&   1.87871 & 0 \\ 
  0        & 0         &   1.8993  & 0&  0&  1.87871  \\ 
  1.96392  & 0         & 0         & 1.86638 &  0& 0  \\ 
  0        & $-1.8993$ & 0         & 0 & $-1.87871$& 0 \\ 
  0        & 0         & $-1.8993$ & 0 & 0& $-1.87871$ \\ 
$-1.96392$ & 0         & 0         & $-1.86638$& 0& 0  \\ 
\hline
 \multicolumn{3}{c}{Ba$_2$YMoO$_6$} \\
 $x$ & $y$ & $z$ \\ 
\hline
0 &  1.96011 &  0.01776  \\
0 &  0.01776 &  1.96011  \\
 1.85123  &  0       &  0 \\
 0 &$-1.96011$&$-0.01776$ \\
 0 &$-0.01776$&$-1.96011$ \\
  $-1.85123$ &  0       &0 \\
\end{tabular}
\end{ruledtabular}
\end{table}

\subsection{APES of $d^1$ double perovskites}
The APES of $\Gamma_8 \otimes (e_g \oplus t_{2g})$ JT model and the APES of $(\Gamma_7 \otimes \Gamma_8) \otimes (e_g \oplus t_{2g})$ JT model were shown in Fig. \ref{Fig:APES}.
The position of minima and energy are listed in Table II in the main text and the Cartesian coordinates of the JT deformed structures are shown in Table \ref{Table:coord}.
The JT distortion of Ba$_2A$OsO$_6$ and that of Ba$_2$YMoO$_6$ are of $e_g$ and $t_{2g}$ types, respectively.
The coordinates of minima of Ba$_2A$OsO$_6$ remain almost the same by the pseudo JT coupling because it mainly modifies the APES around the saddle point. 
On the other hand, the position of minima of Ba$_2$YMoO$_6$ change by pseudo JT coupling ($Q_\theta = 0.24, Q_\zeta = 10.28$ without pseudo JT coupling). 
In the case of Ba$_2$YMoO$_6$, the minima along trigonal deformation ($Q_{E\theta} = Q_{E\epsilon} = 0$ and $Q_{T_2\xi} = Q_{T_2\eta} = Q_{T_2 \zeta}$) were also searched. 
The local minima was found at $Q_{T_2\xi} = Q_{T_2\eta} = Q_{T_2 \zeta} = 10.25$ a.u. with $U = -4.17$ meV.
This indicate, there is only small gap between the minima and saddle in the trough of the APES.

\begin{table}[tb]
\caption{Low-energy vibronic levels and assignments (meV). ``L'' and ``NL'' indicate that the energy levels including only linear and nonlinear vibronic couplings, respectively.}
\label{SMTable:vibro}
\begin{ruledtabular}
\begin{tabular}{cccccccc}
 \multicolumn{4}{c}{Ba$_2$LiOsO$_6$} & 
 \multicolumn{4}{c}{Ba$_2$NaOsO$_6$} \\
 \multicolumn{2}{c}{L} & \multicolumn{2}{c}{NL} &
 \multicolumn{2}{c}{L} & \multicolumn{2}{c}{NL} \\
\hline
$-63.41$& $\Gamma_8$ &$-27.37$& $\Gamma_8$ &$-40.90$& $\Gamma_8$ &$-20.24$& $\Gamma_8$  \\
$-16.68$& $\Gamma_7$ &  18.82 & $\Gamma_6$ &   6.12 & $\Gamma_7$ &  25.06 & $\Gamma_6$  \\
$-14.38$& $\Gamma_8$ &  20.44 & $\Gamma_8$ &   6.81 & $\Gamma_6$ &  26.84 & $\Gamma_8$  \\
$-14.21$& $\Gamma_6$ &  22.07 & $\Gamma_7$ &   6.90 & $\Gamma_8$ &  28.40 & $\Gamma_7$  \\
$-14.16$& $\Gamma_7$ &  25.31 & $\Gamma_8$ &   8.81 & $\Gamma_8$ &  32.82 & $\Gamma_8$  \\
$-13.54$& $\Gamma_8$ &  60.48 & $\Gamma_6$ &  23.98 & $\Gamma_7$ &  69.38 & $\Gamma_6$  \\
\hline
 \multicolumn{4}{c}{Ba$_2$YMoO$_6$ (1)} & 
 \multicolumn{4}{c}{Ba$_2$YMoO$_6$ (2)} \\
 \multicolumn{2}{c}{L} & \multicolumn{2}{c}{NL} &
 \multicolumn{2}{c}{L} & \multicolumn{2}{c}{NL} \\
\hline
$-20.84$& $\Gamma_8$ &$-20.81$& $\Gamma_8$ &$-22.29$& $\Gamma_8$ &$-22.20$& $\Gamma_8$  \\
  11.39 & $\Gamma_7$ &  12.95 & $\Gamma_7$ &   5.85 & $\Gamma_7$ &   7.37 & $\Gamma_7$  \\
  17.35 & $\Gamma_6$ &  16.36 & $\Gamma_6$ &  15.18 & $\Gamma_6$ &  14.22 & $\Gamma_6$  \\
  20.48 & $\Gamma_8$ &  21.12 & $\Gamma_8$ &  18.94 & $\Gamma_8$ &  19.87 & $\Gamma_8$  \\
  33.06 & $\Gamma_8$ &  32.70 & $\Gamma_8$ &  31.17 & $\Gamma_8$ &  30.70 & $\Gamma_8$  \\
  49.38 & $\Gamma_6$ &  52.16 & $\Gamma_6$ &  42.81 & $\Gamma_6$ &  45.67 & $\Gamma_6$  \\
\end{tabular}
\end{ruledtabular}
\end{table}

\begin{table}[h]
\caption{Zeeman splitting under applied field along the positive direction of the $z$ axis in the unit of $\mu_B |\bm{B}|$.
$\mu_\text{B} = 0.0578838$ meV/T.
}
\label{SMTable:Zeeman}
\begin{ruledtabular}
\begin{tabular}{ccccccc}
 & \multicolumn{2}{c}{Ba$_2$LiOsO$_6$} & \multicolumn{2}{c}{Ba$_2$LiOsO$_6$} & \multicolumn{2}{c}{Ba$_2$YMoO$_6$} \\
 & Elec. & Vibro.  & Elec. & Vibro.  & Elec. & Vibro. \\
\hline
$\mp \frac{1}{2}$ & $\pm$0.450 & $\pm$0.459 & $\pm$0.439 & $\pm$0.439 & $\pm$0.358 & $\pm$0.447 \\
$\mp \frac{3}{2}$ & $\pm$0.150 & $\pm$0.158 & $\pm$0.146 & $\pm$0.152 & $\pm$0.119 & $\pm$0.145 \\
\end{tabular}
\end{ruledtabular}
\end{table}

\begin{figure}[htb]
\includegraphics[width=7cm]{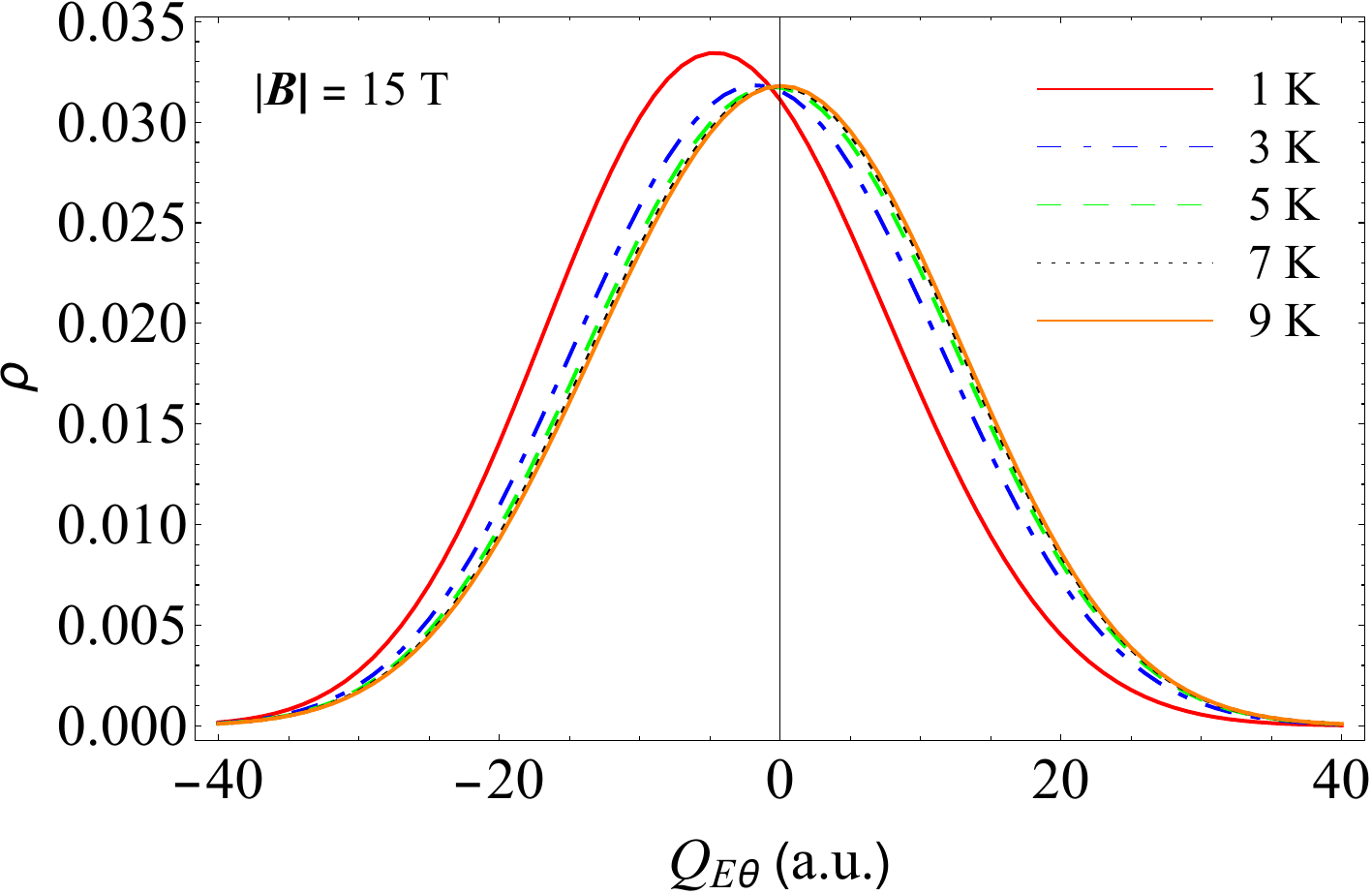}
\caption{$\rho(Q_\theta, T)$ of Ba$_2$NaOsO$_6$ under $\bm{B} =$ (0,0,15 T).}
\label{SMFig:rhoT}
\end{figure}

\section{Vibronic levels}
The low-energy vibronic levels for both linear and nonlinear vibronic Hamiltonians are shown in Table \ref{SMTable:vibro}.
In the case of Ba$_2$YMoO$_6$, to see the effect of covalency, the spin-orbit coupling $\lambda_\text{SO}$ is varied:
The data (1) are with $\lambda_\text{SO}$ from post HF calculations, and the data (2) are with $\lambda_\text{SO} = $ 60 meV. 
The latter is chosen considering the ratio of post HF and DFT $\langle l \rangle$'s.
With the decrease of $\lambda_\text{SO}$, the influence of the pseudo JT coupling increases, and the energy levels vary:
The stabilization energy is enhanced and the first excitation energy with respect to the ground one is slightly reduced. 

\section{Zeeman splitting}
The expectation values of the magnetic moments using the ground $\Gamma_8$ electronic and ground states are listed in Table \ref{SMTable:Zeeman}. 
The temperature evolution of the density of the vibronic states of Ba$_2$NaOsO$_6$ under applied field is shown in Fig. \ref{SMFig:rhoT}.


%